\documentclass{aa}  

\usepackage{graphicx}
\usepackage[varg]{txfonts}
\usepackage{color}
\usepackage{lscape}
\usepackage{rotating}
\usepackage{cancel}
\usepackage{lscape}
\usepackage{ulem}
\usepackage{tabularx}
\usepackage{ltablex,booktabs}
\usepackage{adjustbox}
\usepackage{longtable}
\usepackage{subcaption}
\usepackage{amsmath}
\begin{document}

   \title{Star formation in IC1396:}

   \subtitle{Kinematics and subcluster structure revealed by Gaia \thanks{Tables A.1, D.1, and E.1 are only available in electronic form
at the CDS via http://cdsweb.u-strasbg.fr/cgi-bin/qcat?J/A+A/}}

   \author{Mara E. Pelayo-Baldárrago\inst{1,2}, 
          Aurora Sicilia-Aguilar\inst{2},
          Min Fang\inst{3},  Veronica Roccatagliata\inst{4},\\ Jinyoung Serena Kim\inst{5}, David García-Álvarez\inst{6,7}
          }

   \institute{Departamento de Física Teórica, Facutad de Ciencias, Universidad Autónoma de Madrid, 28049 Cantoblanco, Madrid, Spain\\
              \email{maraelizabeth.pelayo@estudiante.uam.es}
         \and
             SUPA, School of Science and Engineering, University of Dundee, Nethergate, DD1 4HN, Dundee, UK
         \and
            Purple Mountain Observatory, Chinese Academy of Sciences, 10 Yuanhua Road, Nanjing, 210023, PR China
         \and
            University of Pisa, Italy
        \and
            Steward Observatory, University of Arizona, USA  Steward Observatory, University of Arizona, 933 N. Cherry Ave., Tucson, AZ 85721-0065 USA
        \and Instituto de Astrofísica de Canarias, Avenida Vía Láctea, E-38205 La Laguna, Tenerife, Spain
        \and Grantecan S.A., Centro de Astrofísica de La Palma, Cuesta de San José, E-38712 Breña Baja, La Palma, Spain
             }

   \date{Received ; accepted }

 
  \abstract
   {}
   {We investigate the star formation history of the IC1396 region by studying its kinematics and completing the population census.}
   {We use multiwavelength data, combining optical spectroscopy (to identify and classify new members), near-infrared photometry (to trace shocks, jets, and outflows and the interactions between the cluster members and the cloud), along with Gaia EDR3 to identify new potential members in the multidimensional proper motion/parallax space.}
   {The revised Gaia EDR3 distance is 925$\pm$73 pc, slightly closer than previously obtained with DR2. The Gaia data reveal four distinct subclusters in the region. These subclusters are consistent in distance but display differences in proper motion. This, with their age differences, hints towards a complex and varied star formation history. Gaia data also unveil the intermediate-mass objects that tend to evade spectroscopic and disk surveys. Our analysis allows us to identify 334 new members. We estimate an average age of $\sim$4 Myr, confirming previous age estimates. 
   With the new members added to our study, we estimate a disk fraction of 28\%, lower than previous values, due to our method detecting mainly new, diskless intermediate-mass stars. We find age differences between the subclusters, which evidences a complex star formation history with different episodes of star formation.}
   {}


\titlerunning{Star formation history of IC1396}
\authorrunning{Pelayo-Baldárrago et al.}
\maketitle


\section{Introduction}

The IC1396 HII region \citep[also known as S131;][]{Sharpless1959} is part of the large, star-forming Cepheus bubble \citep{Patel1998}, ionized by the multiple system (O5+O9) HD206267 \citep[][]{Peter12,MaizApellaniz2020}.
IC1396 contains the young cluster Tr 37  \citep{Platais1998, Patel1995, Patel1998}, which has been used to estimate the distance to the region, ranging from 870 pc from main sequence (MS) fitting \citep{Contreras2002} to 945$^{+90}_{-73}$ pc based on Gaia DR2 \citep{ Sicilia2019}.
Hundreds of known IC1396 members, mainly young stellar objects (YSOs), have been identified in the region using various techniques. Those include detection of spectral features characteristic of youth, such as Li I absorption or H$\alpha$ emission \citep{Kun1986,Contreras2002,Sicilia2004,Sicilia2005,Barentsen2011, Nakano2012,Sicilia2013}, X-ray emission \citep{Mercer2009A,Getman2012}, near-infrared variability \citep{Meng2019}, infrared \citep{Rebull2013}, and detection of infrared (IR) excess signaling the presence of a disk \citep{Reach2004,Sicilia2006,Morales2009}.

The IC1396 region also contains some bright-rimmed clouds (BRCs) shaped by the UV radiation from HD206267 \citep{Patel1995, Barentsen2011}. These BRCs display evidence of ongoing star formation and are excellent laboratories for studying young stars in different evolutionary stages. The BRCs in the region display a range of velocities (V$_{LSR}=+5$ to -9 km/s) derived from CO molecular maps \citep{Patel1995}. The largest BRCs are IC1396A, IC1396B, and IC1396N \citep[labeled as "E" by][]{Pottasch1956, Patel1995}, all of which contain several tens of solar masses in gas and have different velocities \citep[e.g. $V_{LSR}=-7.9$ for IC1396A,  -5.4 km/s for IC1396B, and 0.6 km/s for IC1396N;][]{Patel1995}. There is also a velocity difference between IC1396A and Tr37 \citep{Sicilia2006_spectro} but, despite of this, the initial Gaia DR2 results revealed no significant discrepancy between the proper motion of the sources physically related to IC1396A and to Tr37 \citep{Sicilia2019}.

The largest globule, IC1396A, contains a moderate population of low-and intermediate-mass stars ranging from a Class 0 source \citep{Sicilia2014, Sicilia2019} to many Class I, II, and III members \citep{Reach2004, Sicilia2006, Getman2007, Barentsen2011}. This has been interpreted as a signature of sequential or triggered star formation \citep{Sugitani1991, Sicilia2005, Sicilia2006, Getman2007, Getman2009, Sicilia2019} related to the expansion of the HII region and its interaction with the surrounding cloud. 
IC1396N also displays recent star formation activity, with many Herbig Haro (HH) objects \citep{Nisini2001}, a large number of H$_2$ jets and knots \citep{Caratti2006, Beltran2009}, and embedded YSOs inside the globule, such as the sources BIMA 1, 2, and 3, two of which are driving molecular outflows \citep{Beltran2009}.

Now, Gaia data \citep{GAIADR22016, GAIADR22018_surveycontent, GAIAEDR32021summary} allows us to explore the astrometry of the known cluster members, which can then be used to identify new members \citep{Lindegren2000, Franciosini2018, Roccatagliata2018, Roccatagliata2020}.
Gaia maps the structure of the region in a multidimensional space, including proper motion and parallax. Using Gaia has the advantage that it can detect objects for which the youth indicators may be less clear (e.g., stars without disks and no accretion, intermediate-mass stars) and can be used to distinguish populations by their astrometry rather than by their location, which is always uncertain due to projection effects. 

In this paper, we use Gaia EDR3 data to provide a new, independent view of IC1396, including the IC1396A, IC1396B, and IC1396N globules. We extract the astrometric properties of the known members, using them to identify new ones.
We also explore the properties of the YSOs, including disk fraction and spatial distribution, and combined 2MASS and WISE data to characterize their disks.
Narrow-band [SII] and H$_2$ photometry are used to trace the presence of shocks, jets, and outflows produced by young embedded members.
Finally, we discuss the advantages and limitations of each method and the improvements of Gaia EDR3 concerning DR2.
In Section \ref{sec:observations} we describe the observations and data reduction. New spectroscopic members are identified in Section \ref{sec:anali_spectra}. In Section \ref{sec:membership}, we study the new members revealed by Gaia and the kinematics of the region. Section \ref{sec:analyphoto} shows a final analysis of the region properties. The results are summarized in Section \ref{sec:conclu}.


\section{Observations and data reduction \label{sec:observations}}

We use ground-based spectroscopy, photometry, and Gaia EDR3 data to study star formation in IC1396. We obtained near-infrared broad- and narrow-band imaging to study the star-cloud interaction and optical spectra to confirm and classify objects that were young star candidates. The details of the observations are given in the following sections.

\begin{figure}
    \centering
    \includegraphics[width=0.47\textwidth]{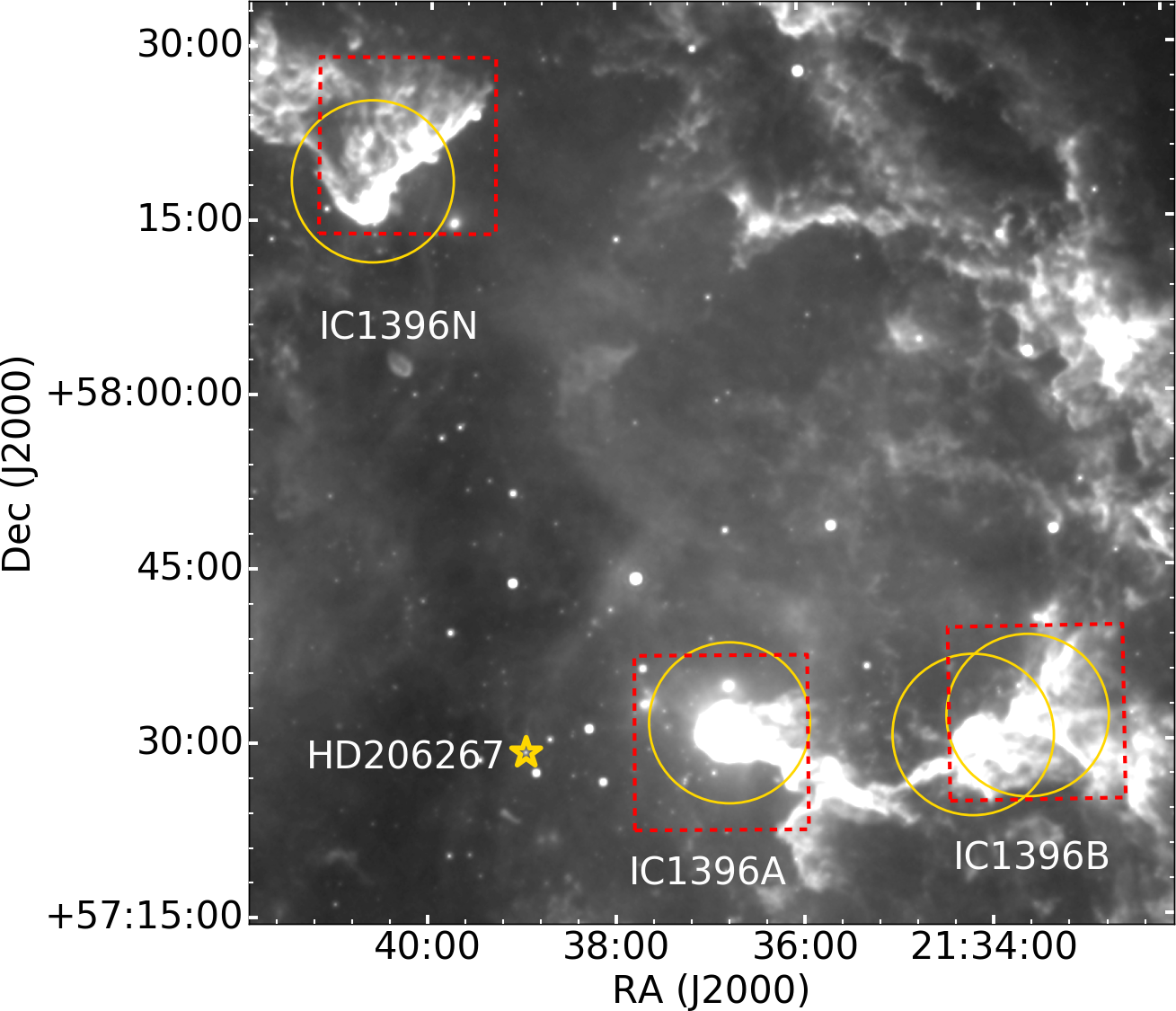}
    \caption{ WISE W4 (22.19 $\mu$m) image of the IC1396 HII region. The BRCs are labeled with their names. The red boxes and yellow circles enclose the fields observed by OMEGA2000 and CAFOS, respectively. Note that the CAFOS FOV has been reduced to account for the fact that the Fabry-Perot etalon has only a small area where the wavelength is accurate. The HD206267 star that is ionizing the region is indicated with a yellow star.}
    \label{IC1396dssred}
\end{figure}

\begin{table*}
    \caption{Summary of CAHA observations}\label{tab:fieldsobserved}
    \centering
    \begin{tabular}{llcclll}
    \hline\hline
     BRCs/    &Instrument  &$\alpha$  & $\delta$ & Date & FOV &Filters\\
     Globule     &   & (J2000)          &  (J2000)      &  & (arcmin)    & \\
     \hline
      IC1396A& OMEGA2000   &21 36 53 & 57 30 29 &2014-09-11/12&15.4' x 15.4' & JHKs, H$_2$, Br$\gamma$ \\
     IC1396A\tablefootmark{*}& CAFOS   &21 36 50 & 57 31 48 &2009-08/09-31/01&16' x 16' & [SII], 6750\\
      IC1396N& OMEGA2000   &21 40 15 & 58 21 03 &2017-06-27/28&15.4' x 15.4'  &Ks, H$_2$\\
             &         &         &          &2017-06-28/29&  &Br$\gamma$ \\ 
             &         &         &          &2017-11-10/12/14& & JHKs\\
      IC1396N& CAFOS   &21 40 43 & 58 17 45 &2017-06-01 & 16'x16' & [SII], 6750 \\
      IC1396B& OMEGA2000   &21 33 33 & 57 32 27 &2017-06-27/28 &15.4' x 15.4' &H$_2$\\
             &         &         &          &2017-06-28/29 & & Ks, Br$\gamma$\\
             &         &         &          &2017-11-09/11/13& & JHKs\\     
      IC1396B& CAFOS   &21 34 13 & 57 30 27 &2017-06-01 & 16'x16' & [SII], 6750\\
      IC1396B& CAFOS   &21 33 37 & 57 31 54 &2017-06-28 &16'x16' &[SII], 6750\\
      \hline 
    \end{tabular}
    \tablefoot{
    \tablefoottext{*} {For CAFOS observations of IC1396A, the data reduction, and results are from \citet{Sicilia2013}.} The [SII] filter (combination of 6717 and 6730 \r{A}) and the line-free continuum at 6750 \r{A} correspond to the Fabry-P\'{e}rot observations.}
\end{table*}


\subsection{Optical and Near-IR Imaging} \label{obs.calaralto}

Since the largest BRCs, IC1396A, IC1396B, and IC1396N, have signs of ongoing star formation, we obtained deep, broad, and narrow-band IR photometry and narrow-band optical observations at the Calar Alto Observatory (Centro Astron\'{o}mico Hispano en Andalu\'{i}a (CAHA), Spain). They included data from proposals F-17-2.2-0.19, F18-3.5-007, and F17-3.5-018 (PI M. Pelayo-Bald\'{a}rrago) and H14-3.5-002 (PI A. Sicilia-Aguilar). The observations are summarized in Table \ref{tab:fieldsobserved}, and the fields covered are shown in Figure \ref{IC1396dssred}. We also include data for IC1396A from \citet{Sicilia2013} to complete the discussion.

We performed JHKs broad-band near-IR photometry using the OMEGA2000 near-infrared wide-field camera \citep{OMEGA2000}\footnote{\scriptsize{http://www.caha.es/CAHA/Instruments/O2000/OMEGA2000\_manual.pdf}} on the 3.5-m telescope.
Its field-of-view (FOV) is large enough to fit each globule within one pointing, achieving a high spatial resolution, sensitivity, and dynamical range ($\sim$9-20mag). We followed the OMEGA2000 pattern of 3 seconds integrations, combining 20 coadds, within a 25-position dither pattern, for a total exposure time of 25 min in the center of the image. The data reduction, including dark and flat subtraction, was done using standard Python Astropy routines \citep{astropy:2013,astropy:2018}. The sky emission was subtracted with Python using sky frames created from the science images. After reduction, images were aligned and combined using AstroImageJ \citep{Collins2017} and the coordinates were derived using Astrometry.net\footnote{\scriptsize{http://nova.astrometry.net/}} \citep{Lang2010A}.

We also obtained narrow-band imaging at 2.122 $\mu$m (H$_2$) and 2.160 $\mu$m (Br$\gamma$) and continuum at 2.144 $\mu$m for all three globules to explore the presence of shock/jets from the embedded population. We obtained 25 dithered positions with 30s exposures and 5 coadds, with a total exposure of 3750s in the center of the image. The sky emission subtraction was from sky frames created from science images. The data reduction was carried out with Python, following the same procedure as for JHKs described above.

To complete the study of feedback in the BRCs, we also obtained narrow-band [S II] images for IC1396N and IC1396B using the Fabry-P\'{e}rot interferometer with the Calar Alto Faint Object Spectrograph (CAFOS) camera\footnote{\scriptsize{https://www.caha.es/es/telescope-2-2m-2/cafos}} at the 2.2 m telescope and following the same setup and procedure used by \citet{Sicilia2013} for IC1396A. The etalon was tuned for a narrow bandwidth of 10 \r{A} around 6717 and 6730 \r{A} (for the [S II] lines) plus an adjacent, line-free continuum at 6750 \r{A}. 

We can only use the central part of the large CAFOS field because the wavelength changes across the Fabry-Pérot FOV. We thus obtained three pointings, one for the IC1396N globule and two for the more extended IC1396B globule. In each case, we obtained three dithered 600s exposures centered on the two [S II] lines and the narrow continuum. Data reduction, including bias and flat fielding, was done with Python (Astropy). We combined all science images with AstroImageJ and assigned the coordinates through Astrometry.net. Finally, we combined both [S II] images at 6717\r{A} and 6730\r{A} to get a better signal-to-noise ratio (S/N) and subtracted the continuum at 6750\r{A} to remove the stellar contribution.


\subsection{MMT/Hectospec Spectroscopy \label{hectospec}}

We obtained the first set of spectra with the multifiber spectrograph Hectospec \citep{Fabricant2005} mounted on the 6.5m Multiple Mirror Telescope (MMT). Hectospec has 300 fibers to assign to individual positions, distributed over a one-degree field of view. It covers a wavelength range from 4000\r{A} to 9000\r{A} with resolution R$\sim$3000.  Our observations include data from proposals 2014b-UAO-S17S and 2015b-UAO-S300 (PI: S. Kim; see Table \ref{tab:Hectoobserved}). We obtained eight fiber configurations, each of which had between $\sim$84 and $\sim$161 fibers for stellar objects and between $\sim$40 and $\sim$80 fibers for sky positions. 

The data were reduced using IRAF routines following standard procedures \citep{Fang2013}. These include flat-fielding and extracting the spectra using dome flats and the IRAF \textit{specred} package (\textit{dofibers} task). The wavelength solution was extracted using the HeNeAr comparison spectra with the IRAF \textit{identify} and \textit{reidentify} tasks and calibrated using IRAF \textit{dispcor} task. For each one of the eight configurations, there were several scientific exposures with exposure times in the range of 450-1500s repeated 3 or 4 times, depending on the magnitude of the stars in the field. Combined those belonging to the same field for the final analysis and sky subtraction (see Table \ref{tab:Hectoobserved}).

\begin{table}
    \caption{Summary of MMT/Hectospec observations}\label{tab:Hectoobserved}
    \centering
    \begin{tabular}{lr}
    \hline\hline
     Observation  & Total integration\\
     Date         & time (s)\\
     \hline
     2014-05-23  & 3060\\
     2014-05-31  & 2700\\
     2014-06-08  & 2700\\
     2015-05-25  & 2700\\
     2015-05-27  & 1800\\
     2015-06-02  & 4800\\
     2015-06-04  & 6000\\
     2015-06-21  & 4000\\
     \hline 
    \end{tabular}
\end{table}

The sky was subtracted using IRAF \textit{sarith}, and following \citet{Sicilia2005, Sicilia2013}. We selected the sky spectra in each setting and combined them using IRAF \textit{scombine}. Because the H II region background is highly variable, we created an average sky spectrum, plus three further templates combining sky spectra visually classified as having bright, intermediate, or faint nebular emission for each configuration. Each sky template resulted from combining $\sim$40-80 spectra for average emission or $\sim$10 for the bright, faint, and medium sky. These low-noise sky templates did a good job for most sources, although in some cases, the sky spectra had to be further re-scaled to remove the H II lines completely. The best result was selected by visual inspection, checking that the forbidden [N II] lines ($\lambda$ $\lambda$ 6548 \r{A}, 6583 \r{A}), which will normally be arising from the HII region, and the telluric 5577 \r{A} line, disappear.
The final outcome of this process was a total of $\sim$900 source spectra with significant S/N, which were later examined (Section \ref{sec:anali_spectra}).


\subsection{GTC / MOS and Long-Slit Spectroscopy \label{gtc}}

The second set of spectra was obtained with the Gran Telescopio CANARIAS \citep[GTC,][]{Rodriguez1998} using the Optical System for Imaging and low Resolution Integrated Spectroscopy \citep[OSIRIS,][]{osiris2003} instrument for doing both long-slit and Multiple Object Spectroscopy (MOS). Our GTC/MOS observations include data from proposals GTC35-14A (PI: M. Fang) and GTC/long-slit observations from proposals GTC55-12A and GTC30-12B (PI: D. Garc\'{i}a \'{A}lvarez)  as part of a campaign to study unknown variable stars.

The GTC long-slit observations were taken in May 2013, and the GTC/MOS were obtained in July 2014. The wavelength range is $\sim$5000-10000\r{A}, and the resolution is R$\sim$2000 for MOS and R$\sim$1900 for the long-slit.
For the MOS spectra, we first cut the data into individual slits. After being extracted, all data reduction was done in the same way for MOS and long-slit spectra, using standard IRAF routines. In total, we extracted 63 spectra from MOS and 18 spectra from the long-slit. Their analysis is presented together with the rest of the spectral data in Section \ref{sec:anali_spectra}.

\subsection{Gaia EDR3 data \label{subsec:DR2EDR3}}

We studied the structure of IC1396 and young star membership using EDR3 data \citep{GAIAEDR32021summary}.
We selected data within a 2$^o$ in radius, centered on HD206267, which contains nearly 1.2 million sources\footnote{See Gaia Archive http://gea.esac.esa.int/archive/}. 
Gaia data is particularly important for the intermediate-mass stars since usual surveys (targeting youth spectral lines features, X-ray, and infrared excesses) have a certain bias against them, which led to them being poorly represented when exploring the initial mass function (IMF) of the region \citep[e.g.][]{Sicilia2005}. We analyzed and compared the Gaia EDR3 and DR2 releases to estimate the differences, focusing the final conclusions on EDR3. 

The membership identification is based on three astrometric parameters: parallaxes ($\varpi$) and proper motions ($\mu_\alpha$, $\mu_\delta$).
We are limited by the reliability of the astrometric solution for faint targets, selecting only those with good-quality astrometry. Gaia EDR3 is complete between G=12 and G=17 mag and has a limit of G=20.7 mag \citep{GAIAEDR32021summary}, having improved slightly for stars fainter than G=18 mag compared to Gaia DR2 \citep{Fabricius2021}. The magnitude limit for reliable data in dense areas is $\sim$18 mag \citep{Luri2018_gaiaparallax}. 
We set a limit on the fractional parallax error of $<$0.25 \citep{BailerJones2015}, so the effective completeness limit drops to G=17 mag. The quality and reliability of the Gaia astrometric data also depend on the Re-normalized Unit Weight Error (RUWE). The RUWE is a statistical indicator obtained from the Unit Weight Error (UWE) or astrometric chi-square that requires a renormalization depending on the magnitude and color of the source \citep[][see their equation 2]{Lindegren2018report}. We only use sources with RUWE<1.4, which have a reliable and consistent astrometric solution \citep{Roccatagliata2020}. Only about 38\% of the known members have Gaia data that are good enough for our analysis. This introduces a bias against detecting new low-mass members but covers the intermediate-mass members that have been less studied.

We also use the Gaia photometry \citep{Riello2021} and the color-magnitude diagrams  \citep{GaiaDR2018HR} and isochrones to constrain the astrometric results and remove further polluting sources. The isochrones used throughout this work are those from the Padova and Trieste Stellar Evolution Code (PARSEC) release v1.2S \citep{Bressan2012}. They were obtained from the CMD 3.6 interactive service from the Osservatorio Astronomico di Padova\footnote{http://stev.oapd.inaf.it/cmd}; using the EDR3 photometric system \citep[filters G, G$_{BP}$, G$_{RP}$;][]{Riello2021}.

\vspace{\baselineskip}
\begin{table*}[]
    \caption{Previously known members now rejected based on Gaia data. Note that they had all been classified as low probability in the literature.}\label{tab:possiblenon-members}
    \centering
    \begin{tabular}{lccl}
    \hline\hline
    Gaia ID & RA (deg) & DEC (deg) & References and comments\\
     & (J2000) & (J2000) & \\
    \hline
    2178443460211017088 & 324.432838 & 57.581183 & SA13 (Probable member), M19\\
    2178384808135636224 & 324.769655 & 57.422599 & SA04, SA05 (Non-member)\\
    2178477304547137536 & 325.250392 & 57.511141 & K86 (weak H$\alpha$), C02 (Non-member), SA06 (star out of the IRAC field). \\
    2178547570213790720 & 324.789941 & 57.906876 & SA13 (Probable non-member)\\
    2178386526122571776 & 324.913273 & 57.461042 & SA13 (Probable non-member)\\
    2178379787317629824 & 325.016674 & 57.314467 & SA13 (Probable non-member)\\
    2178397035896588032 & 324.776032 & 57.469599 & Me09 (Probable non-member)\\
    2178416968852144640 & 324.150531 & 57.392913 & G12 (Diskless)\\
    \hline
    \end{tabular}
    \tablebib{\citet[][K86]{Kun1986}, \citet[][C02]{Contreras2002}, \citet[][G12]{Getman2012}, \citet[][SA04]{Sicilia2004}, \citet[][SA05]{Sicilia2005}, \citet[][SA06]{Sicilia2006}, \citet[][Me09]{Mercer2009A}, \citet[][SA13]{Sicilia2013}, \citet[][M19]{Meng2019}.}
\end{table*}
\vspace{\baselineskip}

\begin{table*}
    \caption{Summary of the total collection of 1536 members of IC1396 found in the literature.}\label{tab:summary_KM}
    \centering
    \begin{tabular}{llr}
    \hline\hline
    Reference      & Methods & Number   \\
                   &        & of sources \\
    \hline
    \citet{Marschall1987}& Astrometric data & 31 \\
    \citet{Kun1990} & H$_\alpha$ emission/optical-IR photmetry& 18  \\
    Sicilia-Aguilar et al (2004, 2005, 2006a,b & Near- and mid-IR observations/optical photometry/spectroscopy/  &435\\
    2010, 2013) & IR excess (Spitzer/2MASS) & \\
    \citet{Barentsen2011} & Photometric H$_\alpha$ survey & 101\\
    \citet{Mercer2009A}  & X-ray and near-IR observations & 29 \\
    \citet{Morales2009}  & Mid-IR variability & 14\\
    \citet{Getman2012}, Getman private com.  & X-ray emission  & 189 \\
    \citet{Meng2019}    & Near-IR variability & 298  \\
    \citet{Nakano2012}   & Wide-field emission-line survey & 408 \\
    \citet{Rebull2013}   & Spitzer mid-IR data  & 13 \\
    \hline
    \end{tabular}
    \tablefoot{Note that several objects have been detected in more than one reference, but we list here the reference providing the most accurate coordinates, which are also the ones listed for sources that do not have Gaia counterparts. The complete collection of 1536 known members (plus two spectroscopic members added from this work) is listed in Appendix \ref{appe:TableKM}, Table \ref{tab:tableKM}.}
\end{table*}

\subsection{Ancillary data \label{subsect:ancillary}}

To study the parallax and proper motion structure of the region, we compiled a collection of objects labeled as cluster members in the literature \citep{Marschall1987,Kun1990,Contreras2002,Mercer2009A,Morales2009,Sicilia2004,Sicilia2005,Sicilia2006,Barentsen2011,Nakano2012,Getman2012,Rebull2013,Sicilia2013, Sicilia2019, Meng2019}. From the original 1544 cluster members, we rejected 8 sources (see Table \ref{tab:possiblenon-members}) that were marked as low-probability members in the literature and/or were in anomalous locations in the parallax/proper motion space. We added to this list our two newly confirmed spectroscopic candidates (see Section \ref{sec:anali_spectra}), summing 1538 known members in total (see Appendix \ref{appe:TableKM}, Table \ref{tab:tableKM}). Summarize the literature member search results in Table \ref{tab:summary_KM}.

We also use 2MASS JHKs data \citep{Cutri2003} to select potential new variable members or characterize the properties of members and mid-IR observations from Wide-field Infrared Survey Explorer \citep[WISE,][]{Cutri2013}, to search for excess emission that could evidence protoplanetary disks. Both datasets were matched to the known member list using 0.6" and 0.7" radii, respectively.


\section{Identification of new spectroscopic members \label{sec:anali_spectra}}

We use the MMT and GTC spectra to identify new members. From an initial collection of nearly 900 spectra, we filtered 121 targets with high S/N that could be potential members. Those are analyzed in more detail below, using four independent membership criteria combined to obtain the final membership. The results derived from each criterion and the final membership for each object are given in Appendix \ref{appe:TableSpectra}, Table \ref{tab:tablespectra}.

\subsection{Characteristic spectral lines of young stars} \label{subsec:spectral_lines}

The most robust youth indicator is the Li I absorption line at 6708 \r{A} \citep{WhiteBasri2003}, but this criterion applies only to fully convective stars, and the line may be hard to confirm in low S/N spectra. Other emission lines indicative of youth are the CaII IR triplet, He I, and strong H$\alpha$ \citep{Hamann1992}. We measured the equivalent widths (EW) of the lines and noise levels using IRAF \textit{splot}.

\begin{figure}
    \centering
    \begin{tabular}{c}
    \includegraphics[height=0.43\textwidth]{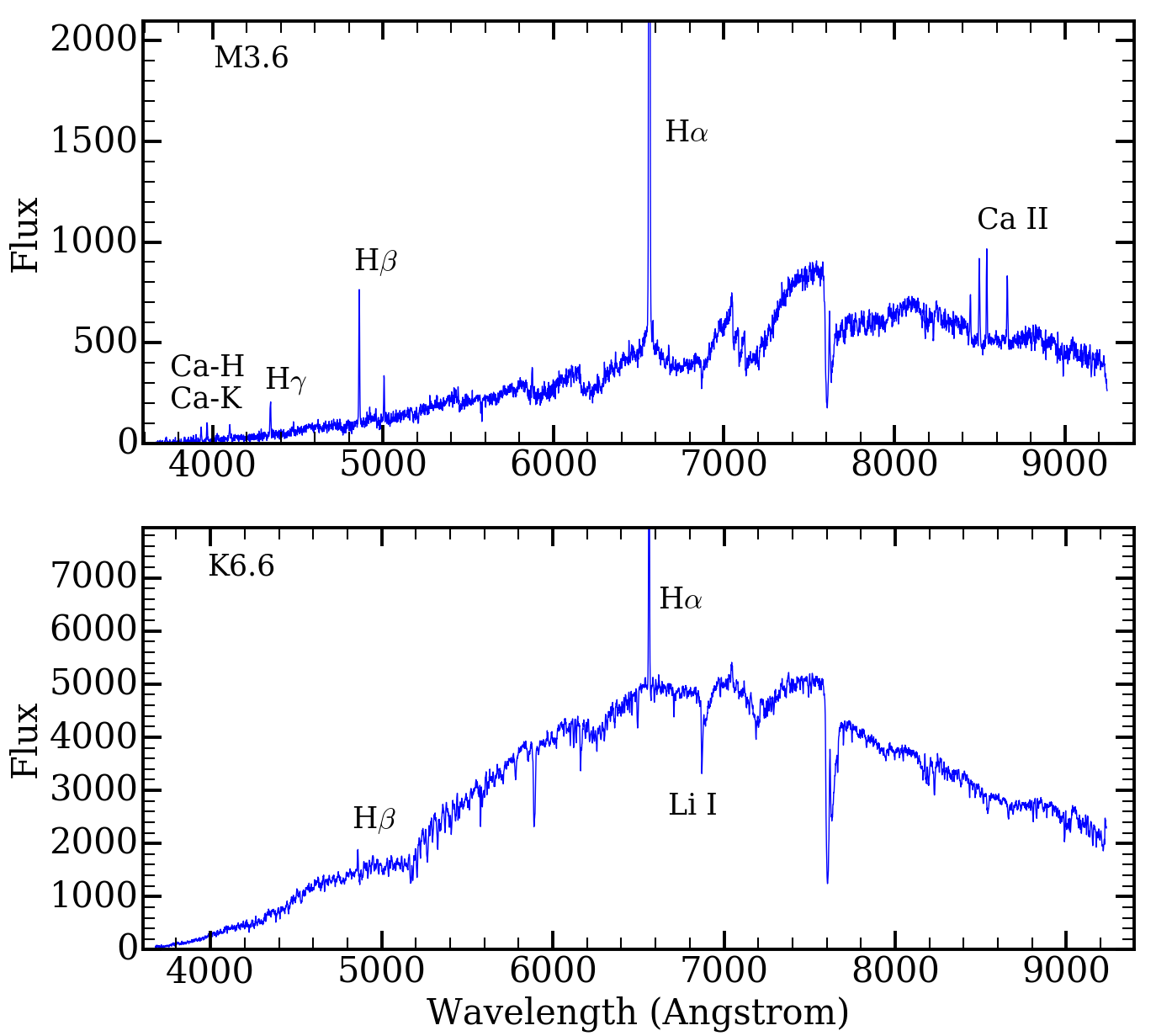}\\
    \end{tabular}
    \caption{MMT (Hectospec) spectra of two confirmed members. The H$\alpha$ emission line is cut off at the top for better visualization. The fluxes are given in arbitrary units.}
    \label{fig:spectra}
\end{figure}

We detected the Li I absorption line and/or Ca II triplet emission lines at $>$3$\sigma$ level in 52 sources. Since the H$\alpha$ emission line is also produced by the HII region, concluding that weak H$\alpha$ emission is of stellar origin is problematic in the fiber spectra \citep[more details in][]{Sicilia2006_spectro}, for which the subtraction of the emission from the HII region may not always be accurate. Therefore, we classify as probable members those objects with large equivalent widths, EW(H$\alpha$)$>$10 \r{A}, plus at least one more membership criterion. This adds up to 49 objects.
The 20 objects with a weak or narrow H$\alpha$ emission, without Li I absorption or Ca II triplet emission lines and no other evidence of membership in the literature (no X-ray emission, no IR excess) are classified as probable non-members. Some spectra examples are shown in Figure \ref{fig:spectra}.
We give a higher priority to this criterion for the final membership classification, so the 52 stars with clear indicators of youth are labeled as confirmed members. 

\subsection{Spectral types and extinction} \label{subsec:ST_A0}

We obtained the spectral type of all 121 candidates following \citet{Fang2020}. We fitted the normalized spectra with two different sets of templates, including observations of pre-main-sequence stars from X-Shooter/VLT \citep[covering a spectral type range from G6 to M9.5, see][]{Min2021}, supplemented by a set of combination of stellar atmosphere models from \citet{Mezaros2012} and \citet{Husser2013}. The spectral templates were artificially veiled and reddened, using the visual extinction law by \citet{Cardelli1989} and assuming R$_V$=3.1. The spectral fit was done considering two cases, with and without veiling. The reduced $\chi{^2}$ is used to estimate the best-fit combination of spectral type and veiling parameter.

We did not use the extinction from the spectral fitting \citep[as pointed out in][]{Fang2020} and instead recalculated the extinction (A$_0$) for the 121 sources using their spectral types and the Gaia colors, which gives more accurate results.
The extinction of the new members is expected to be over a lower limit based on the cluster distance and to follow the extinction distribution of confirmed members \citep{Contreras2002}. Objects outwith the extinction distribution of spectroscopically confirmed members were considered as likely not members (see below). 

We calculated the interstellar extinction, $A_0$ \citep[at $\lambda$= 550 nm,][]{Danielski2018}, using the Gaia photometry and A$_0$=3.1*E(B-V) \citep{GaiaDR2018HR}, so that A$_0$=A$_V$ for the mean extinction law \citep{Cardelli1989}. 
Among 121 candidates the analysis was done only for 119 candidates because two sources lacked complete Gaia EDR3 counterparts. These two sources were labeled as having uncertain extinction (Appendix \ref{appe:TableSpectra}, Table \ref{tab:tablespectra}).
We use the effective temperature vs spectral type relation from \citet[][]{Kenyon1995} and the 4 Myr PARSEC isochrone \citep[since the average age is $\sim$4 Myr,][which we confirm later in Section \ref{subsec:age_mass_NM}]{Sicilia2005, Getman2012}, to obtain the theoretical magnitudes and colors for each spectral type. Those are used to calculate the color excesses E(G$_{BP}$-G$_{RP}$) and E(G-G$_{RP}$). Finally, the interstellar extinction $A_0$ is estimated using the relations from \citet{GaiaDR2018HR} and \citet{Danielski2018}. The final value is obtained as the average of the A$_0$ values derived from the two color excesses for each source, and the uncertainty results from their standard deviation. The values of A$_0$ are consistent for G$<$16 mag, although they show a typical systematic deviation of 0.2 mag. For sources with G$>$16 mag, the $A_0$ values are uncertain due to systematic effects on the Gaia EDR3 bands photometry \citep{Riello2021}. Moreover, there is a trend for the flux in the ${BP}$ band to be overestimated in faint red sources \citep{GAIAEDR32021summary, Riello2021, Fabricius2021}, which appear much bluer in (G$_{BP}$-G$_{RP}$) than they should be. Therefore, their $A_0$ values need to be regarded with care. The correction of these systematic effects in the Gaia bands photometry is beyond the scope of this paper.

\begin{figure}
\centering
\begin{tabular}{c}
\includegraphics[width=0.48\textwidth]{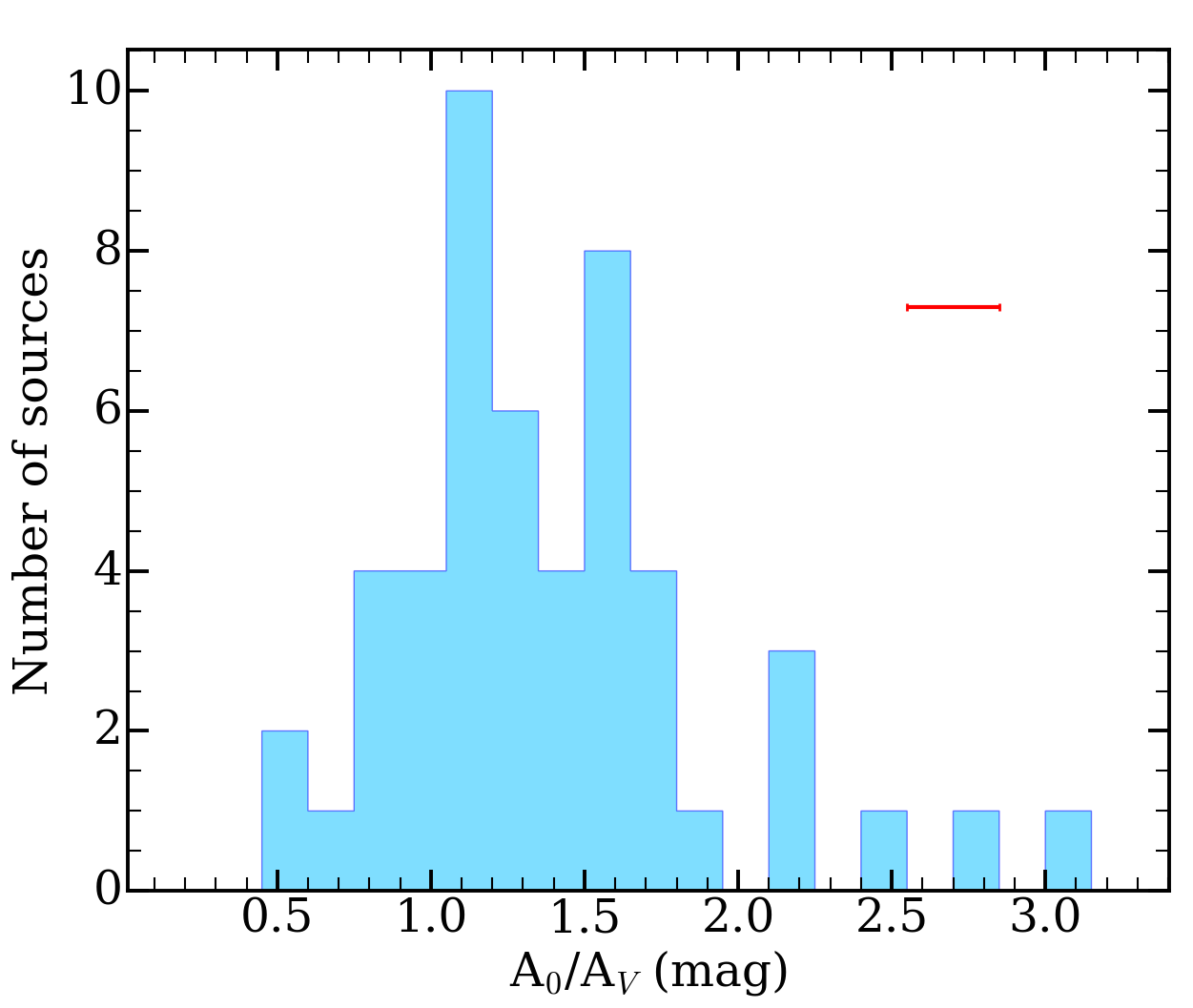}\\
\includegraphics[width=0.48\textwidth]{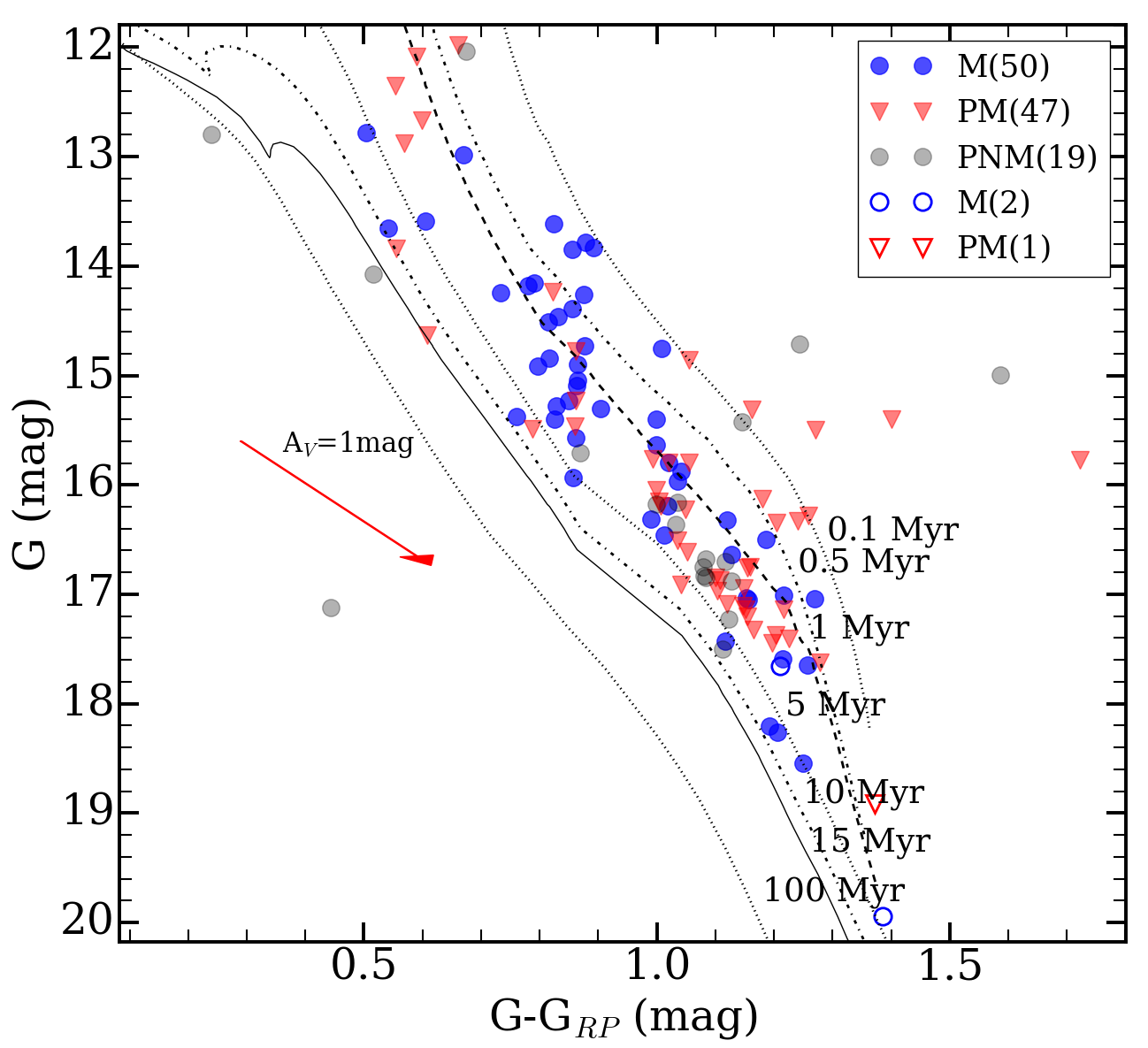}\\
\end{tabular}
\caption{Top: Distribution of the extinction for the spectroscopically confirmed members. The mean error (0.3 mag) is shown with a red bar. Bottom: Color-magnitude diagram for candidates with spectroscopic signatures of youth. Members (M) are marked by blue circles. Probable members (PM) are marked by red triangles. Probable non-members (PNM) are marked by gray circles. The number of objects in each class is also given.
The magnitudes of each object are corrected by their individual extinction (full color symbols), except for objects with non-physical or uncertain extinctions, which have been corrected by cluster average extinction of 1.4 mag (open symbols). The PARSEC isochrones are corrected by the average distance of 925 pc.}
\label{fig:A0_HR}
\end{figure}

We analyzed the extinction distribution for 52 objects with clear Li I and/or Ca II detection (spectroscopically confirmed members, Figure \ref{fig:A0_HR}), to establish the properties of the known members. We obtained an average extinction for the region of $A_0$=1.40 mag with a standard deviation of 0.52 mag (excluding two spectroscopically confirmed members with non-physical extinctions).
This is in good agreement with previous results \citep[A$_V$ $\sim$1.5$\pm$0.5; A$_V$= 1.56$\pm$0.55 for Tr 37; A$_V$ $\sim$1.74 mag, A$_V$
$\sim$1.94$\pm$0.69 mag;][respectively]{Contreras2002, Sicilia2005, Nakano2012, Sicilia2013}. Note that the distribution of extinction values is not Gaussian, extending from 0.5 mag towards larger extinctions up to $<$3.1 mag, as expected due to the presence of dark clouds and to the typical lower limit galactic extinction 1 mag/kpc distance. With this in mind, we consider as members those with extinctions in the range 0.9 mag $<$ A$_0$ $<$ 3.0 mag, which comprises our second membership criterion.

Targets in this extinction range with one additional membership criterion are considered sure members. Without an additional membership criterion, sources with a consistent extinction are considered probable members. Those outside the 0.9-3 mag range without other membership criteria are considered probable non-members. Note that an Anderson-Darling test \citep{Stephens1974} reveals that the extinction distribution of members and probable non-members do not exclude each other. This is likely caused by cluster members and background stars having often similar extinctions in the 1-3 mag range, which makes it hard to rule out objects following the extinction criterion alone.

We use the G vs. G-G$_{RP}$ diagram as a further consistency test (Figure \ref{fig:A0_HR}), correcting the objects by their individual extinctions. As explained before, the G-G$_{RP}$ color offers better results, especially for faint red sources.
We obtained three sources with negative (nonphysical) color excesses, which are labeled as uncertain in Appendix \ref{appe:TableSpectra}, Table \ref{tab:tablespectra}. Two of them are confirmed members (with clear Li I and/or Ca II detection), and another is a probable member. These are faint M-type sources and with Gaia magnitudes G$>$17 mag (before correcting for extinction), beyond the completeness limit of the survey \citep{GAIAEDR32021summary}. Anomalous extinction can be caused by circumstellar material, scattering in the disk, spectral veiling, and/or uncertain spectral types. This is also a further sign that extreme care has to be taken with faint objects when determining ages or masses using the Gaia colors. For these three sources with unreliable extinction, we use the cluster average extinction (A$_0$=1.4 mag) in the color-magnitude diagram (Figure \ref{fig:A0_HR}).

\subsection{Presence of protoplanetary disks}
\label{subsec:disk_spectra}

We use as a third membership criterion the presence of protoplanetary disks or/and accretion, or X-ray emission \citep[indicator of coronal activity, see][]{Feigelson2003, Mercer2009A, Getman2012}. YSOs with a full circumstellar disk have excess emission at $\lambda>$2$\mu$m. However, IR excesses alone need to be treated carefully as it could also correspond to post-main-sequence stars, background quasars, or surrounding nebulosity. We use mid-infrared photometry from the WISE catalog, considering only W1 [3.4$\mu$m] and W2 [4.6$\mu$m] bands because W3 [11.6$\mu$m] and W4 [22.1$\mu$m] are affected by the extended variable background emission for clusters at large distances \citep{Ribas2014}. The WISE data were used in combination with the JHKs 2MASS data. We also check whether any disk signatures have been previously reported in the literature. 

We consider as IR excess $>$3$\sigma$ deviations from the zero-color of stars without disks in both W1-W2 and H-Ks. Only 67 spectroscopic sources have complete IR data with low uncertainties ($<$0.05 mag). The W1-W2 vs H-K diagram (Figure \ref{fig:W1W2_HK_spectra}) reveals 27 targets consistent with having a disk (see Table \ref{tab:tablespectra} in Appendix \ref{appe:TableSpectra}). Five have not been previously classified as disked stars, for another four have confirmed the presence of a disk. The rest had been previously classified as disked stars by other methods \citep{Morales2009, Barentsen2011, Getman2012, Nakano2012, Sicilia2013}, in addition, our work now adds further information such as spectral types and extinction.

Since most disks are expected to accrete, we contrasted this criterion with the veiling value calculated and accretion emission lines found (CaII and a strong H$\alpha$). Among the 27 sources with disk, 52\% (14) show veiling and accretion lines, and 33\% (9) show only accretion lines (strong H$\alpha$ emission, EW(H$\alpha$)$>$10\r{A}). Two sources clearly display the Li I absorption line, and another two sources have weak H$\alpha$ emission. We complete this list with 23 more sources listed as disk-bearing in the literature. From these 23 sources, 11 show veiling plus accretion lines (H$\alpha$/Ca II), and 9 show accretion-characteristic lines only.
Note that the main difference between the disks we found and the literature is that we do not use any wavelength longer than W2, which leaves sources with anemic and/or transitional disks out of our list.

The veiling measurements allow us to estimate the accretion rate for 25 sources. Assuming that the veiling is roughly constant as a function of wavelength \citep{Hartigan1989, Dodin2013}, we can derive a relation between veiling and accretion luminosity \citep[as in][]{Calvet2000} using the data from \citet{Gullbring1998}. As with other accretion estimators, the dispersion is large ($\sim$0.5 dex). We then estimate the accretion rates ($\dot{M}_{acc}$) assuming that the magnetospheric infall radius is 5$R_*$ \citep{Gullbring1998}, and with stellar masses and radii derived from the pre-main sequence (PMS) evolutionary tracks (see Section \ref{subsec:age_mass_NM}).
The results (Table \ref{tab:veiling}) agree with typical accretion rates of Classical T Tauri stars \citep[CTTS; 10$^{-9}$ to 10$^{-8}$ M$_\odot$/yr,][]{Fang2009, Sicilia2010} and with the presence of disks. Among the 25 sources with veiling measurements, 20 are confirmed members with accretion-related lines and disks, and 5 have a low S/N spectrum and/or uncertain spectral types, so their veiling measurements are uncertain, but the overall veiling-disk trend is coherent.

\begin{figure}
    \centering
    \begin{tabular}{c}
    \includegraphics[width=0.48\textwidth]{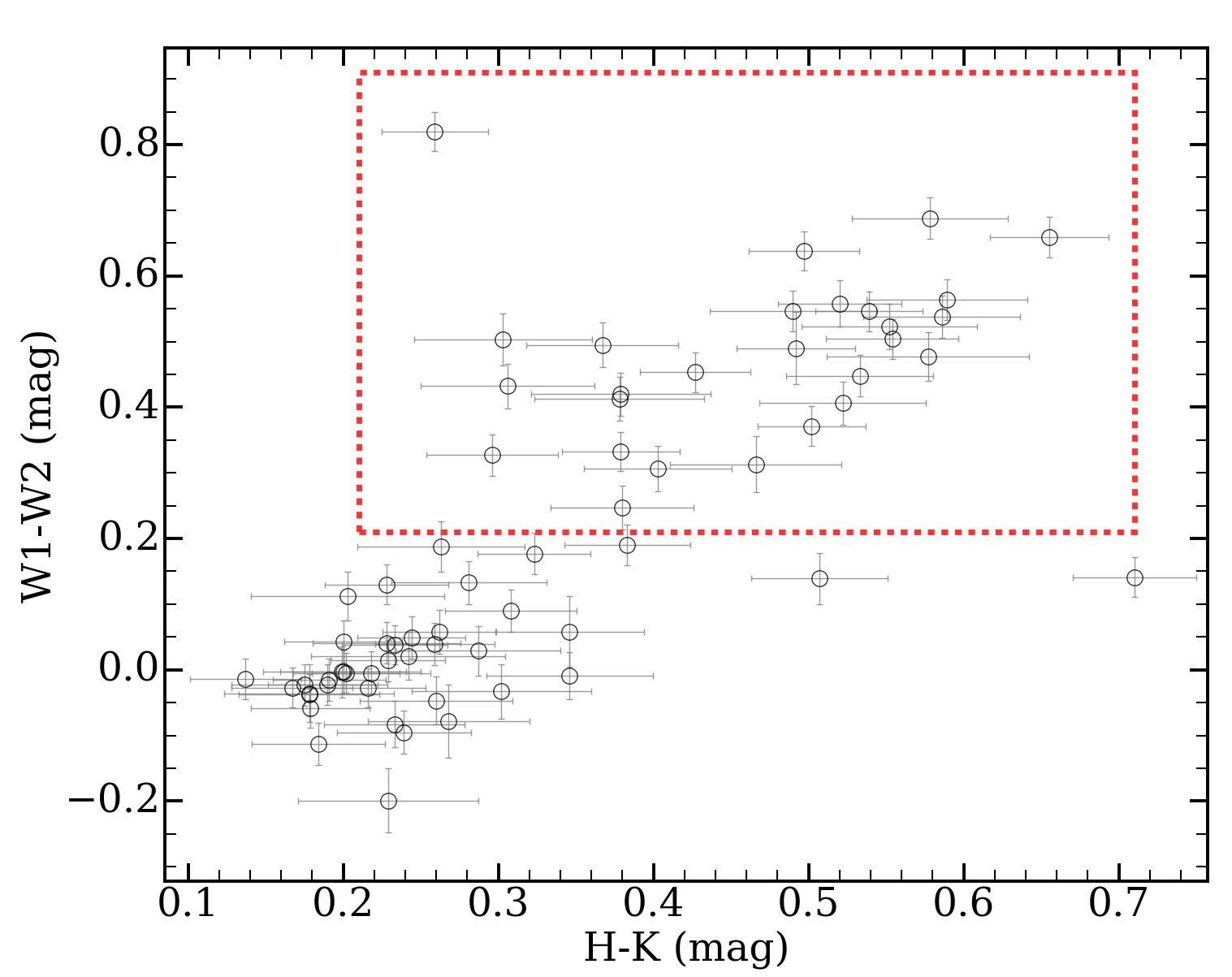}
    \end{tabular}
    \caption{W1-W2 vs. H-K diagram for the 67 spectroscopic candidates with errors $<$0.05 mag. The red box shows the 27 candidates with excess emission in the IR consistent with a disk.
    \label{fig:W1W2_HK_spectra}}
\end{figure}

\begin{table}[]
    \caption{Veiling and accretion rates for spectroscopic candidates.}\label{tab:veiling}
    \centering
    \begin{tabular}{lccc}
    \hline\hline
Gaia ID &Sp.  & Veiling & \.{M}\\
        &Type & r$_{7465}$ & (M$_\odot$ yr$^{-1}$)\\
\hline
2178495549568315776 & K6.6 & 0.41 & 6$\times$10$^{-9}$ \\ 
2178296744121734912 & K7.4 & 0.33 & 4$\times$10$^{-9}$ \\ 
2178450160360276480 & K8.0 & 0.27 & 8$\times$10$^{-9}$ \\ 
2179207998742440960 & K8.2 & 0.09 & 4$\times$10$^{-9}$ \\ 
2178432671252489984 & M2.6 & 0.12 & 6$\times$10$^{-9}$ \\ 
2178458986501139840 & M3.0 & 0.73 & 7$\times$10$^{-9}$ \\ 
2178433805123877504 & M3.6 & 0.23 & 5$\times$10$^{-9}$ \\ 
2178433942562827136 & M3.6 & 0.23 & 6$\times$10$^{-9}$ \\ 
2178547776372232704 & M4.0 & 0.86 & 4$\times$10$^{-8}$ \\ 
2178385838927821440 & M4.2 & 0.15 & 3$\times$10$^{-9}$ \\ 
2179218096198430848 & M4.4 & 0.83 & 9$\times$10$^{-9}$ \\ 
2178434148721240320 & M4.8 & 0.55 & 7$\times$10$^{-9}$ \\ 
2178444177454554496 & M4.8 & 0.64 & 5$\times$10$^{-8}$ \\ 
2178443254045917568\tablefootmark{*} & M5.6 & 1.95 & 4$\times$10$^{-7}$ \\ 
2178385735848603776 & M5.0 & 0.38 & 8$\times$10$^{-9}$ \\ 
2178393771716897536 & M2.2 & 0.26 & 5$\times$10$^{-9}$ \\ 
2178421671829472256 & M6.2 & 0.72 & 2$\times$10$^{-8}$ \\ 
2179285376874962304\tablefootmark{*} & K0.0 & 2.58 & 7$\times$10$^{-7}$ \\ 
2178441604784967168\tablefootmark{**} & K9.6 & 0.09 & 6$\times$10$^{-9}$ \\ 
2179215832763067648 & M2.6 & 0.17 & 4$\times$10$^{-9}$ \\ 
2178398070997784960 & M3.2 & 0.97 & 2$\times$10$^{-8}$ \\ 
2179216824887982592 & M3.6 & 0.30 & 7$\times$10$^{-9}$ \\ 
2178434767196685440 & M4.0 & 0.53 & 1$\times$10$^{-8}$ \\ 
2179224212235705344\tablefootmark{**} & M4.2 & 0.13 & 1$\times$10$^{-8}$ \\ 
2178441428675371136\tablefootmark{**} & M5.2 & 0.13 & 2$\times$10$^{-8}$ \\
\hline
    \end{tabular}
    \tablefoot{
    \tablefoottext{*}{Sources with uncertain veiling measurements. The M5.6 source has a low S/N spectrum, so the veiling measurement is uncertain. The K0.0 source has good S/N, but the spectral type is uncertain, and so is the veiling.}\\  \tablefoottext{**} {Sources with a low S/N spectrum, so their veiling measurements and accretion rates are uncertain.}}
\end{table}

\subsection{Membership constraints from Gaia} \label{subsec:membGaia}

We also used Gaia EDR3 data, with the maximum likelihood and Mahalanobis distance analysis \citep{Mahadistance1936, Mahalanobis1969} as an independent membership test. The method is explained in detail in Section \ref{sec:membership}, here we simply refer to that information to include an additional criterion for the spectroscopic candidates.
Among the 120 candidates with Gaia EDR3 counterparts, we selected those with good quality astrometric data (see section \ref{subsec:DR2EDR3}) and imposed the same conditions as for our probability analysis (see Section \ref{subsec:likelihood}), and the same restrictions in the uncertainties used in Section \ref{subsec:Mahalanobis}, which leaves us with 83 sources.

A total of 38 of them satisfy the Mahalanobis criterion, having a probability of 95\% of belonging to the region. 
Of the remaining 45 objects with good quality astrometric data, 28 are rejected because of their large errors, 3 are out of the age range to be considered as new members, and 14 do not belong to any subcluster of IC1396 (see Section \ref{subsec:likelihood}). Note that the Mahalanobis criterion will lose bona-fide members due to the very stringent constraints imposed to limit contamination (see Section \ref{subsec:Mahalanobis}). Therefore, this criterion is used only to look for extra information for uncertain cases and not to cast doubts on the membership of objects confirmed by spectral lines, disks, or X-ray emission. The criterion can obviously not be applied to the sources with Gaia data below the quality threshold.

\subsection{Combination of all membership criteria }

Combining all four membership criteria for the spectroscopic candidates, we obtain a list with the final membership listed in Appendix \ref{appe:TableSpectra}, Table \ref{tab:tablespectra}. As mentioned before, we give a higher priority to sources with spectroscopic lines indicative of youth. The final membership for the rest of the sources is defined depending on whether two or more criteria confirm the membership. We classify as probable members those sources that satisfy two or more criteria at the probable level. 

From the initial list of 121 candidates, 66 are classified as members, 42 as probable members, and 13 as probable non-members. Two of the confirmed members and three of the probable members had not been previously classified as IC1396 members. The other 116 spectroscopic sources correspond to objects previously listed in the literature. In addition, our spectroscopic analysis supplies information on the previously unknown spectral types for 111 targets, accretion status for 25 targets, and interstellar extinction for 119 sources.

Among the 13 probable non-members, four have spectral types in the A-G type range. They also have uncertain Gaia photometry (G$\geq$18 mag). Nine sources are M-type stars without clear indications of youth (such as weak H$\alpha$ emission or Li I line absorption too close to the noise level), which could be dMe stars. In addition, 4 of these have extinction outside the membership range we consider. From these 13 sources rejected as probable non-members, 6 were previously classified as members, 3 as uncertain, and the remaining one as a non-member.


\section{Characterizing IC1396 with Gaia} \label{sec:membership}

\begin{table*}[]
    \caption{Samples used with Gaia EDR3 data from our IC1396 stellar member analysis.}\label{tab:samples_EDR3}
    \centering
    \begin{tabular}{lclc}
    \hline\hline
    Sample & Nr. sources & Criteria & Code\tablefootmark{*} \\
    \hline
    KM plus 2 new spectroscopic members & 1538 & Listed in the literature +Spectroscopy & KM\\
    KM used for the maximum likelihood analysis & 578 & Fractional parallax error (f$<$0.25)&--- \\
     & & RUWE $<$ 1.4.& \\
     & & Parallax (0.6-1.6 mas) &\\
    New members obtained from the Mahalanobis distance  & 334 & Errors magnitude $<$0.05 mag & NM\\
    & & Errors proper motion $<$0.1 mas/yr & \\
    & & Age cut (0.1-10 Myr) & \\
    \hline
    \end{tabular}
    \tablefoot{
    \tablefoottext{*}{The code will be used throughout the paper for a better understanding of the sample used in the analysis.}}
\end{table*}

In this section, we use Gaia data to study the distance and kinematics of the known members of IC1396, using this information to find new members. We also estimate the age, mass, and the presence of disks among the entire population. The analysis is done using the EDR3 data, although we compare and comment on the differences with DR2.

\begin{table*}
    \caption{Results of the maximum likelihood analysis using the Gaia EDR3 data, showing the best-fit subcluster positions and standard deviations in parallax-proper motion space.}\label{tab:resultslikelihoodDR3}
    \centering
    \begin{tabular}{ccccccccc}
    \hline\hline
    \multicolumn{9}{c}{EDR3}\\
    \hline
     Subcluster/&$\varpi$ &$\sigma_{\varpi}$& $\mu_\alpha$ &$\sigma_{\mu_\alpha}$ &$\mu_\delta$ &$\sigma_{\mu_\delta}$& Nr. of&Distance\\
     Population & (mas)   & (mas)  &(mas/yr)& (mas/yr)   &(mas/yr)&(mas/yr)&Stars &(pc) \\
     \hline
     A & 1.101$\pm$0.006 & 0.130 & -2.432$\pm$0.020 & 0.412 & -4.719$\pm$0.015 & 0.308 & 418 &  908$\pm$73 \\ 
B & 1.098$\pm$0.022 & 0.128 & -2.251$\pm$0.063 & 0.360 & -3.131$\pm$0.044 & 0.252 &  33 &  911$\pm$75 \\ 
C & 1.002$\pm$0.013 & 0.035 & -1.835$\pm$0.034 & 0.091 & -6.603$\pm$0.091 & 0.241 &   7 &  998$\pm$61 \\ 
D & 0.963$\pm$0.030 & 0.090 & -3.876$\pm$0.087 & 0.260 & -2.954$\pm$0.100 & 0.301 &   9 & 1038$\pm$48 \\ 
E & 1.075$\pm$0.016 & 0.060 & -1.374$\pm$0.036 & 0.133 & -3.424$\pm$0.054 & 0.201 &  14 &  931$\pm$62 \\ 
F & 1.092$\pm$0.032 & 0.085 & -3.632$\pm$0.046 & 0.121 & -3.883$\pm$0.054 & 0.143 &   7 &  916$\pm$46 \\ 
G & 0.997$\pm$0.029 & 0.229 & -1.909$\pm$0.645 & 5.159 & -3.124$\pm$0.524 & 4.192 &  64 & 1003$\pm$70 \\ 
    \hline 
    \end{tabular}
    \tablefoot{The parallax and proper motion errors are listed here as the standard errors in the mean for each parameter, which define the mean properties of each subcluster. The $\sigma_{\varpi}$, $\sigma_{\mu_\alpha}$ and $\sigma_{\mu_\delta}$ are the standard deviation or the intrinsic dispersion of the astrometric parameters. Note the large scatter in the parameters of population G, which renders it unusable to identify new members. The final column shows the average distance for each group and its error.}
\end{table*}

\subsection{Selecting the best known members in IC1396 } \label{subsec:best_KM}

The first step is a proper characterization of the parallax and kinematics of the well-known members. A simple proper motion plot revealed deviations among the known cluster members suggestive of several sub-structures \citep[see also][]{Sicilia2019}. We also expected differences in proper motion based on the radial velocity variations between different BRCs and the main Tr37 cluster \citep{Patel1995}. 

We then studied the presence of distinct astrometric groups following \citet{Franciosini2018} and \citet{Roccatagliata2020}. Significant clustering is identified by maximizing the likelihood function for the region known members, assuming that they are distributed in parallax and proper motion space following a number of multivariate Gaussians \citep{Lindegren2000} that represent the position in parallax-proper motion space of each group.
Iterating the procedure for various numbers of clusters allows us to determine which structures are significant and their corresponding parallaxes and proper motions. Details are given in Appendix \ref{appe:MLF}.

The method requires two inputs, a collection of well-known members (rejecting all those that may be uncertain) and an initial subset of potential subclusters with their astrometric parameters (parallaxes and proper motions), their standard deviations, and the number of stars in each subcluster.
We started with the 1538 known members (Table \ref{tab:samples_EDR3}). We cross-matched (using a 0.9" radius and removing sources with multiple matches) them to the Gaia data and imposed the same restrictions as in Section \ref{subsec:DR2EDR3}. We consider as reliable members those with parallaxes between 0.6-1.6 mas, obtaining 578 from EDR3 (see Appendix \ref{appe:TabML_EDR3}, Table \ref{tab:tableML_EDR3}), an increase with respect to the 536 objects with good DR2 data. These represent a very stringent selection of reliable members, which is needed to avoid introducing uncertainties via non-members or sources with unreliable astrometry. They were used to define the properties of the subclusters, but we must keep in mind that they represent only 53\% of the known members with Gaia EDR3 counterparts. Thus, we expect that our search for new members based on them may miss a similar number of cluster members with poorer Gaia data. From the 578 known members, the mean parallax is 1.081$\pm$0.006 mas, and the average distance to the region falls to 925$\pm$73 pc. This value is consistent with the Gaia DR2 result of 943$\pm$82 pc, although the difference could reflect the known biases in the DR2 dataset \citep{Stassun2018}.

The main difference between DR2 and EDR3 is that the uncertainties in the EDR3 data are significantly smaller \citep{GAIAEDR32021summary}, especially on the proper motion. Proper motion uncertainties are typically reduced by half compared to DR2. We also gain 6\% (85) members, and 5\% of the DR2 sources are rejected because they fail to meet the goodness criteria (e.g., RUWE$>$1.4, parallax values outside the range of stated values, or fractional error higher to 0.25). This number reflects the potential contamination of the members listed in the literature, although the fraction is small.

The initial subcluster parameters were obtained by examining the 3D astrometric space positions for the stringent collection of known members. The visually-identified structures allowed us to create an initial conditions file with 10 potential subclusters, the significance of which is then refined by maximizing the likelihood function. The likelihood function is computed by shifting the positions of these initial subclusters and revising them according to the data. Thus, if the initial subcluster parameters are reasonable and the sampled parameter space is large enough, this method will correct any imprecision in the initial subcluster positions. Subclusters that are not significant will also be automatically removed, as explained below.

\subsection{ The Maximum Likelihood algorithm for clustering} \label{subsec:likelihood}

The significance of the subcluster structure is determined via maximizing the likelihood function using a customized Python routine. The algorithm uses the initial subclusters file, where each j-th subcluster is defined, in a the multidimensional space, by seven astrometric parameters: the mean parallax ($\varpi_{j}$), the mean proper motion in right ascension and declination ($\mu_{\alpha,j}$, $\mu_{\delta,j}$), their intrinsic dispersion for the stars in the subclusters ($\sigma_{\varpi,j}$, $\sigma_{\mu_\alpha,j}$, $\sigma_{\mu_\delta,j}$ ), and the fraction stars that belong to each subcluster (fs$_j$).  We follow the same formulation used by  \citet{Roccatagliata2018, Roccatagliata2020} where the probability of each i-th star belongs to the j-th subcluster is given by
\begin{equation}
P_{i,j}=fs_{j} \frac{L_{i,j}}{L_i}
\end{equation}
where L$_{i,j}$ is the likelihood function for the i-th star belonging to the j-th subcluster (see below), L$_i$ is the total likelihood, and fs$_j$ is the fraction stars of belonging to the j-th subcluster. The individual likelihood L$_{i,j}$ is calculated as 
\begin{equation}
    L_{i,j}=(2\pi)^{2/3}|C_{i,j}|^{1/2}exp[-\frac{1}{2}(a_i-a_{j})^TC_{i,j}^{-1}(a_i-a_{j})],
    \label{eq_likelihood}
\end{equation}
and the total likelihood is obtained by summing all the individual ones multiplied by their corresponding fraction of stars for the total of n subclusters, i.e.
\begin{equation}
    L_i=\sum_{j=1}^n fs_jL_{i,j}.
\end{equation}
In equation (\ref{eq_likelihood}), $C_{i,j}$ is the covariance matrix, $|C_{i,j}|$ is its determinant, $a_i$ corresponds to the vector for the i-th star in the multiparameter parallax/proper motion space, $a_i$=[$\varpi_{i}$ $\mu_{\alpha,i}$ $\mu_{\delta,i}$], and $a_{j}$ is the equivalent vector for the subcluster parameters, $a_{j}$=[$\varpi_{j}$ $\mu_{\alpha,j}$ $\mu_{\delta,j}$]. Transposition is denoted by $^T$. Further details are given in Appendix \ref{appe:MLF}. 
Once the multidimensional location of the clusters is known, it can be used to identify new members using the Mahalanobis distance technique, which we describe in Section \ref{subsec:Mahalanobis}.

We calculate the likelihood for the well-known cluster members by varying the initial subcluster positions over a grid of initial parallax and proper motions distributed over $\pm$2$\sigma$ around the initial subcluster values \citep{Lindegren2000}. We tested various grid samplings, ranging from 14 to 34 points. In each loop, we first use the values of the initial parameter file as a guess solution, modifying them around the grid mentioned above to estimate the likelihood. The guess value is then refined once to allow for differences in the number of stars in each cluster \citep{Franciosini2018} and to revise the central position of each subcluster parameter for objects that belong to the cluster. All the combinations of parallax and proper motion over the $\pm$2$\sigma$ space are tested, and each position of the subcluster is revised in turn. Therefore, as long as the initial parameter file is sensible and the number of grid points is reasonable, the final values are not significantly dependent on the initial cluster positions nor on the number of points in the grid. 

\begin{figure*}
    \centering
    \begin{tabular}{cc}
    \includegraphics[width=0.46\textwidth]{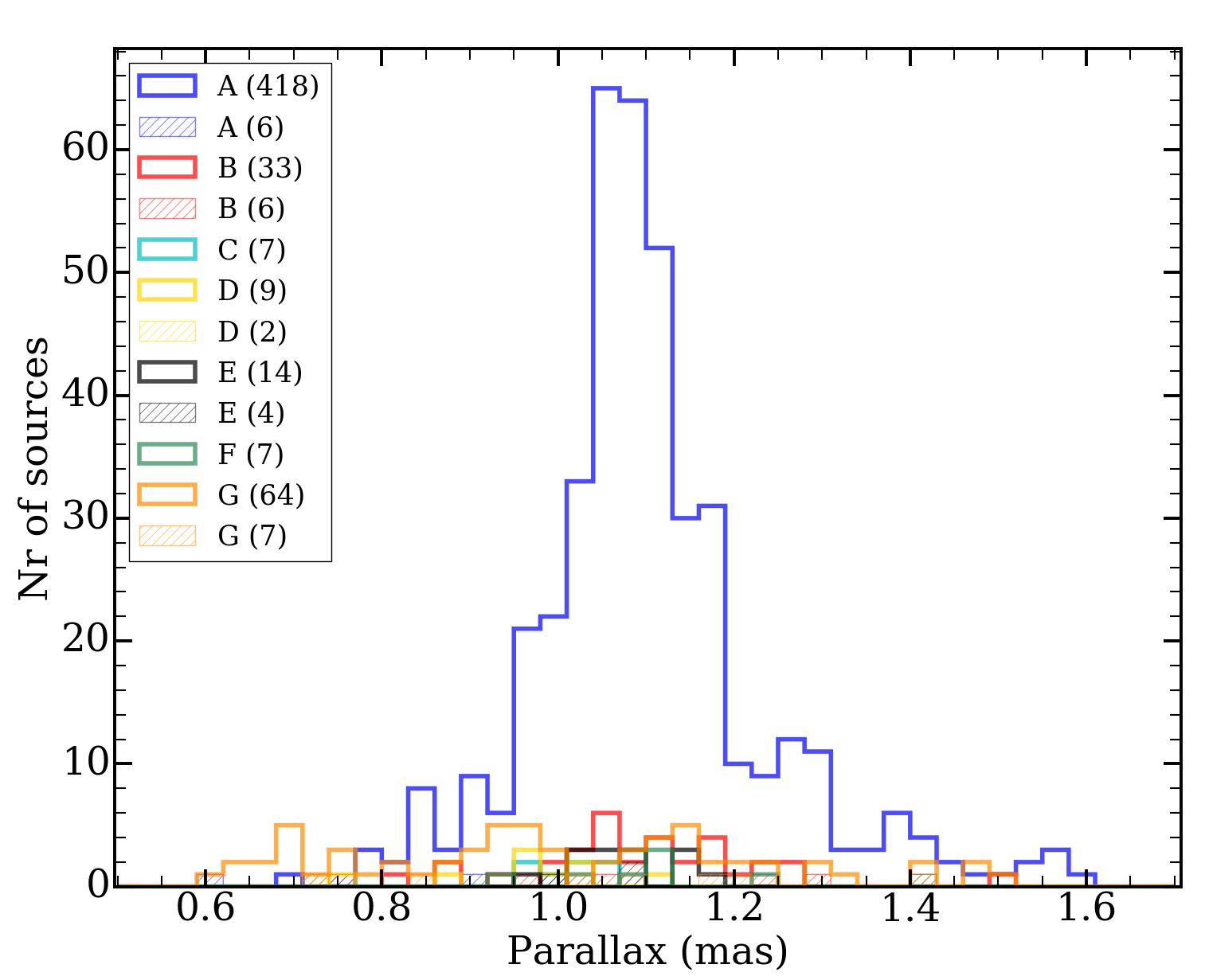}
    \includegraphics[width=0.46\textwidth]{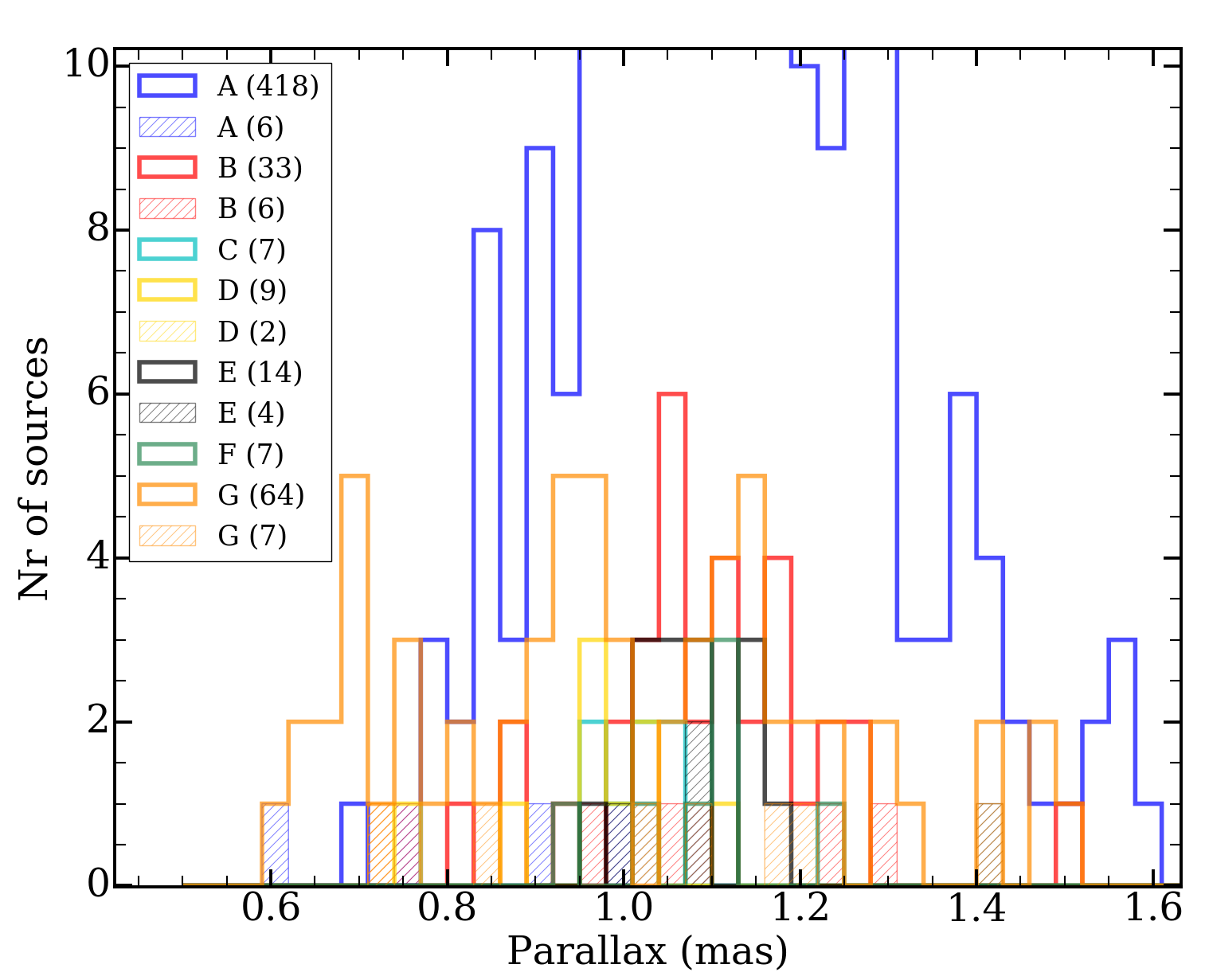}\\
    \includegraphics[width=0.46\textwidth]{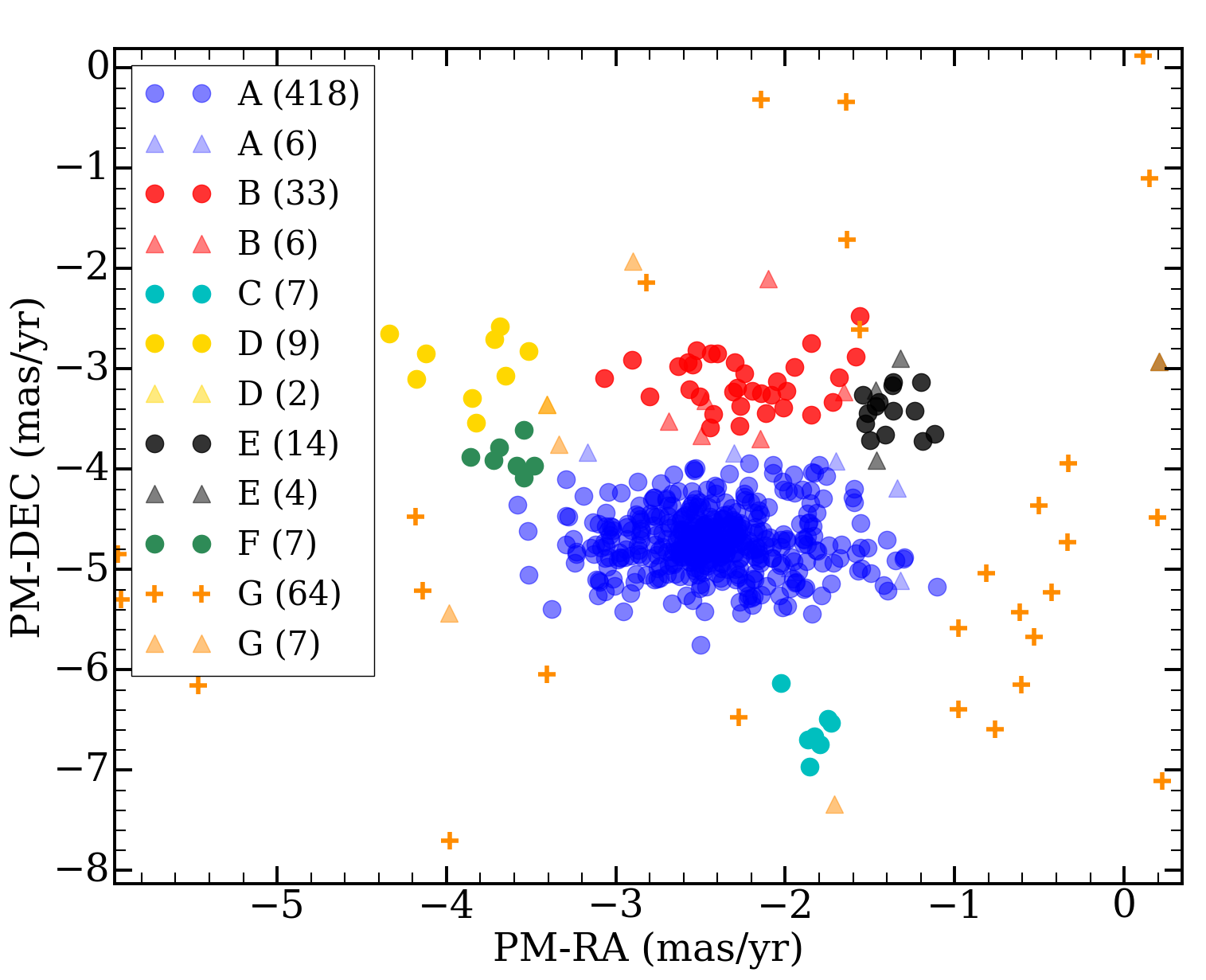}
    \includegraphics[width=0.46\textwidth]{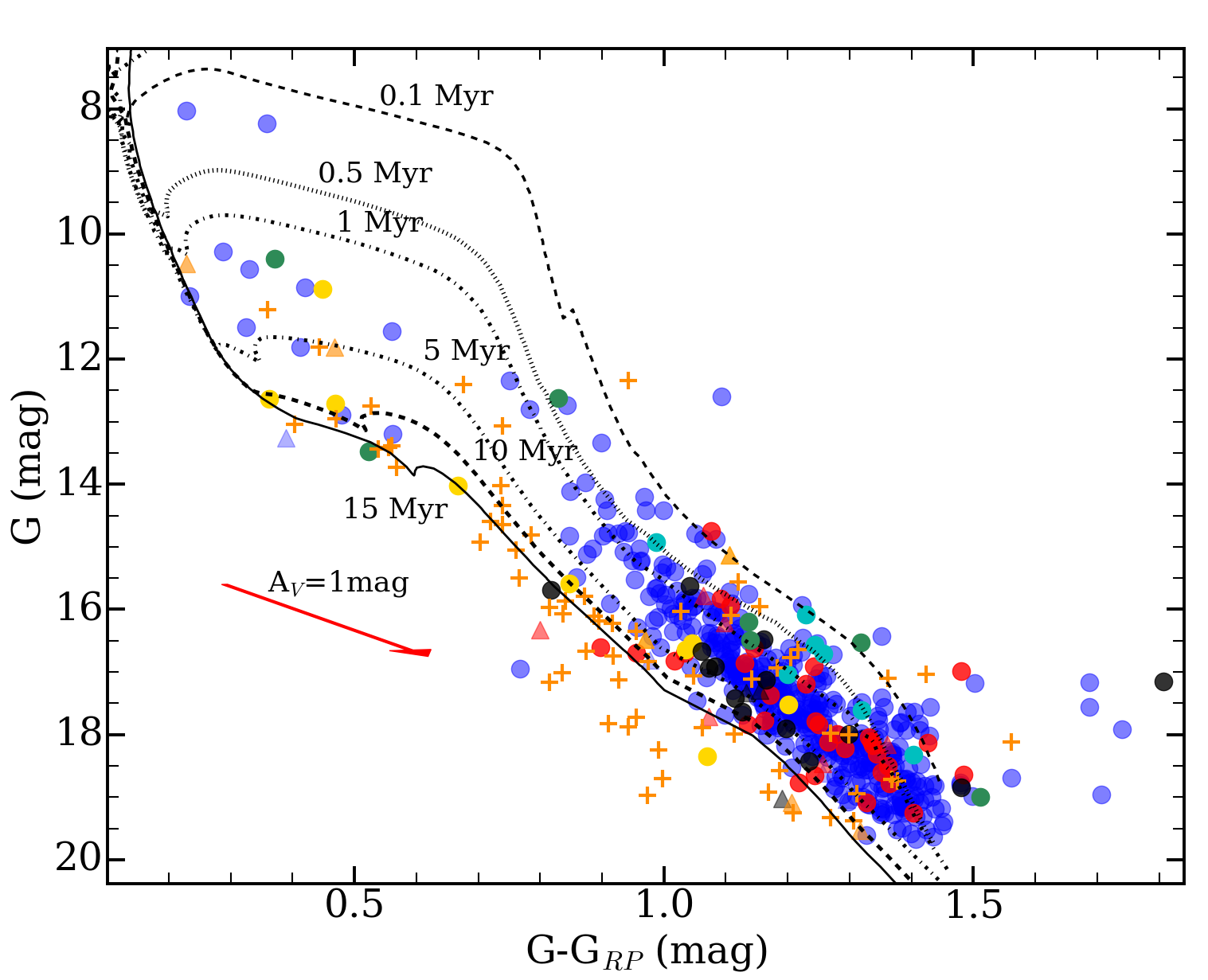}\\
    \end{tabular}
    \caption{Parallax histogram (top, including the full histogram on the left and a zoom of the subclusters on the right), proper motion and color-magnitude diagram (bottom) of the known members of the subclusters found by the maximum likelihood analysis with the Gaia EDR3 data. The symbols and colors mark the various subclusters using circles, triangles, and capital letters (see legend). The number of subcluster members is given in parentheses. Full circles represent stars with a membership probability $>$80\%. Triangles represent objects with probabilities $<$80\%. The theoretical PARSEC isochrones on the color-magnitude diagram have been corrected for a distance of 925 pc and a lower limit extinction A$_V$=1 mag to reject foreground stars. The extinction vector for 1 mag is indicated with a red arrow.}
    \label{fig:probabilityclustersDR3}
\end{figure*}

With this procedure, subclusters that are not significant disappear since the probability of a star belonging to that subcluster vanishes as the probability becomes negligible when all stars are too far from the center parameters. Non-significant clusters also tend to merge with others with more members. For instance, we started with 10 initial subclusters that merged into only 7 statistically significant subclusters. Note that the size of these 7 subclusters does not exceed the limit imposed by \citet{Kounkel2018} to avoid linking physically unrelated structures or contaminating field stars.

The combination of position and number of subclusters with the maximum likelihood is kept as the best solution for the structure of the region, with the standard deviation of the parameters of all stars assigned to each subcluster representing the distribution of subcluster members. For the final result, we consider those stars having a probability higher than 80\% of belonging to one subcluster as subcluster members. We performed several additional tests by changing the subclusters initial conditions and the number of subclusters, always converging to the same results. 

\begin{table}
    \caption{Anderson-Darling test results, for parallax ($\varpi$) distribution, between each pair of subclusters [i,j]. The number of members per subcluster, [n$_i$,n$_j$], is also given to assess significance.}\label{tab:AD_test_DR3}
    \centering
    \begin{tabular}{ccc}
    \hline\hline
     Subclusters  & Nº members &$\varpi$      \\
     {[i,j]}      & {[n$_i$,n$_j$]} & Significance   \\
     \hline
      {[A,B]} & {[418,33]} & ---     \\
      {[A,C]} & {[418,7]} & $<$1\%  \\
      {[A,D]} & {[418,9]} & $<$1\%  \\
      {[A,E]} & {[418,14]} & ---     \\
      {[A,F]} & {[418,7]} & ---     \\
      {[B,C]} & {[33,7]} & $<$1\%  \\
      {[B,D]} & {[33,9]} & $<$1\%  \\
      {[B,E]} & {[33,14]} & ---     \\
      {[B,F]} & {[33,7]} & ---     \\
      {[C,D]} & {[7,9]} & ---     \\
      {[C,E]} & {[7,14]} & $<$1\%    \\
      {[C,F]} & {[7,7]} & 2.5-5\% \\
      {[D,E]} & {[9,14]} & $<$1\%  \\
      {[D,F]} & {[9,7]} & 1-2.5\%  \\
      {[E,F]} & {[14,7]} & ---     \\
     \hline 
    \end{tabular}
    \tablefoot{The cases where the differences are not statistically significant are marked with "--".  The populations are statistically different if the significance level is $<$1\%. Note that we do not consider the G extend population in this comparison due to its large standard deviations.}
\end{table}

Table \ref{tab:resultslikelihoodDR3} shows the resulting best-fit parameters for the 7 surviving subclusters, which we called with letters A through G. A total of 552 stars (among the 578 known members with good astrometric data, see Table \ref{tab:samples_EDR3}) have a probability $>$80\% of belonging to one of these subclusters. This does not mean that objects with a lower probability are less likely to be members of IC1396, but they may not be so clearly associated with one of the particular subclusters. For instance, we identify 26 sources with intermediate probabilities, 4 of them with probabilities $>$40\% of belonging to two subclusters (which are considered probable members of both) and the rest with probabilities less than 80\% (considered probable members of one of the particular subclusters). Figure \ref{fig:probabilityclustersDR3} shows the distribution of parallax and proper motion of the stars associated with each subcluster and their positions in the color-magnitude diagram (see also Table \ref{tab:resultslikelihoodDR3}).  

The distances of subclusters A, B, E, and F are all consistent (although subcluster F has only seven members). This is also true for the distances of groups C, D, and G.
Population G (64 sources) is poorly defined in parallax and proper motion (e.g., it is extended from $\mu_{\alpha}$=-20 mas/yr to +15 mas/yr). It also has a consistently larger age than the rest, with 58\% of its members below the 10 Myr isochrone (Figure \ref{fig:probabilityclustersDR3}). Therefore, the extended population G is not considered further to search for new members, as it could include too much contamination. However, many of its members are bona-fide cluster members, with clear indicators of youth, such as disks or accretion. This shows that IC1396 contains a significant number of members that are very extended in the proper motion plane.

Subcluster D has 9 members, of which 4 are intermediate-mass (B, A, F spectral types) stars considered as probable members  \citep{Contreras2002}, and one additional source identified by \citet{Nakano2012} with low EW(H$\alpha$) $<$ 10 \r{A}, which makes them less reliable. Since its proper motions overlap with the densest part of the galactic plane, it is likely that many, if not most, of the objects associated with subcluster D, are contaminants previously thought to be members. Subcluster D also has a different parallax (Table \ref{tab:resultslikelihoodDR3}), being more distant than the rest of the subclusters, followed by the extended population G and subcluster C. The fact that subclusters C, D, and G distances are consistently higher suggests that these groups could also contain a significant number of contaminants. Therefore, we do not use subclusters C and D to search for new members either.

Following on the previous, we checked if the distance distribution of the sources belonging to each subcluster is significantly different using the Anderson-Darling test \citep{Stephens1974}. This non-parametric test examines the total difference in the cumulative distribution. We use the $scipy.stats$ module of the Anderson-Darling test for k-samples \citep{Scholz1987k, SciPy-NMeth2020}. The test returns the test value, the critical values, and the significance levels (25\%, 10\%, 5\%, 2.5\%, 1\%), which can reject the null hypothesis.
We consider the samples are statistically different if the significance level is less than 1\%. Table \ref{tab:AD_test_DR3} shows the test results. Although subclusters C, D, and F have few members, limiting the significance of the test, we find a significant difference in distance between subclusters A and D; A and C; B and C; B and D; C and E, and D and E. The distances of subclusters C and D are not significantly different, but both C and D are more distant than the rest, while subclusters A, B, E, and F all share a similar distance. For comparison, the proper motions of the subclusters are different from expected since, otherwise, they would have been merged by the algorithm.

We also conclude that the EDR3 data has made the identifications of the DR2 subclusters more reliable. The maximum likelihood analysis with Gaia EDR3 data confirms and recovers previously identified subclusters (and the extended population) with the Gaia DR2 data. The reliable subclusters A, B, E, and F, are essentially identical, but the results with Gaia EDR3 data make them more compact and thus better defined in the parallax-proper motion space. In contrast, extended populations D and G do not improve.
The objects associated with subcluster C have a lower spread in proper motion in EDR3 than in DR2. However, the number of members of subcluster C remains so low that a further independent study (including more confirmed members around this region) is required before its status as a subcluster can be confirmed.

\begin{figure}
    \centering
    \begin{tabular}{c}
    \includegraphics[width=0.48\textwidth]{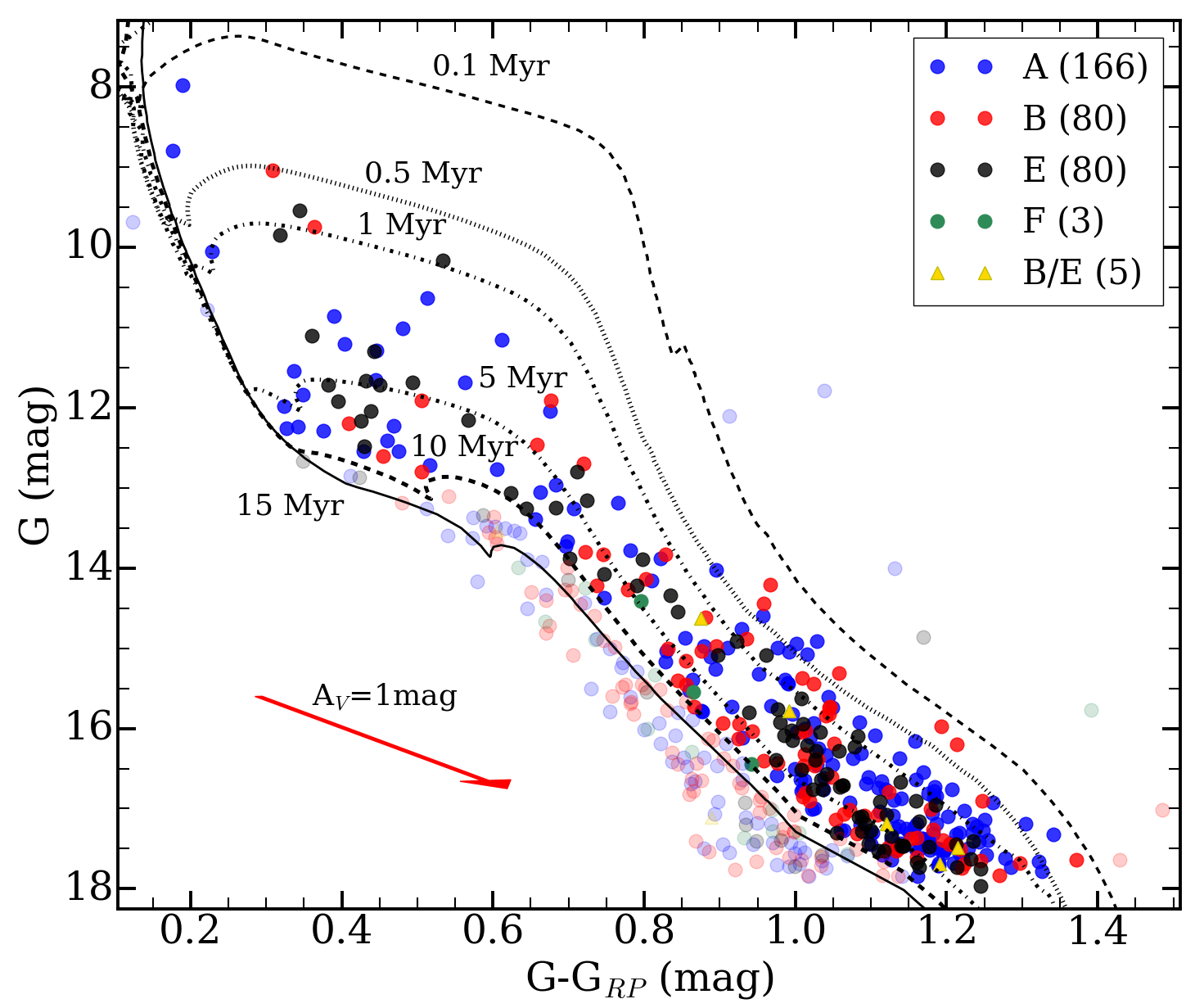}
    \end{tabular}
    \caption{Color-magnitude diagram for the new members obtained from Mahalanobis distance analysis with the Gaia EDR3 data. The 334 new members between 0.1 and 10 Myr are in full color, candidates with ages $>$10 Myr are rejected and shown in semi-transparent color. The theoretical PMS tracks have been corrected for a distance of 925 pc and a lower limit of extinction of A$_V$=1 mag to reject foreground stars. Five members are plotted as triangles and belong to both subclusters B and E. Note there is an empty space between the 10 Myr and 15 Myr isochrones, which reinforces the criterion of considering as members only those up to 10 Myr. A vector extinction is indicated by a red arrow.
    \label{fig:HR_NM_EDR3}}
\end{figure}

\subsection{ New candidates identified via the Mahalanobis distance} \label{subsec:Mahalanobis}

\begin{figure*}
    \center
    \includegraphics[width=0.9\textwidth]{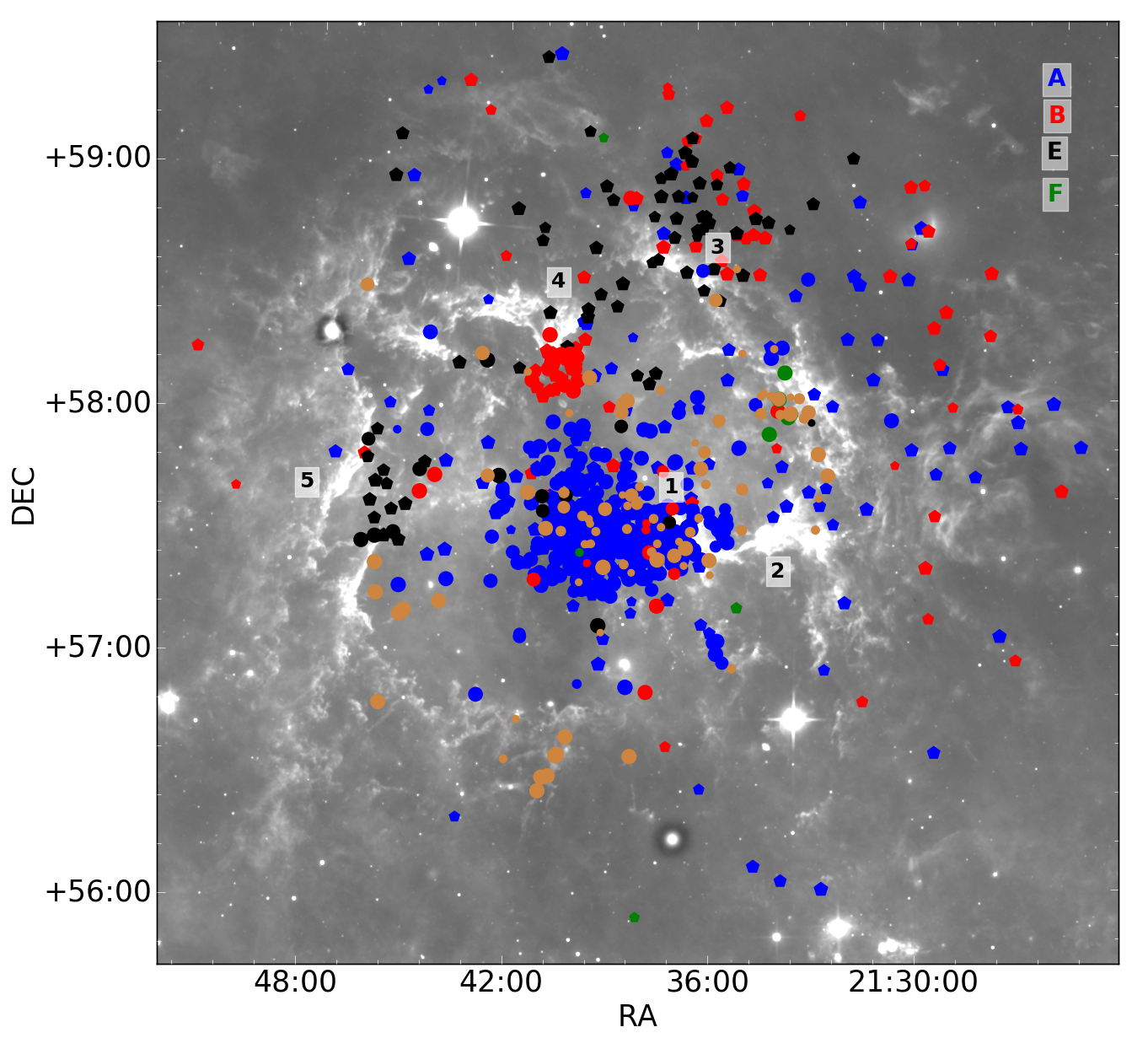}
    \caption{Spatial distribution of the known (full-color circles) and new (pentagon) members. The colors represent the four subclusters (A, B, E, F, see legend). Orange circles mark the rest of the known members not belonging to any subcluster. The age of members (derived in Section \ref{subsec:age_mass_NM}) is represented by the symbol sizes, with the youngest members having large symbols ranging from 0.1 to 20 Myr. Black numbers indicate the main BRCs, 1=IC1396 A, 2=IC1396 B, 3=IC1396 D, 4=IC1396 N, 5=IC1396 G. The background image is a  $\sim$4$^o$ x $\sim$4$^o$ mosaic built from WISE 22.19 $\mu$m.}
    \label{fig:spatialdist_EDR3}
    \center
\end{figure*}

The Mahalanobis distance technique can be used to identify objects that belong to the same subclusters as the known members \citep{Mahadistance1936, Mahalanobis1969}. The Mahalanobis distance is equivalent to the Gaussian probability, generalized to standard Euclidean distance in a transformed, multidimensional space (see Appendix \ref{appe:MLF}).
Sources with a high probability of belonging to the same distribution than the known members are considered new members. For a collection of i stars and j subclusters, we estimate the probability of each i-th source to be a member of each j-th subcluster, calculating the multidimensional distance between the star position and the subclusters centers in the parallax-proper motion space (Mahalanobis distance), using the Python routine $scipy.spatial.distance.mahalanobis$.
As it happens in a 1-D Gaussian space, a Mahalanobis distance lower than 2 corresponds to a $\sim$95\% confidence interval regarding the membership of the object to any subcluster with the multivariate position considered.

We used the Mahalanobis distance to find new members of IC1396.
For this, we used the reliable subclusters (A, B, E, F) from the maximum likelihood analysis to estimate the membership probability of other stars with good quality Gaia data. Our first tests showed that the parallax positions derived from the maximum likelihood method were poorly defined due to the typical large uncertainties for objects at nearly 1 kpc, which resulted in excessive contamination. We thus readjusted the parallax distribution with a Gaussian function to the best-defined subcluster (A). The fitting produced a mean parallax of $\varpi$=1.100 mas, with standard deviation $\sigma$=0.054.
Since there are no significant differences in the distances of the rest of the well-defined subclusters, we used this value for all.

We explored the membership of the sources in a 2$^o$ radius field around HD206267. Imposing the same restrictions, we obtain $\sim$68k sources with good quality astrometric data and in the parallax range (0.6-1.6 mas) of known members. 
We calculated the Mahalanobis distance and selected those within the 95\% confidence interval as members. We further reduced contamination by restricting the errors of the magnitudes (G, and G$_{RP}$) to less than 0.05 mag, and the errors on proper motion to less than 0.1 mas/yr, to reduce contamination from those sources with large errors\footnote{Note that sources with large uncertainties in proper motion or parallax may be technically consistent with the clusters but are statistically very unlikely to be real cluster members}. In addition, we imposed an age cut between 0.1 Myr and 10 Myr, assuming a fixed extinction A$_V$=1.0 mag, avoiding excessive contamination by those objects at $\geq$1 kpc with relatively large uncertainties. This is particularly important for IC1396 since the proper motions of the members are not so different from those of the galactic background. According to our maximum likelihood results and considering only the four reliable subclusters (Table \ref{tab:newmemberstotalDR3}), $\sim$91\% of known members are in this age range, so there is no substantial evidence of an older population. This also implies an approximate loss of $\sim$9\% of new members, to add to these what we excluded from the extended population (G), representing 11\% of total known members. Although our final new members list is incomplete, the limits are required to avoid too much contamination by sources that are unlikely members.

The procedure described above resulted in a total of 334 new members (see Table \ref{tab:samples_EDR3}) belonging to the subclusters (see Appendix \ref{appe:TableMahalanobis}, Table \ref{tab:tableMD_EDR3}). Among these, 5 sources belong to
both subclusters B and E (with probabilities $>$30-60\%). Figure \ref{fig:HR_NM_EDR3} shows the new members distribution in the color-magnitude diagram. Table \ref{tab:newmemberstotalDR3} shows the final result of the total number of members per subcluster from the Mahalanobis distance analysis and from the maximum likelihood analysis.

\begin{table}
  \caption{Membership analysis for subclusters found with EDR3.}\label{tab:newmemberstotalDR3}
  \centering
  \begin{tabular}{cccccc}
  \hline\hline
  \multicolumn{3}{c}{Maximum likelihood}&
  \multicolumn{3}{c}{Mahalanobis result}\\
  \hline
   Group & KM    & KM        &NM     &NM & Total\\
     & Total & 0.1-10Myr & Total &0.1-10 Myr &Rejected\\
  \hline
      A &418 & 388 & 234 & 166 & 15\%\\
      B & 33 &25 & 148  &80 & 42\% \\
      E & 14 & 11 & 92 & 80 & 14\%\\
      F & 7 & 4 & 16 & 3 & 70\%\\
      B/E & - & - & 7 & 5 & -\\
  \hline 
    Total& 472& 428 & 497 & 334 &\\
  \hline
  \end{tabular}
  \tablefoot{Column 1: Subclusters obtained from the maximum likelihood considered in the Mahalanobis analysis. Columns 2 and 3: Total known members (KM) per cluster and those between  the 0.1 Myr and 10 Myr isochrones. Columns 4 and 5: Total new members (NM) from the Mahalanobis analysis and those between the 0.1 Myr and 10 Myr isochrones. Column 6: Percentage of sources outwith the 0.1-10 Myr isochrones, including both the known and new members.}
\end{table}

\begin{figure*}
    \centering
    \begin{tabular}{cc}
    \includegraphics[width=0.5\textwidth]{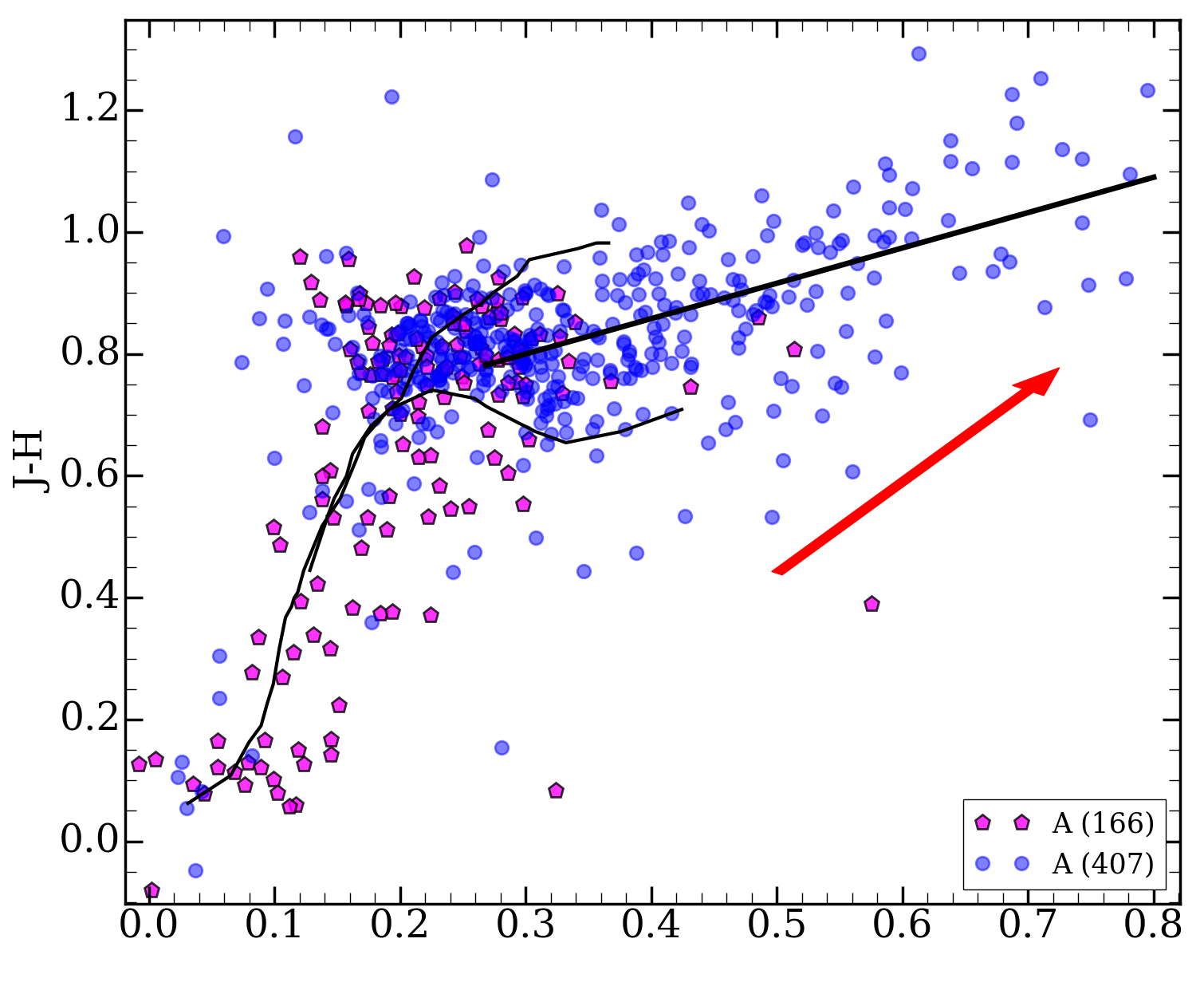}
    \includegraphics[width=0.5\textwidth]{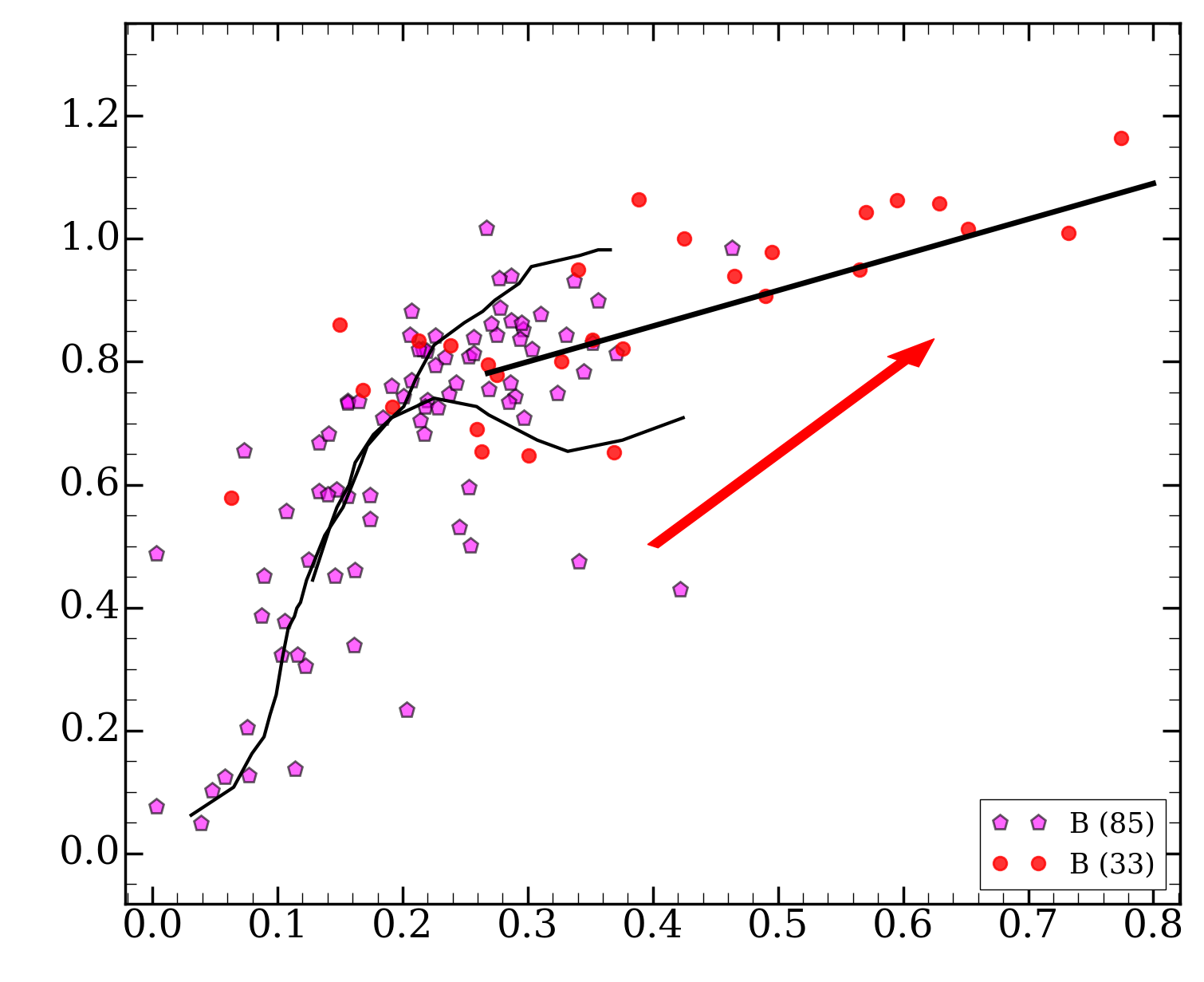}\\
    \includegraphics[width=0.5\textwidth]{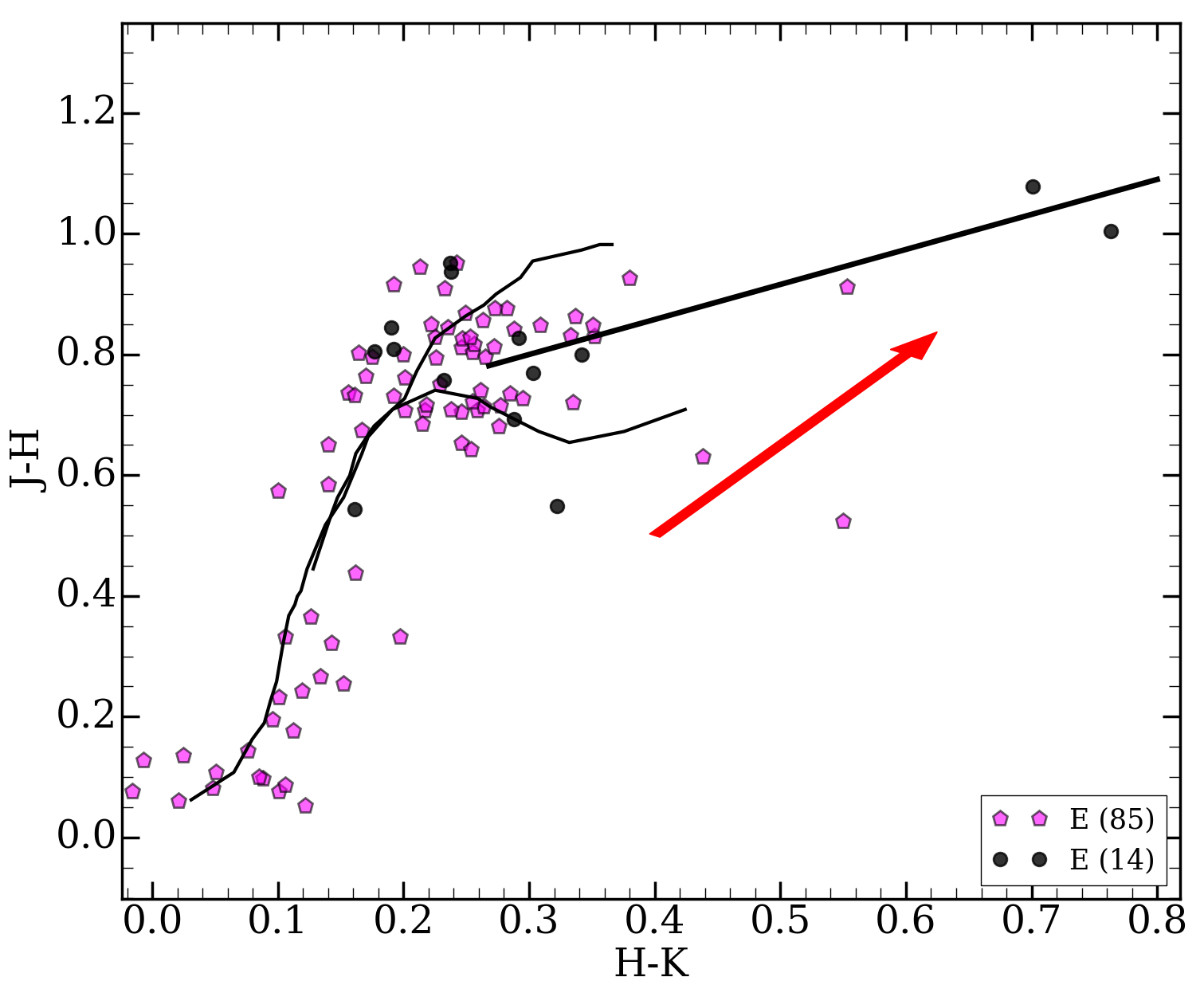}
    \includegraphics[width=0.5\textwidth]{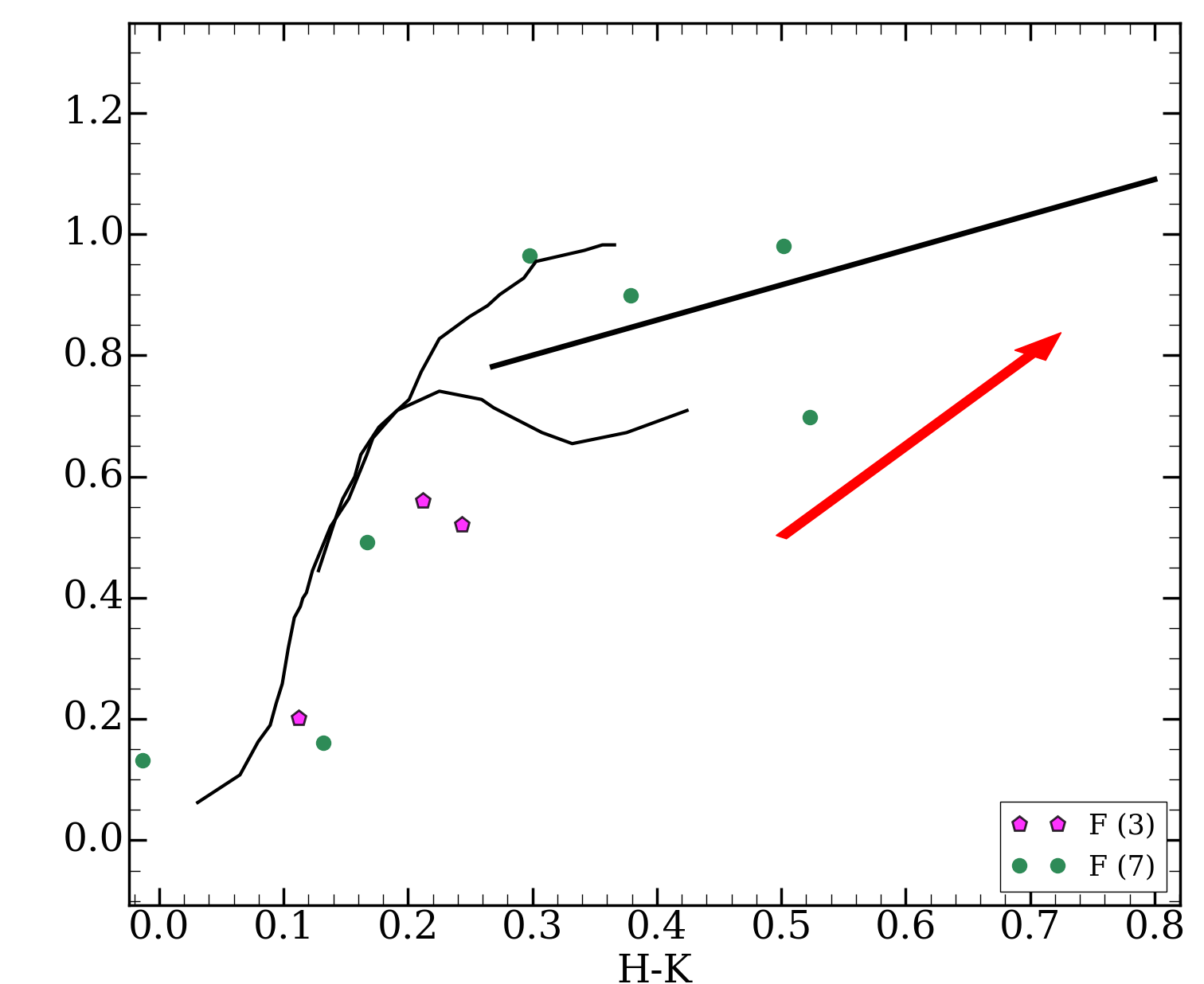}\\
    \end{tabular}
    \caption{J-H vs. H-K diagrams for known members associated with the various subclusters obtained from the Maximum likelihood analysis (colored circles) and new members obtained from the Mahalanobis distance analysis (magenta pentagons). The number of sources with complete JHK photometry is indicated in the legend. Subclusters B and E include the new members with an intermediate probability of belonging to both subclusters. The MS and the giant branch are indicated by thin solid lines \citep{BessellBrett1988}. The CTTS locus \citep{Meyer1997} is indicated by a thick solid line. The theoretical tracks are corrected for extinction using the threshold A$_V$=1 mag. A reddening vector for A$_V$ = 3 mag is shown as a red arrow.}
    \label{fig:JHKdiagrams}
\end{figure*}


\begin{figure*}
    \center
    \includegraphics[width=0.8\textwidth]{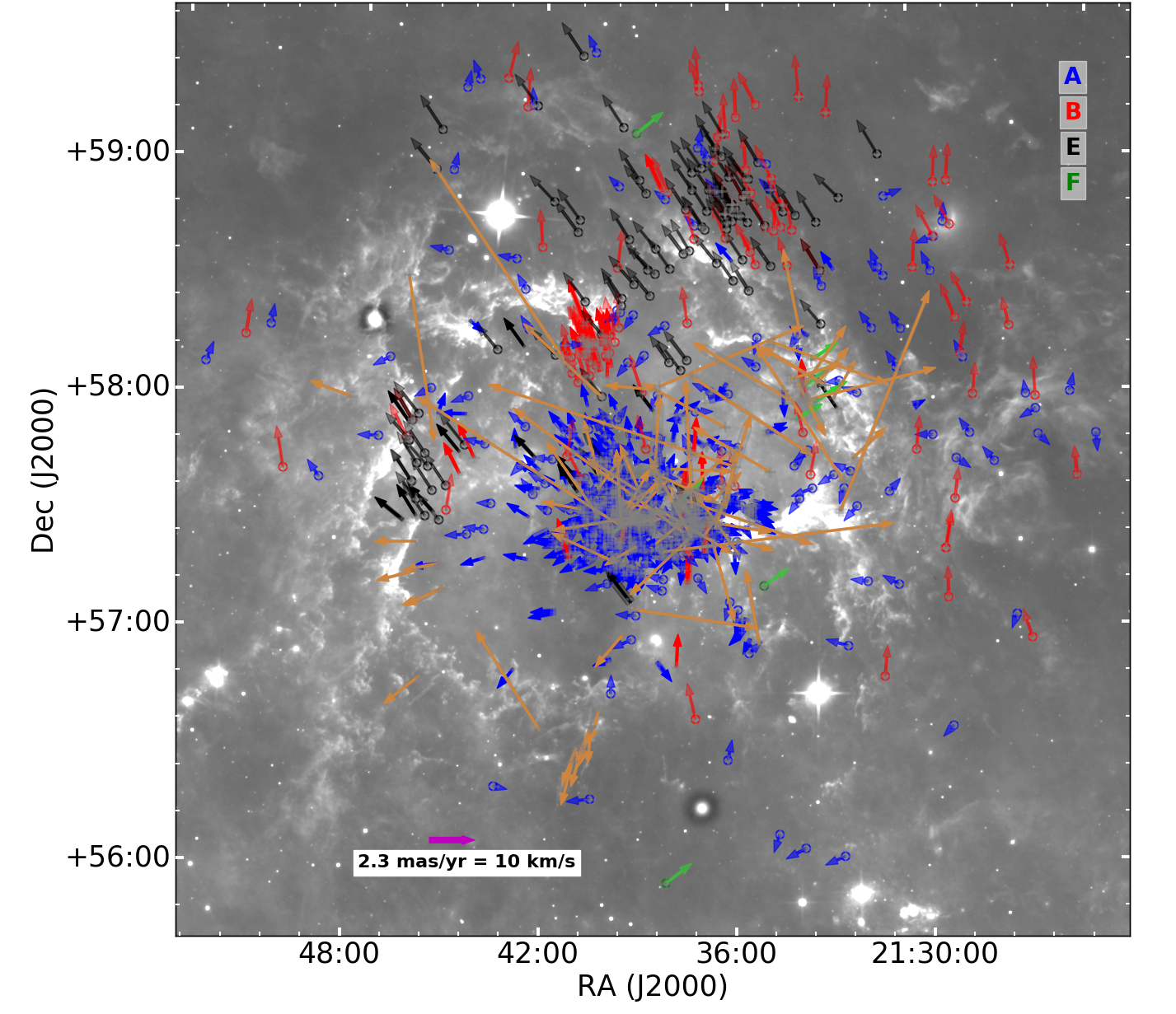}
    \caption{Map of the velocity dispersion in proper motion of known and new members. The arrows mark the direction and size of the proper motion vectors relative to the average proper motions of subcluster A. They are color-coded by subcluster (see legend). Known members not associated with any subcluster are shown in orange. The long orange arrows are mostly from the extended population G. New members are marked by open circles. The magenta arrow in the bottom left corner indicates the physical velocity scale (at 925 pc).}
    \label{fig:pm_EDR3_refcenter}
    \center
\end{figure*}

We analyze the spatial distribution of all known and new members of IC1396 (Figure \ref{fig:spatialdist_EDR3}). Most of the known members are clustered around HD206267, while most new members are near the BRCs and on the filaments around the region.
Although the coordinates have not been used to identify the subclusters, most of the subclusters A, B, and E members are not uniformly distributed in space. This spatial structure and clustering, together with the differences in proper motion between subclusters, are additional confirmations that they have different physical origins. Subcluster A is the best defined, located in the center of the region, and it has few new members because the IC1396A globule and the Tr37 cluster have been extensively studied. The spatial distribution of subclusters B and E members are clearly different from subcluster A, suggesting various star-forming episodes.

Subcluster B mostly runs along the northern part of the region, besides IC1396N, with some objects extending towards the northwest behind BRC IC1396D.
The members of subcluster E are east of HD206267, spreading to the northwest past the ionized edge of the bubble. Subcluster F has too few members to discuss its spatial location meaningfully despite being well-defined in the parallax-proper motion space.  
In contrast, the members of subcluster D and the extended population G are not spatially clustered, which is expected from groups that are not real, are heavily contaminated, or are not well-defined in astrometric space. The lack of clear astrometric and spatial signatures further confirms that, even if a fraction of confirmed IC1396 members belong to these groups, we cannot use them to identify further members (Section \ref{subsec:likelihood}). However, subcluster C, also rejected, is spatially clustered in the south of the region, but as we mentioned before, it has too few members to be confirmed.

We used the 2MASS J-H vs. H-K diagrams to explore the properties of the new members. Figure \ref{fig:JHKdiagrams} shows that most of the new members are consistent with diskless intermediate-mass and massive T Tauri stars, in contrast with the known members, which are mostly low mass T Tauri stars (see Section \ref{subsec:age_mass_NM} for a detailed mass analysis). This is likely a consequence of the requirements imposed on the astrometry quality. It also reflects that the most efficient methods used to identify members (X-ray emission, H$\alpha$ excess from photometry or spectroscopy, Li I absorption, and the presence of near- and mid-IR excesses) are biased towards revealing solar-and low-mass stars. For instance, we found 43 new members (23 in subcluster A, 6 in subcluster B, and 14 in subcluster E) at (J-H)$<$0.2 mag. From these 43 sources, 33 are known stars with spectral types A0-B1  \citep[][]{Alknis1958, Marschall1987, Kun1986, Patel1998, Getman2007}, which are now found to be cluster members.

Figure \ref{fig:pm_EDR3_refcenter} shows the velocity dispersion in the plane of the sky for all members relative to the mean proper motion of subcluster A. The relative motions also indicate that the subclusters members are dispersing. Further analysis of the velocity structure of the region is discussed in Section \ref{subsec:VSAT_EDR3}.

Comparing with Gaia DR2, the Gaia EDR3 errors are significantly lower, especially for the proper motion (see Table \ref{tab:resultslikelihoodDR3}). This produces a tangible improvement in membership selection. Gaia EDR3 allows us to reject 62 objects which had been previously identified as new members with the Gaia DR2 data. Eight of them now have RUWE$>$1.4, 30 have photometric errors $>$0.05 mag, and the rest were border-line cases for which the probability of belonging to the subclusters has now dropped below the 95\% confidence levels due to the reduction in the uncertainties. The fact that the Gaia EDR3 data have smaller uncertainties and, still, the number of members increases is an additional confirmation that the subgroups are real. Non-existing groups affected by significant uncertainties would tend to disappear when the uncertainties are reduced. 

Recent studies of the west of Tr37 \citep{Silverberg2021} found over 400 members via X-rays, some of which were previously known members. We did a very brief test finding that only 45 of them have Gaia EDR3 counterparts, and 10 were already in our collection of known members. Finally, from these, only 22 sources have good quality astrometric Gaia data. We added these 22 sources to our 578 reliable known members (see Table \ref{tab:samples_EDR3}) to check the robustness of our method, we calculated the maximum likelihood function again and obtained the same distribution of subclusters in the entire region. The only difference was a slightly greater overlap in proper motion between cluster B and cluster E with respect to previous results since some more of the new members have an intermediate probability of belonging to both subclusters B and E. This happened to a lesser degree before, and it may indicate that these two subclusters are, in fact, part of the same structure that spans a continuous range of proper motions. To test the physical relation between both subclusters, we would need more objects in each group, detected by means independent of their proper motions, which is left for future research. Looking for new members, we did not find changes with respect to our previous analysis since the subcluster locations do not change significantly, proving that our study is robust against adding small numbers of new sources.

\subsection{The velocity structure of IC1396 \label{subsec:VSAT_EDR3}}
\begin{figure}
    \centering
    \begin{tabular}{c}
    \includegraphics[width=0.33\textwidth]{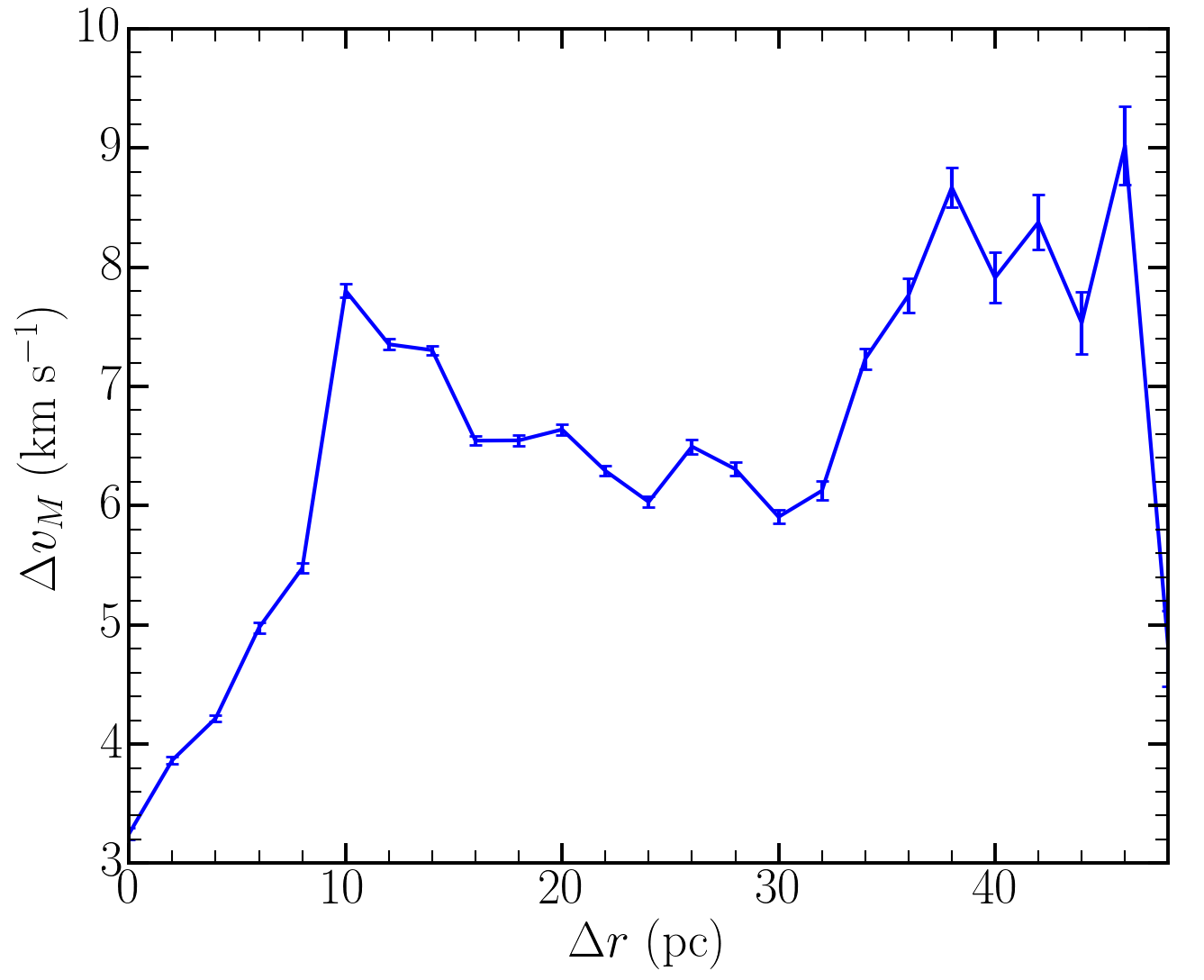}\\
    \includegraphics[width=0.33\textwidth]{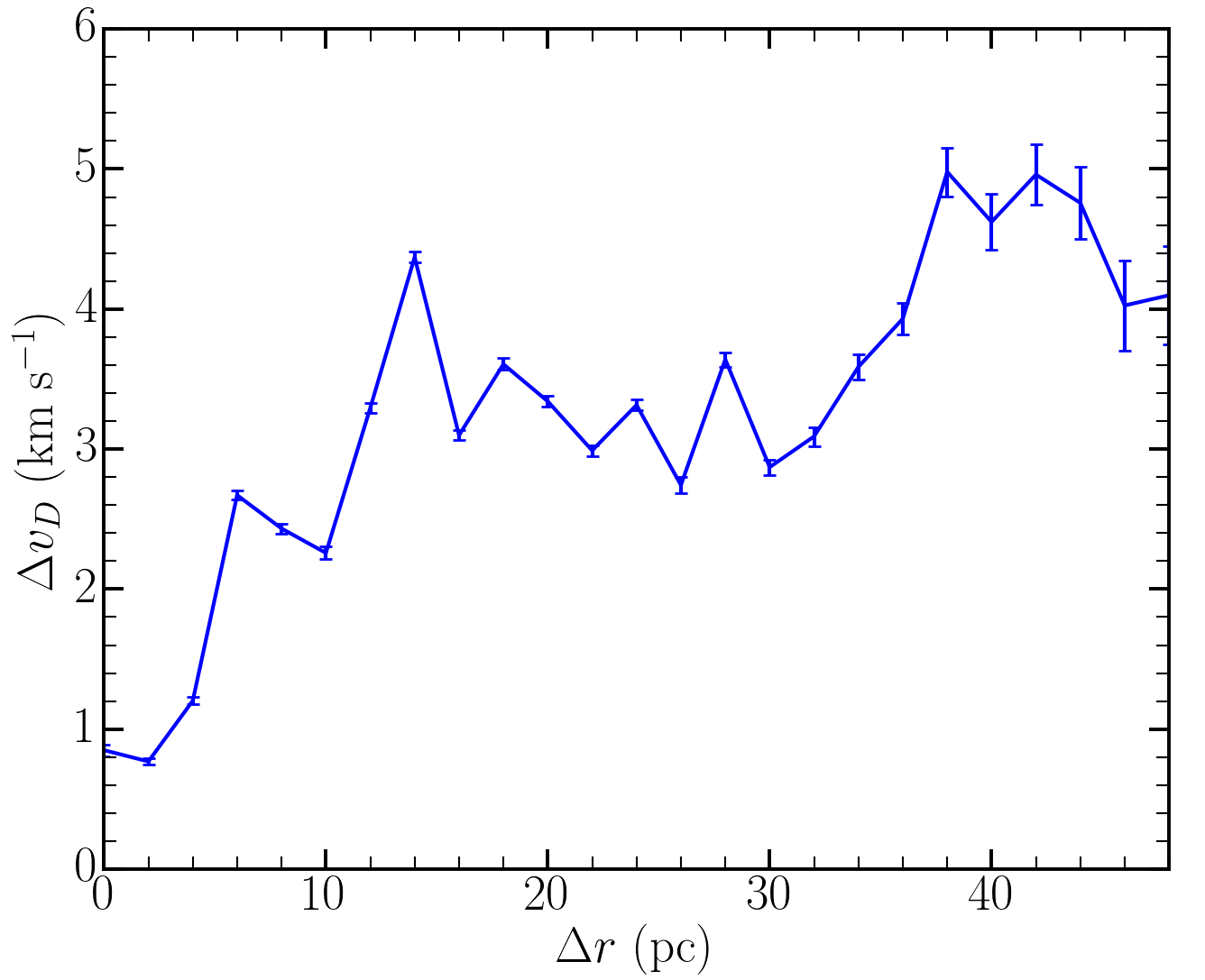}\\
    \includegraphics[width=0.33\textwidth]{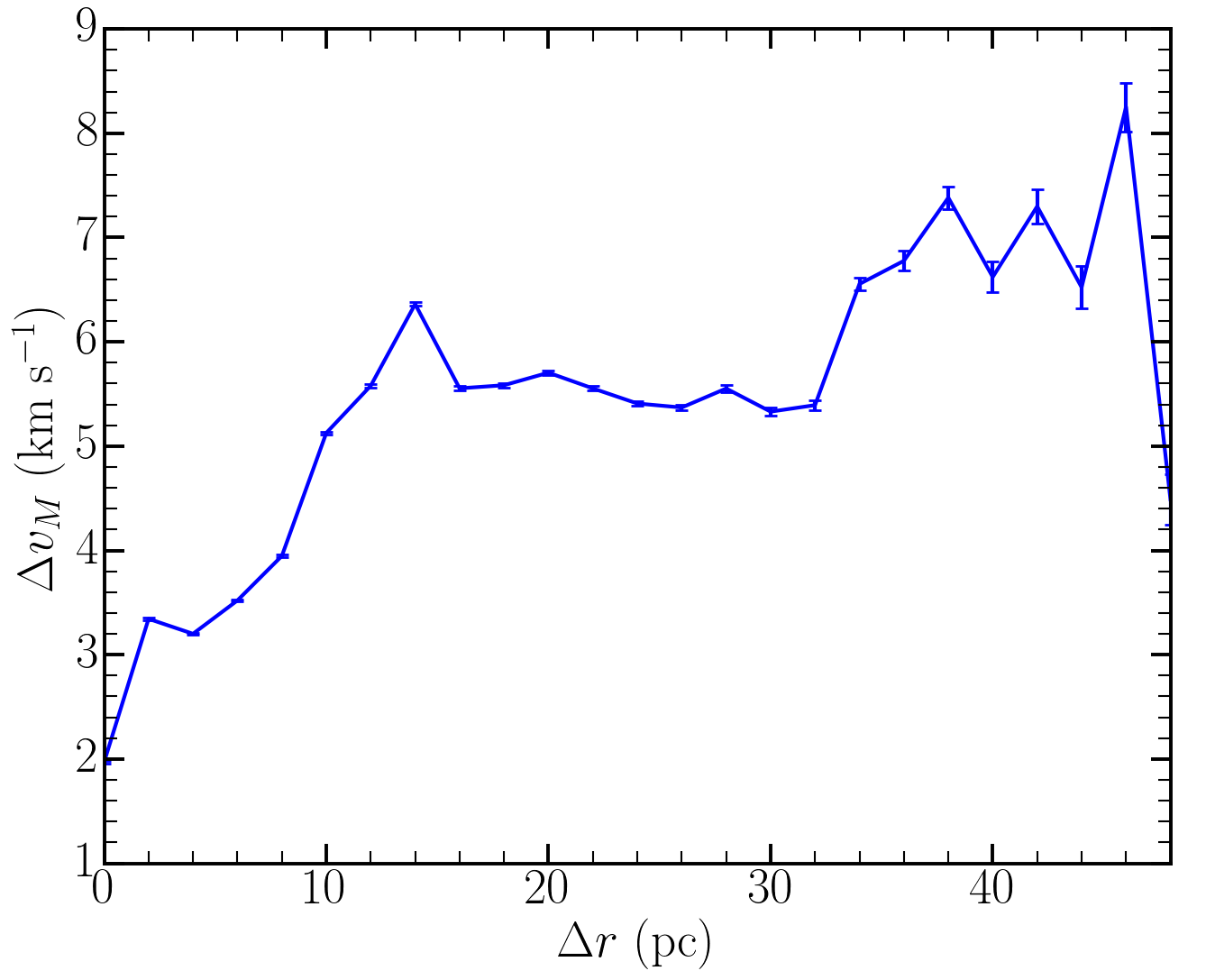}\\
    \includegraphics[width=0.33\textwidth]{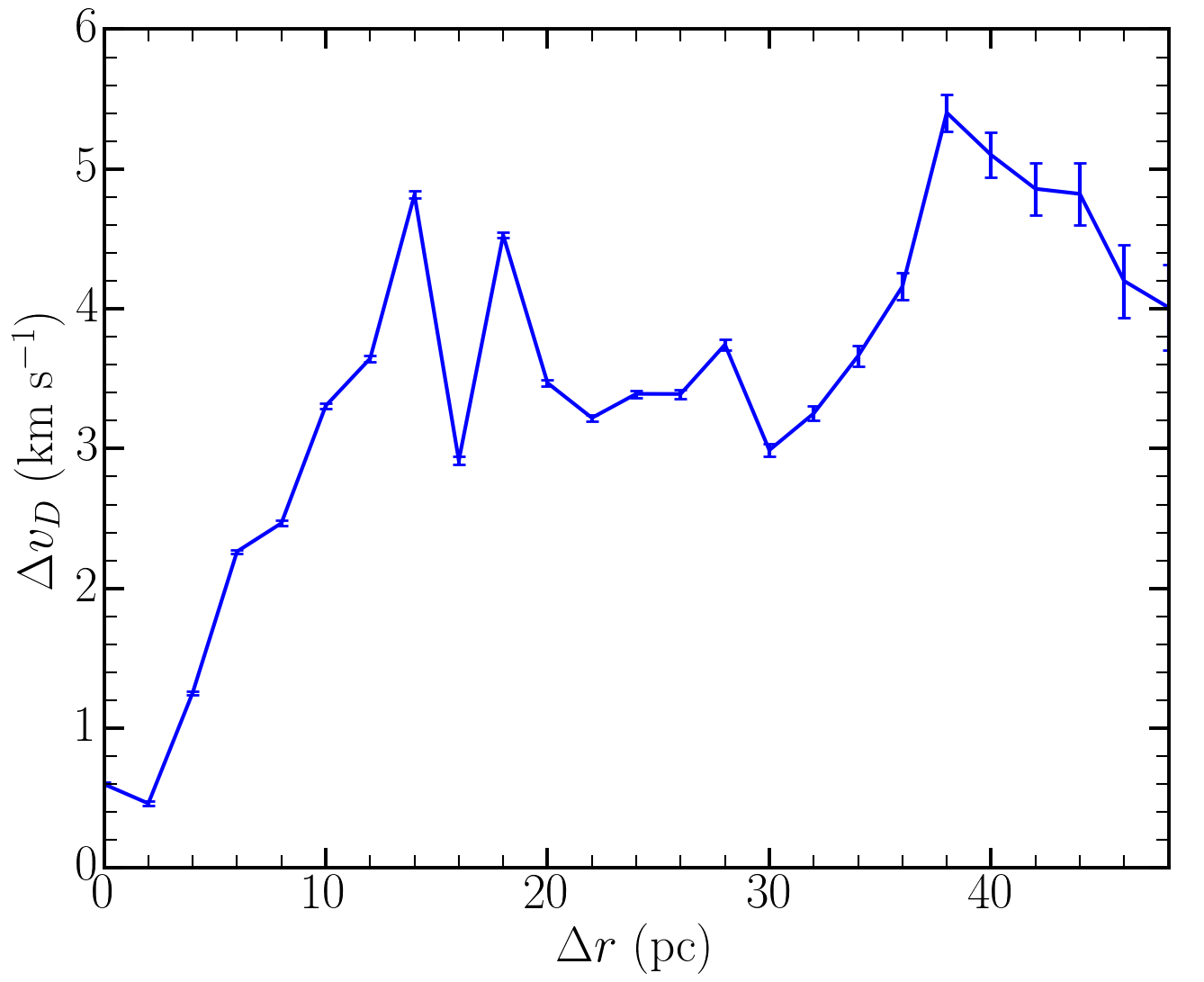}\\
    \end{tabular}
    \caption{Velocity structure in the IC1396 region, shown in terms of the mean difference between the velocity magnitude $\Delta$v$_M$ (left panels) and the directional velocity $\Delta$v$_D$ (right panels) versus the difference distance in parsec ($\Delta$r; see text). The two top panels show the region velocity structure considering all known and new members, including the extended population (G), which has considerably larger velocities and velocity dispersion than the rest of clustered members. The two bottom panels include only the clustered part of the population, excluding subcluster G, and thus offer a better view of the evolution and dispersal of the clustered members.}
    \label{fig:vsat_EDR3}
\end{figure}

In this section, we present a complementary analysis of the kinematic structure of IC1396 using the Gaia EDR3 data for all members. This analysis also helps us to give a more quantitative explanation of the apparent expansion trend observed in Figure \ref{fig:pm_EDR3_refcenter}. We examine the velocity structure of IC1396 using the 578 known members (Appendix \ref{appe:TabML_EDR3}, Table \ref{tab:tableML_EDR3}) and 334 new members (Appendix \ref{appe:TableMahalanobis}, Table \ref{tab:tableMD_EDR3}). Since only very few objects have reliable radial velocities \citep[and they belong to the central part of subcluster A;][]{Sicilia2006_spectro}, it is not possible to analyze the kinematics in 3D, so we concentrate on the velocity structure on the plane of the sky using the proper motions.

The velocity structure is explored using the {\textit{Velocity Structure Analysis Tool}} \citep[$VSAT$,][]{ArnoldBecky2019}. $VSAT$ is a Python routine that estimates the projected velocity differences ($\Delta$v$_{ij}$) between any pair of stars (i,j) in the region, together with its corresponding projected source-to-source distance ($\Delta$r$_{ij}$). We examine the relation between $\Delta$v$_{ij}$ and $\Delta$r$_{ij}$, which gives us information about velocity gradients to reveal bulk motions within the cluster. The object pairs are binned in the projected source-to-source distance ($\Delta$r), calculating the average projected velocity difference ($\Delta$v) per bin. 
Provided that the number of pairs in each bin is large enough, the width of the bin should not affect the results. Observational errors ($\sigma_{\Delta v_{ij}}$) are propagated when calculating each $\Delta$v$_{ij}$, and each $\Delta$v$_{ij}$ is weighted accordingly ($\omega$=1/$\sigma_{\Delta v_{ij}^2}$). The errors in the spatial coordinates are negligible compared with the uncertainties in the projected velocities. The code returns $\Delta$v($\Delta$r) showing the average $\Delta$v values per bin, against their corresponding $\Delta$r. 
The method tracks two aspects of the projected velocity structure of the region, the magnitude ($\Delta$v$_M$) and the directional ($\Delta$v$_D$) projected velocity difference. Here we include a brief explanation since details are in \citet{ArnoldBecky2019}. If each pair of stars (i,j) has vector velocities ({v$_i$, v$_j$}), then the magnitude of the projected velocity difference ($\Delta$v$_M$) is |v$_i$-v$_j$|, which is always positive. This definition is a general measurement of how similar or different the velocity vectors are. In a two-dimensional space (x,y), $\Delta$v$_M$ is calculated as
\begin{equation}
    \Delta v_{ijM}=\sqrt{(v_{xi}-v_{xj})^2+(v_{yi}-v_{yj})^2}.
\end{equation}

On the other hand, the directional projected velocity difference, $\Delta$v$_D$ is the rate at which the projected source-to-source distance ($\Delta$r) between a pair of stars changes, which helps to understand if the region is expanding or collapsing. If the projected distance between stars ($\Delta$r) increases and the bulk of stars are moving away from each other, $\Delta$v$_D$ is positive; otherwise, $\Delta$v$_D$ will be negative. Therefore, $\Delta$v$_D$ is calculated as
\begin{equation}
    \Delta v_{ijD}=\frac{(x_i-x_j)(v_{xi}-v_{xj})+(y_i-y_j)(v_{yi}-v_{yj})}{\Delta r_{ij}}.
\end{equation}

VSAT does not require the center of the region or its radius to be defined, and the frame of reference is irrelevant since all measurements are relative. The results are not strongly affected by the observational uncertainties and biases of individual objects, which are washed out by binning the data. For better visualization, we convert RA and DEC coordinates to parsecs and the proper motions to km/s (see Figure \ref{fig:vsat_EDR3}). The velocity structure of the region is quite clean (bottom panels) if we do not consider the members of the extended population G, which has very different (large) proper motions (often very large) compared to the clustered populations. Population G could include ejected members, although it is also likely affected by contamination at a higher level than the clustered members. If we consider the extended population (top panels), although the tendency is the same as before because the number of objects in the extended population is small, the noise increases due to the larger velocity differences. In any case, there are signs of expansion up to distances $\Delta$r$<$12 pc.

Analyzing the velocity structure (Figure \ref{fig:vsat_EDR3}, bottom panels) without the extended population, the magnitude of projected velocity difference $\Delta$v$_M$($\Delta$r)  shows an increase in the velocity differences between nearby stars, rising from 2km/s to 6 km/s for stars with $\Delta$r  up to $\sim$12-14 pc (Figure \ref{fig:vsat_EDR3}, left-side panels). This shows that nearby stars have similar velocity vectors, so $\Delta$v$_M$ is small. Stars that are more distant from each other are less connected, as evidenced by their very different velocity vectors and large $\Delta$v$_M$. The region shows a clear velocity structure until $\Delta$r $\sim$12-14 pc.
At relative projected distances $\Delta$r$>$12 pc and up to $\Delta$r$\sim$30-32 pc, $\Delta$v$_M$ remains almost constant at $\sim$5.5 km/s, not revealing any significant structure at these scales. When comparing stars separated by even larger distances ($\Delta$r$>$30-32 pc), $\Delta$v$_M$ appears to increase again, but this result is uncertain since large distance bins have fewer stars and noise increases.

For the directional projected velocity difference $\Delta$v$_D$($\Delta$r), (Figure \ref{fig:vsat_EDR3}, right panels), we find a positive correlation between $\Delta$v$_D$ and $\Delta$r, which means that the stars are moving radially away from the center of their velocities distribution. We observe expansion between stars up to separations $\Delta$r $\sim$ 12-14 pc, and the expansion velocity increases from $\sim$0.6 km/s to $\sim$4.8 km/s respectively. Those projected velocities exceed what we could expect from gravitational attraction due to the mass in the region, a sign that the region is not gravitationally bound.
This is in agreement with the values obtained by CO maps \citep[][ring radius=12 pc, and V$_{exp}$=5 $\pm$1 km/s] {Patel1995}, also suggested by \citet{Getman2012} and \citet{Sicilia2019}. Since we are using all the stars, this gives us a better perspective of the velocity of the entire region. Stars separated more than $\sim$12 pc belong to physically unrelated populations. This is in agreement with the stellar crossing times for sources with typical dispersion velocities 2-3 km/s \citep[see also][]{Sicilia2006_spectro}, which would travel between 8 and 12 pc in a cluster lifetime of 4 Myr. 
At separations between $\sim$12 pc and $\sim$30-32 pc, $\Delta$v$_{D}$ does not show any significant trend that could hint expansion or contraction, probably due to noise.  Beyond $\sim$30-32 pc distance, the noise level increases even further (see Figure \ref{fig:vsat_EDR3}, right panels). 

In summary, considering $\Delta$v$_M$, we observed a velocity structure on distance scales up to 12 pc.  Considering $\Delta$v$_D$, the region is also expanding until up $\Delta$r$\approx$12 pc. This 12pc scale is similar to the distance of the compressed molecular ring around the periphery of the IC1396 region \citep{Patel1995}. There is no significant velocity structure on scales beyond that.


\section{Properties of the IC1396 members} \label{sec:analyphoto}

\subsection{Ages and masses for new and known members} \label{subsec:age_mass_NM}

\begin{figure*}
    \centering
    \begin{tabular}{cc}
    \includegraphics[height=0.43\textwidth]{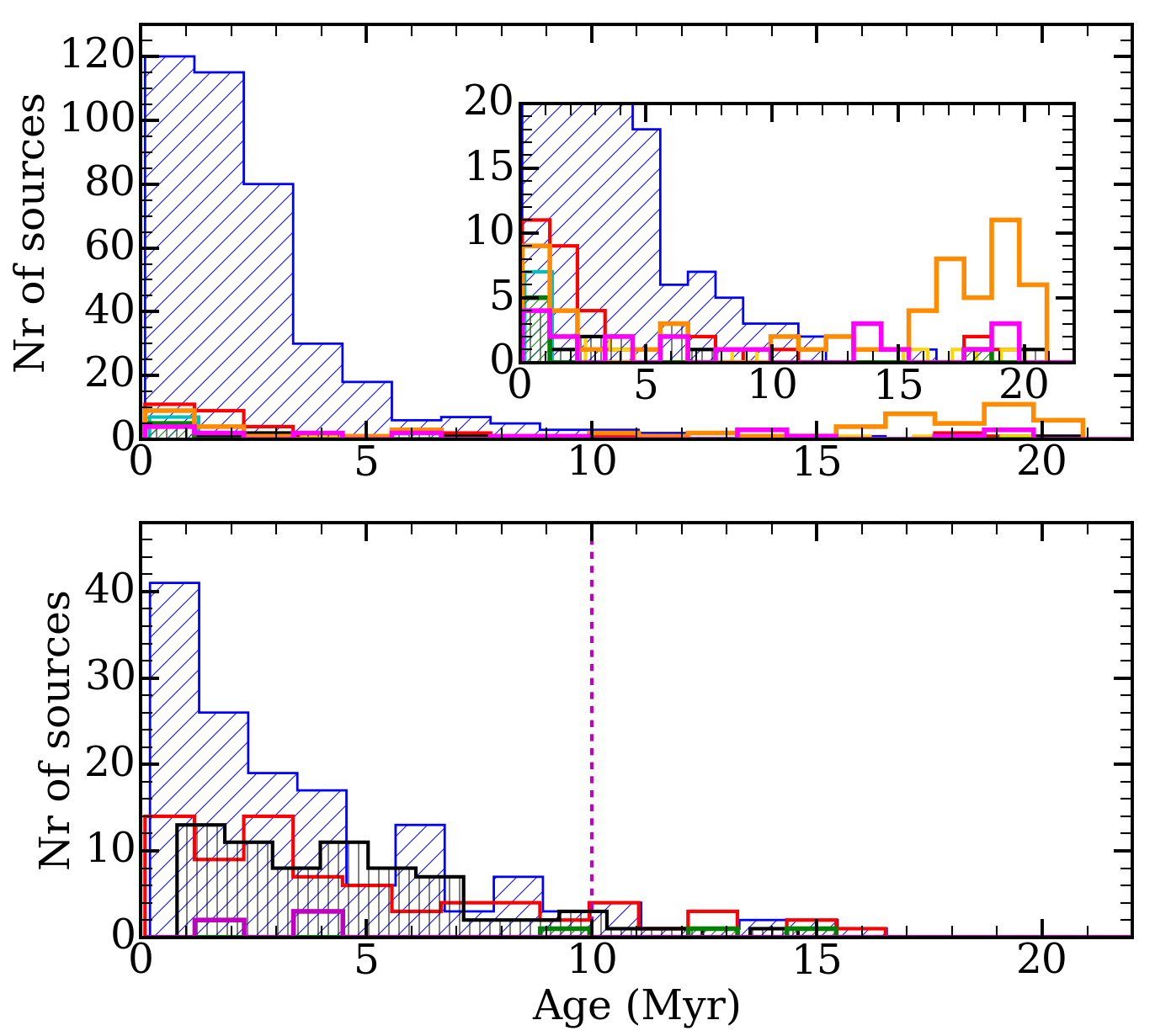}&
    \includegraphics[height=0.43\textwidth]{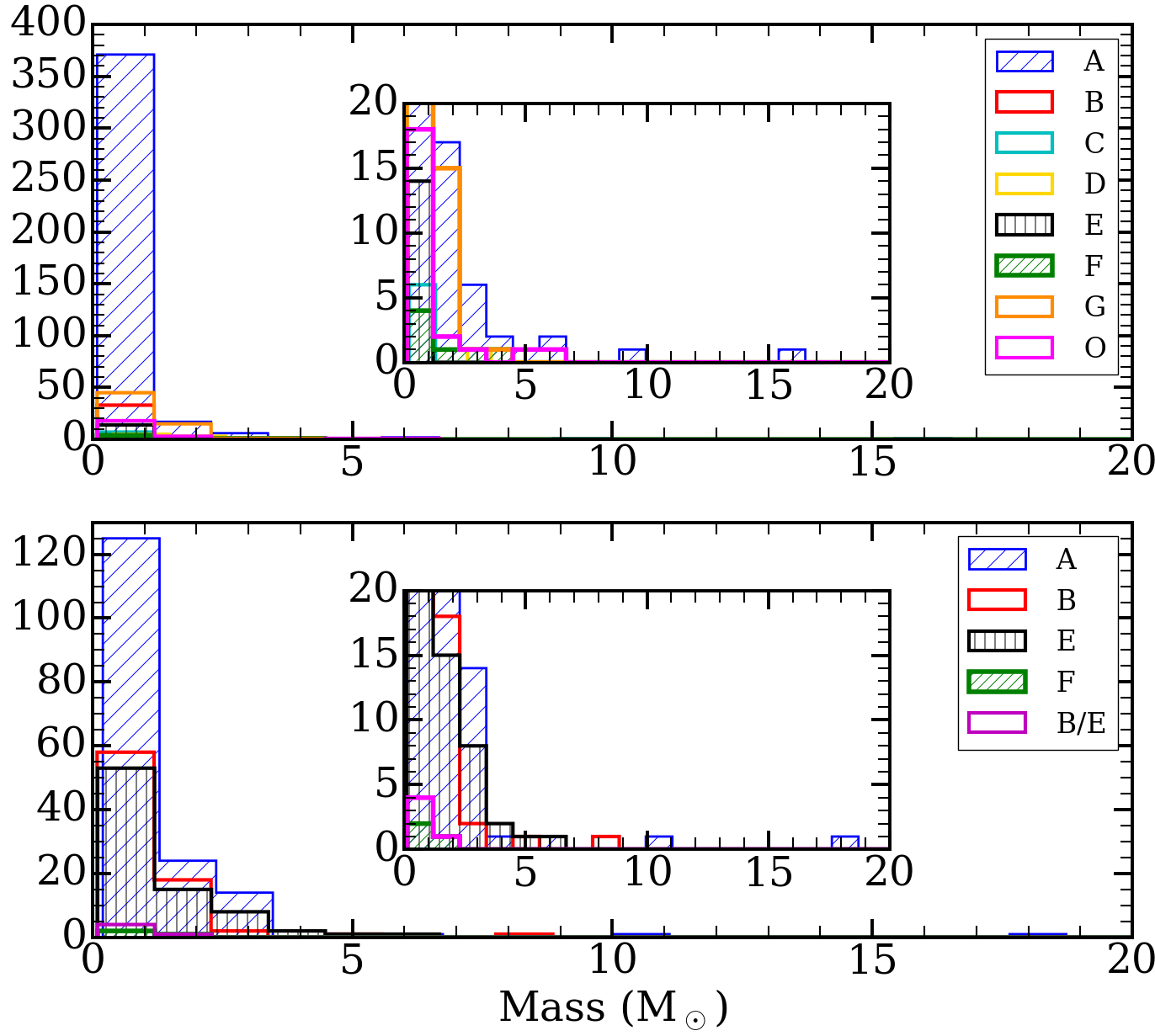}\\
    \end{tabular}
    \caption{Age (left) and mass (right) distribution of known members (top panels) and new members (bottom panels) from different subclusters. We have included those known members that do not belong to any subcluster and those that belong to one more subcluster (labeled as others "O" in the legend). Some new members belong to two subclusters, marked as "B/E". Objects with ages 20 Myr and over are all included in the last bin.}
    \label{fig:age_mass}
\end{figure*}

We derived the age and mass for new and known members (see Figure \ref{fig:age_mass}). The magnitudes were corrected using the average extinction, A$_V$=1.40$\pm$0.52 mag, and its standard deviation provides the uncertainties. As in previous sections, we used the PARSEC isochrones, corrected by the 925 pc mean distance. We derived individual ages and masses by interpolating the isochrones from 0.1 to 20 Myr, using a spacing of 0.1 Myr, to fit the color-magnitude data (G vs. G-G$_{RP}$). This offers a better solution than G$_{BP}$-G$_{RP}$ since the G$_{BP}$ magnitudes are often underestimated, leading to an offset in age, although the relative age values are maintained in both cases.

For the age calculation, we excluded 52 objects with G$\lesssim$13 mag since they are already in the main sequence and the separation between isochrones in this age range is significantly less than the uncertainties in magnitude and extinction. Among the known members, we consider only those with magnitude errors $<$0.05 mag (537). The mean age is $\sim$4.0 Myr, with a standard deviation of $\sim$5.0 Myr. The mean age for the 296 new members is also $\sim$4.0 Myr (standard deviation $\sim$4.0 Myr), although we note that we imposed an age cut for selecting new members\footnote{Some new members have a final age $>$10 Myr despite the age cut because the individual ages are estimated using the average extinction, while the age cut is done for the minimum extinction significant ($A_0=1$ mag) for a cluster at 925 pc.}.

We also obtained the mean ages for each subcluster: A ($\sim$2 Myr), B ($\sim$4 Myr), E ($\sim$4 Myr), and the extended population G ($\sim$13 Myr). The older age for G, together with its very large age spread, suggests once again contamination. Subcluster F has only six known members, so its mean age is uncertain. An age histogram is shown in Figure \ref{fig:age_mass}. 
Since we had imposed a limit on the age for the new member selection, there is a bias against older ages. Considering only the known members, we found that subcluster A has 2\% members with ages$>$10 Myr, subcluster B has 12\% members with ages$>$10 Myr, and subcluster E has 7\%. Therefore, it is to be expected that the new member determination is worse for the older regions. 

The mean masses are 0.7 M$_\odot$ and 1.0 M$_\odot$ for the known- and the new members, respectively, so the region is dominated by low mass stars. 
There are only 4\% intermediate-mass stars ($>$5M$_\odot$) among the known members, but 14\% of the new members fall in this range. This confirms that we are preferentially identifying more massive sources, an expected bias considering that most surveys were more sensitive to lower-mass stars and the Gaia quality data requirements. We find six new members with M$>$5 M$_\odot$ previously identified as MS stars with spectral types B1-B8. The most massive stars are HD 239689 \citep[B2,][]{Patel1998} for which we find $\sim$11 M$_\odot$, and LS III +57 12 \citep[B1,][]{Marschall1987}, with $\sim$18 M$_\odot$. Other massive members confirmed by Gaia are BD+56 2622, HD 239731, BD+57 2363, and BD+58 2294. Among the previously-known members, the most massive stars are HD 239729 \citep[B0,][$\sim$10 M$_\odot$]{Patel1998,Mercer2009A} and LS III +57 14 \citep[B1,][$\sim$16 M$_\odot$]{Mercer2009A}.
Figure \ref{fig:age_mass} also shows the final distribution of masses for known and new members. 

We use the Anderson-Darling test to check the statistical differences of the age and mass distributions of known and new members. Here, we must note that although we can compare the known members of different subclusters between themselves since we have imposed a cut in the isochrone for the final selection of new members, comparisons of the age differences between known and new members can only be performed after making the same cut for both collections.

First, we test the age/mass distribution of known members in subclusters A, B, E, and G. We do not consider subcluster F because it has too few members. Subcluster E has only 11 members, which limits the reliability of the test. The results are shown in Table \ref{tab:AD_agemass_KM}, the age distributions are statistically different between subcluster A and subclusters B, G, and probably E. There are no statistically significant differences between the mass distributions, except for the subclusters B and G, but this is because population G has many members with larger masses than subcluster B, which again evidences contamination. The age differences between subclusters A and B, and E (despite their low number of members), further confirms that the IC1396 region is very complex and contains various star formation episodes.


\begin{table}
    \caption{Anderson-Darling test results for age and mass distributions of all the previously known members.}\label{tab:AD_agemass_KM}
    \centering
    \begin{tabular}{cccc}
    \hline\hline
     Subcluster & Nº members & Age    & Mass   \\
     {[i,j]} & {[n$_i$,n$_j$]}& significance & significance \\
     \hline
      {[A,B]} & {[252,22]} & $<$1\% & 5-10\%   \\
      {[A,E]} & {[252,11]} & $<$1\% & ---      \\
      {[A,G]} & {[252,32]} & $<$1\% & 1-2.5\%  \\
      {[B,E]} & {[22,11]} & ---    & 10-25\%  \\
      {[B,G]} & {[22,32]} & 2.5-5\% & $<$1\%  \\
      {[E,G]} & {[11,32]} & 5-10\% & 10-25\%  \\
     \hline 
    \end{tabular}
    \tablefoot{The cases where the differences are not statistically different are marked with "--". Statistically different results have $<$1\%. To test the ages, we do not consider the massive members (see discussion on text). The number of members reflects those used for the age test.}
\end{table}

\begin{table}
    \centering
    \caption{Anderson-Darling test for the age and mass distributions for each subcluster, including the total population (KM+NM).}
    \begin{tabular}{cccc}
    \hline\hline
     Subcluster & Nº members & Age  & Mass  \\
     {[i,j]} & {[n$_i$,n$_j$]}& significance & significance\\
     \hline
      {[A,B]} & {[365,80]} & $<$1\%  & 5-10\%  \\
      {[A,E]} & {[365,73]} & $<$1\%  & $<$1\%  \\
      {[B,E]} & {[80,73]} & $<$1\%  & 5-10\%   \\
     \hline 
    \end{tabular}
    \label{tab:AD_agemass_KM_NM}
    \tablefoot{Populations are statistically different if have $<$1\%. The number of members reflects those used for the age test, for which we do not consider the massive members.}
\end{table}
We also ran an Anderson-Darling test for the total population in subclusters A, B, and E. For this, the known members are cut the same way as new members, adopting the fixed extinction A$_V$=1 mag and those between isochrones 0.1 - 10 Myr. Table \ref{tab:AD_agemass_KM_NM} shows that the age distributions of subclusters A, B, and E are statistically different. 
There are no significant differences in the mass distribution other than those induced by poor statistics in the clusters with few members.
Our results confirm age differences between subclusters A, B, and E, suggesting various episodes of star formation giving rise to the subclusters. This adds to the previous evidence of age differences and triggered and sequential star formation \citep{Patel1995, Reach2004, Sicilia2006, Sicilia2019, Getman2012}.

Analyzing subclusters B and E together with the spatial distribution of their members, we find that they are not smoothly distributed and cluster in front of the BRCs. Subcluster B members tend to appear in front of IC1396N, and subcluster E members group in two regions, in front of IC1396G in the east and behind IC1396D towards the north-west (see Figure \ref{fig:spatialdist_EDR3}). We tested the age distribution of the subclusters B and E members in spatial proximity to the BRCs versus the rest of the non-clustered B and E members for the entire population (known and new members).
The Anderson-Darling test shows that for subcluster B, the clustered and non-clustered members are statistically different at a significance level $<$1\%. For subcluster E, the age distributions between the two clustered populations and the spread members are not significantly different. Therefore, there is some evidence that clustered objects are younger, on average, than the spatially extended population, despite both belonging to the same proper motion group. This is, to some extent, what is also observed between the IC1396A and the Tr37 population, which have similar proper motions but different ages \citep{Sicilia2019}.  Nevertheless, the picture is not complete because although there are clearly younger and less evolved sources in the bright-rimmed clouds \citep{Beltran2009, Reach2004, Sicilia2006, Sicilia2014} and we confirm this for subclusters A and B \citep[populations related to the BRC IC1396A and IC1396N, see also][]{Sicilia2019}, Gaia cannot detect embedded objects, so we are missing part of the population. 

The global ages observed in the region, comparing the center and the outskirts, do not point to star formation triggered by HD 206267 since subcluster A is, on average, younger. Age distributions where the central part of the cluster is younger than the outskirts have been observed \citep[][in NGC 2024, Orion nebula cluster, respectively]{Kuhn2014, Kuhn2015, Getman2014}. More complex, non-triggered structures may result from the coalescence or different star formation processes \citet{Smilgys2017}. 

\subsection{Disk fraction among known and new members \label{subsec:disk_KM}}

\begin{figure}
    \centering
    \begin{tabular}{c}
    \includegraphics[width=0.48\textwidth]{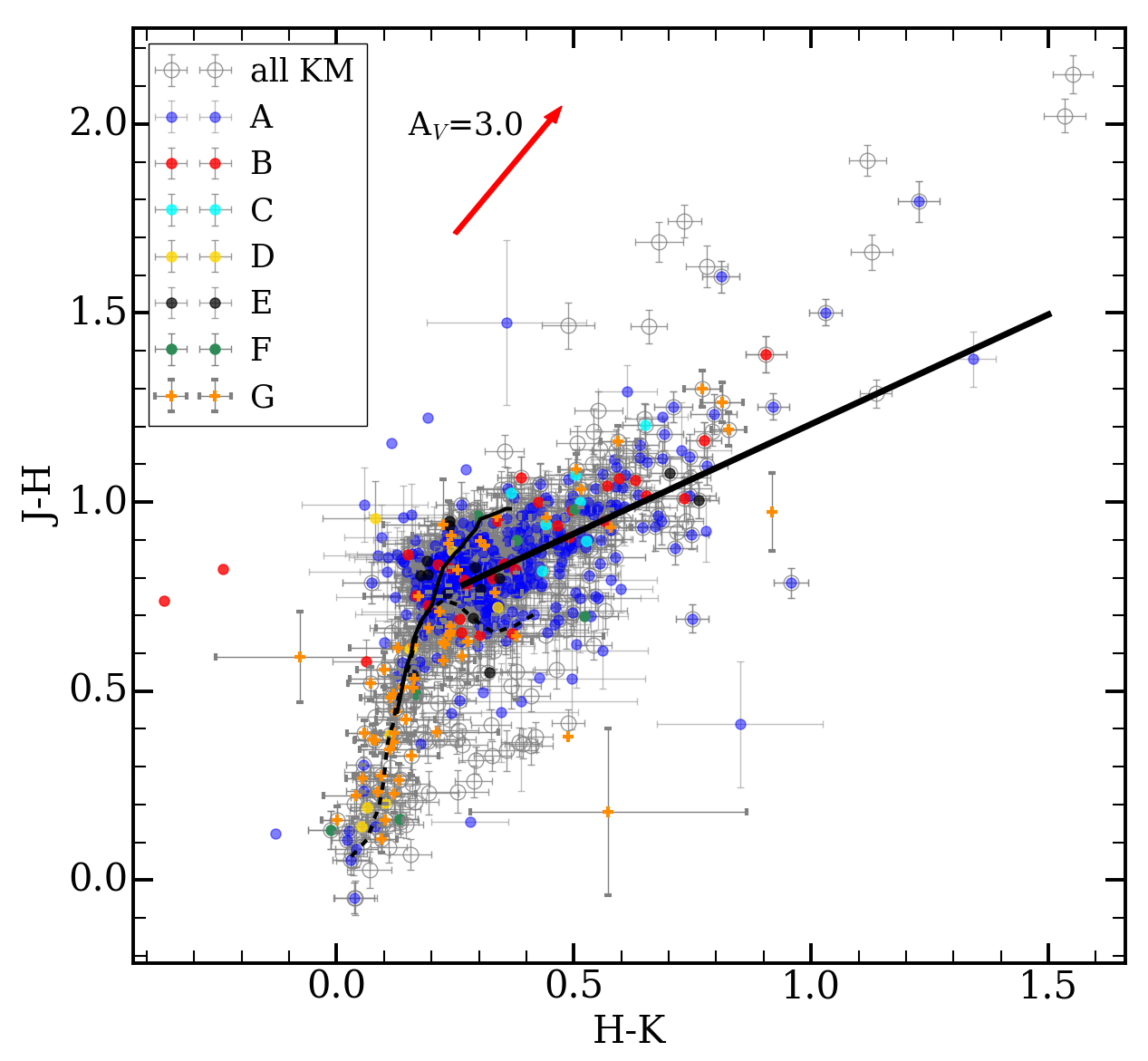}
    \end{tabular}
    \caption{J-H vs. H-K diagram for all known members (KM) with magnitude uncertainties $<$0.05 mag. The MS and the giant branch \citep{BessellBrett1988}, as well as CTTS locus \citep{Meyer1997}, are indicated by a solid line. The theoretical tracks are corrected for extinction in JHKs \citep{Cardelli1989}, using the threshold A$_V$=1 mag. The extinction vector is indicated with a red arrow. KM in full color belong to the populations (A, B, C, D, E, F, and G) found from the maximum likelihood analysis.}
    \label{fig:JH_HKallKM}
\end{figure}

Since the known members had been identified with different techniques (X-rays emission, H$\alpha$ excess from photometry or spectroscopy, and near- and mid-IR excess), it is necessary to characterize the presence of disks uniformly so that we can compare those properties with the newly identified members.
We built a J-H vs. H-K diagram using the JHKs photometry from the 2MASS survey to characterize as many members as possible. After collecting 1538 known members, we obtained 591 JHKs counterparts (with a matching radius of 0.6") and photometric errors $<$5\%. The color-color diagram is consistent with intermediate- and low-mass members (see Figure \ref{fig:JH_HKallKM}). A significant part of the objects lies within the locus of the CTTS \citep{Meyer1997} and likely harbor primordial disks, but longer wavelength data is required to verify the disk fraction independently of extinction effects. 


\begin{figure}
    \centering
    \begin{tabular}{c}
    \includegraphics[width=0.48\textwidth]{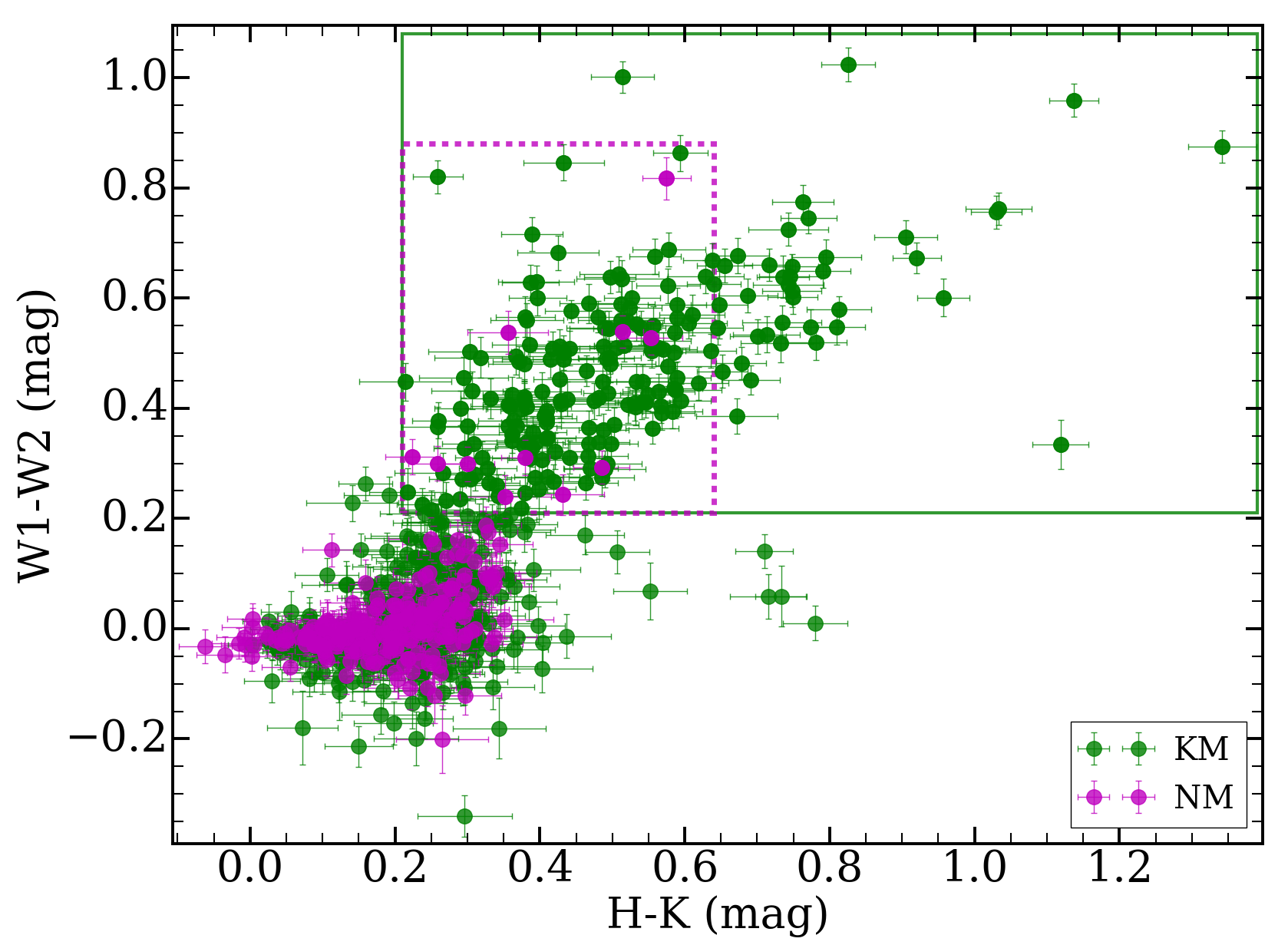}
    \end{tabular}
    \caption{W1-W2 vs. H-K diagram for all members with magnitude errors $<$0.05 mag. Known members (KM) are shown in green and new members (NM) in magenta. The green and magenta boxes enclose the known and new members, respectively, with color excesses consistent with disk emission.}
    \label{fig:W1W2_HK_allmemb}
\end{figure}

We further revised this classification using the WISE data combined with 2MASS, obtaining 518 known members with complete, good-quality data in the W1, W2, H, and Ks bands and errors up to 0.05 mag. The W1-W2 vs. H-Ks diagram (Figure \ref{fig:W1W2_HK_allmemb}) reveals 200 sources with excess emission $>$3$\sigma$, consistent with protoplanetary disks, or a disk fraction of $\sim$39\%. This fraction is slightly lower than what has been derived from other studies \citep[e.g., 45\% from][]{Sicilia2006}, which likely arises from the fact that we are not able to detect disks with low excess or inner gaps.  We tested this classification using the 148 objects in common with \citet{Sicilia2006, Sicilia2013}, which were classified using the multi-wavelength spectral energy distribution (SED), including optical and Spitzer data. With our method, we recover all the diskless objects. All those we detect as harbouring disks correspond to disky sources, and we only fail to detect as disks 10\% of the weakest disks classified via the SEDs, consistent with transitional or settled/depleted disks.
This means that our methods agree very well, so the disk populations, excluding transition and depleted disks, are comparable. Also note that we study a more extended region, where ages/disk fractions could differ.

We also investigated the presence of disks around the 334 new members (see Table \ref{tab:samples_EDR3}) identified with Gaia, of which 240 have W1-W2 and H-K data. Figure \ref{fig:W1W2_HK_allmemb} reveals 11 sources consistent with excess emission larger than 3$\sigma$. Therefore, the disk fraction for new members is 5\%, significantly lower. This could be due to the fact that with Gaia, we are preferentially identifying  more massive stars, which are thought to lose their disks faster \citep{Contreras2002, Hernandez2005}. In addition, the known members were preferentially identified using methods that favor lower-mass and disked stars (e.g. photometry, spectroscopy, infrared excesses), while Gaia identifies sources independently from their disk status.

The final disk fraction for stars with all masses, putting together the objects, is 28\%.
We also revised the disk fraction for subcluster A, the most populous one, which is 30\%. This result agrees with the disk fraction calculated by methods unbiased towards diskless sources, such as X-ray, e.g., the $\sim$39\% fraction derived by \citet{Mercer2009A}, the 29\% derived by \citet{Getman2012} and the 33\% fraction from \citet{Silverberg2021}.


\subsection{Feedback evidence in the BRCs \label{subsec:feedbackBRCs}}

We searched for evidence of interaction and feedback between the stars and the cloud in IC1396A, IC1396B, and IC1396N. 
For this purpose, we combined the [SII], H$_2$, and the Ks band images to unveil the places of interaction between the stars and the clouds at different depths identifying shocks, knots, and jets. We also compared [SII] (which results from combining the 6716 and 6730 \r{A} images) with 6750 \r{A} and H$_2$ with Br$\gamma$ to distinguish them from continuum emission.

We detect substantial emission in H$_2$ and [S II] at the edge of all the BRCs (Appendix \ref{appe:3colorimages}, Figures \ref{fig:F1}, \ref{fig:F2}, \ref{fig:F3}). The combination of these two emission lines and the fact that the extinction in the globules is larger than A$_V \sim$1 mag, the limit for significant FUV irradiation \citep{Burton1992}, rule out the possibility of fluorescent emission caused by far-UV field from HD206267. Some fluorescent emission may nevertheless be present at the edges of the globules.
For IC1396N, we clearly detect several knots and chains of knots in the deep continuum-subtracted H$_2$ images, which match with previous observations \citep{Nisini2001,Caratti2006,Reipurth2003,Beltran2002,Beltran2009,Rebull2013}. A strong chain of knots \citep[labeled B1, B2, and B3,][]{Beltran2002} is also clearly visible in the JHKs images, and the weakest H$_2$ features are also observed in Ks band due to the width of the filter. The H$_2$ knots lie in the same direction than the outflows found by \citet[][]{Beltran2002,Beltran2012}. For the N and S outflows \citep[driven by the C and I sources, respectively;][]{Codella2001}, the trajectories followed by the knots suggest the collision of both outflows, as recently pointed out by \citet{Lopez2022}.

We also detect shocks, possibly the result of strong outflows coming from YSOs inside and breaking the wall of the cloud, impacting the material surrounding IC1396N. Bow shock 3 \citep[HH 777,][]{Reipurth2003} is more intense than the others, and possibly driven by the near infrared (NIR) source 331A \citep[21:40:41.4 +58:16:38.4;][]{Reipurth2003}. The same source could be responsible for the other bow shock just below bow shock 3. Shock 2 could be driven by the 1.3 mm continuum source IC1396N-C \citep[source C from][]{Codella2001}, and it would be associated with the chain of knots around the source. We detected 2 weaker shocks, numbered 1 and 4. All shocks are shown in Appendix \ref{appe:3colorimages}.

In IC1396A, we detected several known members inside and outside the globule, and few new members, mainly scattered around the globule, since most members in the globule were already known. We observe a small jet close the "eye" of IC1396A (Appendix \ref{appe:3colorimages}, Figure \ref{fig:F1}). In IC1396B, we observe mainly the shocked material on the small irregular edges of the cloud by the combination of H$_2$ and [SII] emission. Note that most of the known members are around the globule, and we do not find any new members, suggesting a lower rate of embedded star formation, as mentioned before. In IC1396B, most of the known members are found around the globule, and we did not find any new embedded sources with our study, which suggests a lower rate of embedded star formation. However, our study has a bias against very low mass and highly embedded stars. For this globule, most of the excited and shocked material we observe in the figure is probably caused by external sources and appears along the irregular edges of the cloud (Appendix \ref{appe:3colorimages}, Figure \ref{fig:F3}).


\section{Summary and conclusions} \label{sec:conclu} 

In this paper, we study the star formation history and the global structure of IC1396, including the Tr37 cluster, and the properties and kinematics of their young stars. For this purpose, we first identify, classify and confirm new members using optical spectroscopy. We then compile the complete collection of known members, including those from the literature, and use their parallaxes and proper motions to characterize the region. This allows us to identify further stars consistent with cluster members. The revised collection of members allows us to study the disk fraction, ages, and masses more uniformly.

The Gaia data reveal a complex star formation history. IC1396 is composed of several substructures that are clearly distinct in proper motion and, some of them, in age. This, together with the spatial distribution of members, suggests a multi-episodic formation process. We also find that the region is not gravitationally bound and is expanding as a whole. Expansion is also evident in subcluster A. Gaia EDR3 data confirms a distance of 925$\pm$73 pc, with the bulk of members being distributed around 113 pc (standard deviation), slightly closer than the value derived from DR2 (943$\pm$82 pc). 
A detailed summary of the results is provided below.

We obtained optical spectra for 121 sources, deriving their spectral types, interstellar extinction, and accretion rates (from veiling), and reviewed their membership. Among those, we confirm 66 as members (2 of which are new members) based on their spectral features, extinction, presence of disks, and/or kinematics. An additional 42 are classified as probable members. We also rejected 13 sources as probable non-members, of which six were previously classified as members. For 111 of them, there were no previous estimates of the spectral type. The veiling measurements allow us to estimate accretion rates for 25 sources, which are in the range of 10$^{-9}$-10$^{-8}$ M$_\odot$/yr. These accreting sources consistently display strong emission lines and have IR excesses from disks. We derive individual extinction values and obtain an average extinction of A$_V$=1.4 mag for the region.

Using Gaia EDR3, we confirm that IC1396 contains four independent subclusters (A, B, E, F), significantly different in proper motion but not in parallax. The various populations are located in clearly distinct areas around the bubble, with subcluster A filling most of the center, around HD206267. Subclusters B and E run along the ridges and BRCs associated with IC1396N, IC1396D, and IC1396G. Subcluster E is a new subcluster entirely identified in our analysis. In addition, we also observe a population very extended in proper motion. Although part of the extended population could correspond to contaminants, it also clearly contains bona-fide members, showing that IC1396 has some high proper motion members.
       
We identify 334 new members using the multidimensional (parallax and proper motion) information provided by Gaia in a region $\sim$2$^o$ in radius. We confirm the mean age of the region to be 4 Myr. We use 2MASS and WISE data to re-calculate the disk fraction in a self-consistent way for all sources. We obtain a disk fraction for all known members of 39\%, consistent with previous values excluding anemic and transitional disks. Including the newly-identified members, the disk fraction is 28\%, lower than previous studies, due to our method detecting mainly more massive and less embedded sources.

The color-magnitude diagrams show that most new members are intermediate- to solar-mass stars. This is reasonable since previous studies have identified members using techniques biased towards the lowest-mass members and confirm that Gaia data can help complete the stellar content of young regions for objects that are otherwise more difficult to identify. 

Our results reveal a clear velocity structure in IC1396 on scales up to $\Delta$r$<$ $\sim$12-14 pc (which is approximately the edge of the ionized cloud). The expansion is a sign that the region is not gravitationally bound. We do not find any velocity structure on scales $\sim$12$<\Delta$r$<$30 pc, as expected from the velocity dispersion in the region, $\sim$2-3 km/s, for a cluster age of $\sim$4 Myr. The velocity structure in the region, along with the spatial distribution of all members, also suggests different episodes of star formation.

We find differences between the ages of the subclusters, with the populations in the outskirts of subclusters B and E being older (on average) than subcluster A, except for the subcluster B objects spatially related to IC1396N. We confirm that the members of IC1396A do not show any significant difference in proper motion compared to the Tr37 members, which form most of subcluster A, which makes it a case similar to subcluster B. This is counterintuitive considering some triggered star formation models \citep[e.g.][]{Mintz2021} and indicates that even physically related groups may experience several star-forming events. Nevertheless, one limitation to further delving into this issue is that most objects physically associated with the globules are too embedded and not detected by Gaia.

The kinematics, age, and evolutionary state differences in groups of sources within the same cloud suggest that IC1396 has suffered more than one-star formation episode over time from various mechanisms, although radial velocity differences mean that triggering is not obvious \citep{Sicilia2019}.
This, together with the spatial distribution of their members, indicates multiple episodes of star formation, which may be indicative of other processes such as coalescence \citep{Smilgys2017}. The importance of Gaia data to complete cluster surveys, including intermediate-mass stars, is also demonstrated.

\begin{acknowledgements}
We thank the anonymous referee for providing valuable comments that helped to clarify this paper. 
M.P.B is supported by the funding of the Peruvian National Council of Science, Technology and Technological Innovation (``Consejo Nacional de Ciencia, Tecnología e Innovación Tecnológica-CONCYTEC)'') via a predoctoral grant, contract ``033-2016-FONDECYT (Peru)''.\\
We thank Konstantin Getman for his valuable comments and for providing his unpublished list of X-ray sources for complete the collection of known members.\\
Observations reported here include data obtained at the MMT Observatory, a joint facility of the University of Arizona and the Smithsonian Institution, and observations made with the Gran Telescopio Canarias (GTC), installed at the Spanish Observatorio del Roque de los Muchachos of the Instituto de Astrofísica de Canarias, on the island of La Palma. This work is (partly) based on data obtained with the instrument OSIRIS, built by a Consortium led by the Instituto de Astrofísica de Canarias in collaboration with the Instituto de Astronomía of the Universidad Autónoma de México. OSIRIS was funded by GRANTECAN and the National Plan of Astronomy and Astrophysics of the Spanish Government. 
This work is based on observations collected at the Centro Astron\'{o}mico Hispano-Alem\'{a}n (CAHA) at Calar Alto, operated jointly by Junta de Andaluc\'{i}a and Consejo Superior de Investigaciones Cient\'{i}ficas (IAA-CSIC).
This work has made use of data from: the European Space Agency (ESA) mission {\it Gaia} (\url{https://www.cosmos.esa.int/gaia}), processed by the {\it Gaia} Data Processing and Analysis Consortium (DPAC, \url{https://www.cosmos.esa.int/web/gaia/dpac/consortium}).  Funding for the DPAC has been provided by national institutions, in particular, the institutions participating in the {\it Gaia} Multilateral Agreement; from the Two Micron All Sky Survey, which is a joint project of the University of Massachusetts and the Infrared Processing and Analysis Center/California Institute of Technology, funded by the National Aeronautics and Space Administration and the National Science Foundation, and from the Wide-field Infrared Survey Explorer, which is a joint project of the University of California, Los Angeles, and the Jet Propulsion Laboratory/California Institute of Technology, funded by the National Aeronautics and Space Administration. 
We also have made use of data products of the SIMBAD database, of the VizieR catalog access tool, operated at CDS, Strasbourg, France, and of Astropy (\url{http://www.astropy.org}), a community-developed core Python package for Astronomy \citep{astropy:2013, astropy:2018}.
\end{acknowledgements}

This paper has been typeset from a TEX/LATEX file prepared by the author.

%
   \bibliographystyle{bibtex/aa} 
   \bibliography{bibtex/ic1396} 
%


\begin{appendix} 

\section{Collection of known members} \label{appe:TableKM}

Table \ref{tab:tableKM} shows the information on known members of IC1396 obtained from the literature.

\begin{table*}
   \caption{Total collection of known members from the literature.} \label{tab:tableKM}
    \centering
    \begin{tabular}{clll}
     \hline\hline
ID &  RA & DEC & Refs.\\
& (deg) & (deg)  &    \\
     \hline
2178430162991379712 & 323.5407187877424 & 57.49864031973727 & \citet{Sicilia2004}\\ 
2178420537960651264 & 323.81782973275574 & 57.47280785409621 & \citet{Sicilia2005}\\
2179307568959248512 & 323.154629032842  & 58.5449016845008 & \citet{Barentsen2011}\\
2179194048688570240 & 323.32405097814495 & 57.80372284131545 &\citet{Barentsen2011}\\
2178391920604542336 & 324.629408138989 & 57.466769136483826 & \citet{Mercer2009A}\\
2178397864836164096 & 324.63688539342223 & 57.48841744730858 & \citet{Mercer2009A}\\
------------------- & 324.235458 & 57.531 & \citet{Morales2009}\\
------------------- & 324.262667 & 57.513528 & \citet{Morales2009}\\
2178420778472686976 & 324.01622674909845 & 57.45336624201454 & \citet{Getman2012}\\
2178418235859643264 & 324.05927136332025 & 57.46595890845687 &\citet{Getman2012}\\
2179200680117948800 & 323.08768907947183 & 57.80498084917999 & \citet{Meng2019}\\
2179204317946731008 & 323.09246758140864 & 57.88204975557804 & \citet{Meng2019}\\
2179262458929142784 & 323.3001563983867 & 58.20147621775045 & \citet{Nakano2012}\\  
-------------------  & 323.314167 & 57.995333 & \citet{Nakano2012}\\
\hline
    \end{tabular}
    \tablefoot{The complete collection of known members is available in electronic format via CDS. Coordinates are from Gaia EDR3, except for objects without Gaia counterparts, for which we list the coordinates of the corresponding reference.}
\end{table*}


\section{Spectroscopic members} \label{appe:TableSpectra}

Table \ref{tab:tablespectra} shows the 121 spectroscopic candidates from MMT and GTC telescope observations, the results of the analysis of the different membership criteria, and the final membership derived from them all.

\longtab[1]{
\begin{landscape}
{\tiny
\setlength{\tabcolsep}{3pt}
\begin{longtable}{ccccccccccccccclll}
\caption{Spectroscopic candidates from MMT and GTC telescope observations.}\label{tab:tablespectra}\\
\hline\hline
Gaia EDR3 ID & RA & DEC & EW & EW & EW & Telesc. & Sp. & Veiling & RUWE & A$_0$ $\pm$ std & Mem & Mem & Disk? & Mem  & Final  & Refs. & Comments \\
 & (deg) & (deg) &  (H$\alpha$) & (Li I) & (Ca II) &  & Type & r$_{7465}$ &  &  (mag) & (lines) & (A$_0$) &(Ours/Ref) &(Gaia) & Mem &  &  \\
\hline
\endfirsthead
\caption{continued.}\\
\hline\hline
Gaia EDR3 ID & RA & DEC & EW & EW & EW & Telesc. & Sp. & Veiling&RUWE & A$_0$ $\pm$ std & Mem & Mem & Disk? & Mem  & Final  & Refs. & Comments \\
 & (deg) & (deg) &  (H$\alpha$) & (Li I) & (Ca II) &  & Type & r$_{7465}$&  &  (mag) & (lines) & (A$_0$) &(Ours/Ref) &(Gaia) &Mem &  &  \\
\hline
\endhead
\hline
\endfoot
\hline
\endlastfoot
2178434561038259840 & 323.960316 & 57.596456 & -0.4 & 0.3 & ---    & MMT & G7.8 & 0.00 & 0.95  & 1.8 $\pm$ 0.0 & Y & Y & ND/- & Y & Y & 7 & A \\ 
2178397315080376064 & 324.765688 & 57.494891 & ---- & 0.4 & ---    & MMT & K0.2 & 0.00 & 0.95  & 1.6 $\pm$ 0.1 & Y & Y & -/ND & Y & Y & 4 & A \\ 
2178395597093124992 & 324.384830 & 57.490160 & -0.3 & 0.5 & ---    & MMT & K4.2 & 0.00 & 1.05  & 1.5 $\pm$ 0.0 & Y & Y & ND/ND & Y & Y & 7 & A \\ 
2178397452519326976 & 324.788327 & 57.513955 & ---- & 0.5 & ---    & MMT & K4.6 & 0.00 & 1.00  & 1.2 $\pm$ 0.1 & Y & Y & ND/ND & Y & Y & 4 & A \\ 
2178394497581292800 & 324.446132 & 57.434589 & -0.2 & 0.5 & ---    & MMT & K4.8 & 0.00 & 1.05  & 0.9 $\pm$ 0.0 & Y & Y & ND/ND & Y & Y & 7 & A \\ 
2179212328069456512 & 323.315766 & 57.996542 & -0.8 & 0.4 & ---    & MMT & K5.2 & 0.00 & 1.22  & 2.2 $\pm$ 0.2 & Y & Y & -/- & - & Y & 8 & --- \\ 
2178397658677755008 & 324.712455 & 57.478499 & -1.7 & 0.5 & ---    & MMT & K5.2 & 0.00 & 1.02  & 1.3 $\pm$ 0.1 & Y & Y & ND/ND & Y & Y & 4 & A \\ 
2178387556914944896 & 325.008316 & 57.565332 & -0.9 & 0.5 & ---    & MMT & K5.2 & 0.00 & 1.02  & 0.8 $\pm$ 0.1 & Y & P & ND/- & Y & Y & EDR3 & A \\ 
2178289459857084160 & 325.610689 & 57.497822 & -6.3 & 0.5 & ---    & MMT & K5.4 & 0.00 & 1.05  & 1.1 $\pm$ 0.1 & Y & Y & ND/Y & - & Y & 6 & NoV \\ 
2178402400321563904 & 324.566584 & 57.615719 & -1.4 & 0.5 & ---    & MMT & K5.4 & 0.00 & 1.02  & 1.1 $\pm$ 0.1 & Y & Y & ND/ND & Y & Y & 10 & A \\
2178398414591976704 & 324.706992 & 57.532099 & ---- & 0.6 & ---    & MMT & K6.2 & 0.00 & 1.08  & 1.1 $\pm$ 0.1 & Y & Y & ND/ND & Y & Y & 4 & A \\ 
2178441089388920832 & 324.333159 & 57.479476 & -0.8 & 0.3 & ---    & MMT & K6.2 & 0.00 & 1.03  & 1.4 $\pm$ 0.1 & Y & Y & ND/ND & - & Y & 7 & --- \\ 
2178445453075794432 & 324.197551 & 57.579164 & -1.2 & 0.5 & ---    & MMT & K6.2 & 0.00 & 4.39  & 1.6 $\pm$ 0.6 & Y & Y & -/ND & - & Y & 7 & --- \\ 
2178443769448645376 & 324.298756 & 57.558096 & -1.7 & 0.5 & ---    & MMT & K6.2 & 0.00 & 1.17  & 1.6 $\pm$ 0.3 & Y & Y & -/ND & - & Y & 7 & --- \\ 
2178397757459759360 & 324.710058 & 57.501377 & -1.3 & 0.6 & ---    & MMT & K6.4 & 0.00 & 0.95  & 1.2 $\pm$ 0.1 & Y & Y & ND/ND & Y & Y & 4 & A \\ 
2178398105354352128 & 324.775995 & 57.517859 & -0.3 & 0.4 & ---    & MMT & K6.4 & 0.00 & 1.06  & 0.8 $\pm$ 0.0 & Y & P & ND/ND & Y & Y & 4 & A \\ 
2178398070994613888 & 324.748486 & 57.502202 & -7.3 & 0.5 & ---    & MMT & K6.6 & 0.00 & 1.04  & 1.2 $\pm$ 0.0 & Y & Y & Y/Y & Y & Y & 4 & LiI/A \\ 
2178396868403780096 & 324.742690 & 57.470825 & -2.0 & 0.5 & ---    & MMT & K6.6 & 0.00 & 1.01  & 1.1 $\pm$ 0.1 & Y & Y & -/ND & Y & Y & 4 & A \\ 
2178495549568315776 & 325.249122 & 57.734327 & -43.5 & 0.5 & -5.13/-5.26/-4.95 & MMT & K6.6 & 0.41 & 14.02  & 0.6 $\pm$ 0.2 & Y & P & Y/Y & - & Y & 6 & AL+V \\ 
2178286985955579904 & 325.350781 & 57.393637 & -2.7 & 0.7 & ---    & MMT & K6.6 & 0.00 & 1.09  & 0.8 $\pm$ 0.1 & Y & Y & ND/- & - & Y & 8 & --- \\
2179212259349982336 & 323.333867 & 57.981397 & -3.2 & 0.6 & ---    & MMT & K7.4 & 0.00 & 1.17  & 2.2 $\pm$ 0.3 & Y & Y & Y/--- & Y & Y & 7up & AL/F \\
2178296744121734912 & 325.569965 & 57.606092 & -49.4 & 0.5 & ---    & MMT & K7.4 & 0.33 & 1.03  & 0.9 $\pm$ 0.1 & Y & Y & ND/Y & Y & Y & 6 & H$_\alpha$+LiI+V/A \\ 
2178397624318007424 & 324.678487 & 57.481829 & -1.7 & 0.7 & ---    & MMT & K7.6 & 0.00 & 3.36  & 1.8 $\pm$ 0.5 & Y & Y & ND/ND & - & Y & 4 & --- \\ 
2178397143281700992 & 324.857156 & 57.495962 & -7.8 & 0.3 & ---    & MMT & K8.0 & 0.00 & 1.15  & 1.1 $\pm$ 0.1 & Y & Y & Y/Y & Y & Y & 4/10/12 & AL/A \\ 
2178396937123251840 & 324.734111 & 57.482892 & -1.4 & 0.5 & ---    & MMT & K8.0 & 0.00 & 1.06  & 1.1 $\pm$ 0.4 & Y & Y & -/ND & Y & Y & 4 & A \\ 
2178450160360276480 & 324.607917 & 57.701334 & -29.6 & 0.5 & ---    & MMT & K8.0 & 0.27 & 1.04  & 1.0 $\pm$ 0.4 & Y & Y & -/Y & Y & Y & 6/12 & H$_\alpha$+LiI+V/A \\
2179207998742440960 & 323.481364 & 57.914035 & -16.4 & 0.4 & ---    & MMT & K8.2 & 0.09 & 1.01  & 1.3 $\pm$ 0.2 & Y & Y & Y/--- & Y & Y & 10 & AL+V/F \\
2178397693040671744 & 324.727699 & 57.490203 & -0.3 & 0.5 & ---    & MMT & K8.4 & 0.00 & 1.13  & 1.5 $\pm$ 0.8 & Y & Y & -/ND & - & Y & 4 & --- \\ 
2178421779215464704 & 323.887185 & 57.479608 & -4.8 & 0.5 & ---    & MMT & K8.6 & 0.00 & 4.94  & 1.2 $\pm$ 0.3 & Y & Y & ND/- & - & Y & 14 & --- \\ 
2178395489703989504 & 324.481621 & 57.492245 & -3.2 & 0.6 & ---    & MMT & K9.2 & 0.00 & 0.96  & 1.1 $\pm$ 0.2 & Y & Y & ND/ND & Y & Y & 7/10 & A \\
2178442841735735168 & 324.411907 & 57.547055 & -2.2 & 0.5 & ---    & MMT & M1.0 & 0.00 & 1.09  & 1.6 $\pm$ 0.2 & Y & Y & ND/ND & Y & Y & 7 & A \\ 
2178432671252489984 & 323.792562 & 57.529835 & -43.8 & 0.6 & ---    & MMT & M2.6 & 0.12 & 1.73  & 1.6 $\pm$ 0.3 & Y & Y & Y/Y & - & Y & 6 & H$_\alpha$+LiI+V \\ 
2178458986501139840 & 324.118403 & 57.714324 & -112.8 & ---  & -16.83/-15.87/-14.33 & MMT & M3.0 & 0.73 & 1.00  & 0.7 $\pm$ 0.1 & Y & P & Y/Y & - & Y & 6 & AL+V \\ 
2178433805123877504 & 323.851714 & 57.502340 & -125.5 & ---  & -5.62/-5.16/-4.18 & MMT & M3.6 & 0.23 & 1.00  & 1.0 $\pm$ 0.2 & Y & Y & Y/Y & - & Y & 6 & AL+V \\ 
2178433942562827136 & 323.887034 & 57.533188 & -72.2 & ---  & -0.99/-1.11/-0.8 & MMT & M3.6 & 0.23 & 1.03  & 1.6 $\pm$ 0.3 & Y & Y & Y/Y & - & Y & 6 & AL+V \\ 
2178440642705389824 & 324.239399 & 57.458397 & -129.5 & ---  & -14.08/-14.38/-11.58 & MMT & M4.0 & 0.00 & 0.98  & 2.1 $\pm$ 0.8 & Y & Y & -/Y & - & Y & 6 & AL \\ 
2178547776372232704 & 324.911801 & 57.914236 & -77.3 & ---  & -2.09/-2.42/1.73 & MMT & M4.0 & 0.86 & 1.15  & 2.8 $\pm$ 1.4 & Y & Y & Y/Y & - & Y & 6 & AL+V \\ 
2178385838927821440 & 324.983837 & 57.452166 & -152.0 & ---  & -21.34/-21.30/-16.28 & MMT & M4.2 & 0.15 & 1.05  & 1.0 $\pm$ 0.4 & Y & Y & -/Y & - & Y & 6 & AL+V \\ 
2179218096198430848 & 323.868379 & 57.973070 & -178.1 & ---  & -37./-45.03/-36.94 & MMT & M4.4 & 0.83 & 0.98  & 0.6 $\pm$ 0.5 & Y & P & -/Y & - & Y & 6 & AL+V \\ 
2178434148721240320 & 323.820348 & 57.545050 & -198.3 & ---  & -29.93/-31.54/-26.44 & MMT & M4.8 & 0.55 & 1.03  & -0.2 $\pm$ 0.2 & Y & U & Y/--- & - & Y & 6 & AL+V \\ 
2178444177454554496 & 324.309163 & 57.604939 & -46.0 & ---  & -8.54/-8.12/-7.06 & MMT & M4.8 & 0.64 & 0.99  & 3.1 $\pm$ 1.3 & Y & P & Y/Y & - & Y & 7 & AL+V \\ 
2178443254045917568 & 324.442181 & 57.574461 & ---  & ---  & -3.18/-2.43/-2.41 & MMT & M5.6 & 1.95 & 0.93  & 1.9 $\pm$ 0.5 & Y & Y & -/Y & - & Y & 6 & CaII+V \\ 
2178385735848603776 & 324.966179 & 57.449502 & -20.4 & ---  & 2.123/1.818/0.949 & MMT & M5.0 & 0.38 & 1.03  & 1.5 $\pm$ 0.3 & Y & Y & -/Y & - & Y & 6 & AL+V \\ 
2178440814510995456 & 324.224779 & 57.466286 & ---  & 0.6 & ---    & GTC & G5.5 & 0.00 & 1.01  & 1.7 $\pm$ 0.0 & Y & Y & -/ND & Y & Y & 7 & A \\ 
2178446002831454848 & 324.067038 & 57.580120 & -2.0 & 0.8 & ---    & GTC & K1.9 & 0.00 & 1.02  & 2.5 $\pm$ 0.1 & Y & Y & ND/ND & - & Y & 7 & --- \\ 
2178369930368001536 & 324.206879 & 57.374168 & -9.7 & 0.5 & ---    & GTC & K3.9 & 0.00 & 1.07  & 1.3 $\pm$ 0.1 & Y & Y & ND/Y & Y & Y & 7 & NoV/A \\ 
2178392848313851392 & 324.368435 & 57.390246 & -1.3 & 0.5 & ---    & GTC & K4.8 & 0.00 & 1.09  & 1.4 $\pm$ 0.1 & Y & Y & ND/ND & Y & Y & 7 & A \\ 
2178393947825444864 & 324.320866 & 57.438297 & -3.5 & 0.5 & ---    & GTC & K5.6 & 0.00 & 1.08  & 1.6 $\pm$ 0.0 & Y & Y & Y/Y & Y & Y & 7 & LiI/A \\ 
2178395356574973056 & 324.420927 & 57.466933 & -5.2 & 0.5 & ---    & GTC & M0.6 & 0.00 & 1.03  & 1.3 $\pm$ 0.1 & Y & Y & -/ND & - & Y & 7 & --- \\ 
2178393771716897536 & 324.339668 & 57.436753 & -146.4 & ---  & -27.62/-27.85/-23.29 & GTC & M2.2 & 0.26 & 1.06  & 1.1 $\pm$ 0.7 & Y & Y & -/Y & Y & Y & 7 & AL+V/A \\ 
2178440741484235136 & 324.158434 & 57.449372 & ---  & ---  & -12.83/-13.07/-10.39 & GTC & M5.0 & 0.00 & 1.06  & 1.7 $\pm$ 1.0 & Y & Y & -/Y & - & Y & 5 & CaII \\ 
2178421671829472256 & 324.052480 & 57.523963 & -237.0 & ---  & -14.13/-14.82/-13.21 & GTC & M6.2 & 0.72 & 1.01  & 1.6 $\pm$ 1.8 & Y & U & -/Y & - & Y & 5 & AL+V \\ 
2178397040202477952 & 324.769910 & 57.471413 & ---  & 0.3: & ---    & MMT & G9.0 & 0.00 & 1.05  & 1.8 $\pm$ 0.0 & P & P & -/ND & Y & P & 4 & A \\ 
2178385838928060672 & 324.985907 & 57.459288 & ---  & 0.5: & ---    & MMT & K8.2 & 0.00 & 1.13  & 1.7 $\pm$ 1.0 & P & P & -/ND & - & P & 10 & ---- \\
2179209476211255552 & 323.411501 & 58.007403 & ---  & 0.8: & ---    & GTC & K4.6 & 0.00 & 0.96  & 1.7 $\pm$ 0.2 & P & P & -/- & Y & P & 7up & B \\
2178446140270407040 & 324.103477 & 57.608952 & -1.0 & 0.4: & ---    & GTC & K5.4 & 0.00 & 0.98  & 1.1 $\pm$ 0.1 & P & P & ND/ND & - & P & 7 & --- \\ 
2178441261187598208 & 324.289150 & 57.494766 & -0.4 & 0.3: & ---    & MMT & G8.0 & 0.00 & 0.93  & 1.7 $\pm$ 0.0 & P & P & -/ND & Y & P & 7 & A \\ 
2178398105354353664 & 324.764166 & 57.508008 & ---  & 0.3: & ---    & MMT & G8.2 & 0.00 & 1.76  & 1.7 $\pm$ 0.0 & P & P & ND/ND & - & P & 4 & --- \\ 
2178398895628585984 & 324.858822 & 57.529840 & ---  & 0.3: & ---    & MMT & K0.0 & 0.00 & 1.62  & 1.3 $\pm$ 0.0 & P & P & ND/ND & - & P & 4 & --- \\ 
2179285376874962304 & 323.455601 & 58.226037 & -16.2 & ---  & ---    & MMT & K0.0 & 2.58 & 1.08  & 2.4 $\pm$ 0.1 & P & Y & Y/--- & Y & Y & 10 & H$_\alpha$+V/A \\
2178391920604542336 & 324.629408 & 57.466769 & ---  & 0.3: & ---    & MMT & K0.4 & 0.00 & 1.09  & 1.7 $\pm$ 0.1 & P & P & -/ND & Y & P & 4 & A \\ 
2178394016544915328 & 324.294569 & 57.436194 & ---  & 0.4: & ---    & MMT & K1.8 & 0.00 & 1.08  & 1.0 $\pm$ 0.0 & P & P & ND/ND & - & P & 7 & --- \\ 
2178421813575206400 & 323.905550 & 57.478271 & -3.0 & 0.4: & ---    & MMT & K6.4 & 0.00 & 1.16  & 0.9 $\pm$ 0.1 & P & P & ND/- & - & P & 7 & --- \\ 
2178441055029170688 & 324.297757 & 57.491271 & -3.1 & 0.3: & ---    & MMT & K7.4 & 0.00 & 1.04  & 1.8 $\pm$ 0.1 & P & P & ND/ND & Y & P & 7 & A \\ 
2178400922852856448 & 324.629784 & 57.530438 & ---  & 0.3: & ---    & MMT & K8.4 & 0.00 & 0.98  & 1.1 $\pm$ 0.2 & P & P & -/- & - & P & EDR3 & --- \\ 
2178461872719263744 & 324.214778 & 57.801842 & -45.3 & ---  & ---    & MMT & K9.2 & 0.00 & 1.00  & 3.0 $\pm$ 1.4 & P & P & Y/Y & - & P & 6 & H$_\alpha$ \\ 
2178441604784967168 & 324.160031 & 57.488165 & -26.4 & ---  & ---    & MMT & K9.6 & 0.09 & 1.03  & 3.7 $\pm$ 0.8 & P & PN & ND/Y & - & P & 5/10 & H$_\alpha$+V \\
2178475792718532992 & 325.260374 & 57.451900 & ---  & 0.4: & ---    & MMT & M0.0 & 0.00 & 1.00  & 0.8 $\pm$ 0.1 & P & P & -/ND & - & P & 10 & --- \\
2178393054472271872 & 324.378334 & 57.416702 & -3.9 & 0.2: & ---    & MMT & M0.6 & 0.00 & 1.30  & 2.0 $\pm$ 0.7 & P & P & -/ND & - & P & 7 & --- \\ 
2178389618498468352 & 324.394164 & 57.346490 & -28.1 & 0.3: & ---    & MMT & M1.0 & 0.00 & 0.91  & 0.9 $\pm$ 0.1 & P & P & -/- & - & P & EDR3 & --- \\ 
2179248916888239744 & 323.882254 & 58.466785 & -58.2 & ---  & ---    & MMT & M1.2 & 0.00 & 0.97  & 2.5 $\pm$ 0.3 & P & Y & ND/Y & - & Y & 6 & H$_\alpha$ \\ 
2178397899195900032 & 324.656023 & 57.505707 & -38.8 & ---  & ---    & MMT & M2.4 & 0.00 & 0.99  & 1.2 $\pm$ 0.1 & P & P & -/P & - & P & 4/7 & --- \\ 
2178397658666232832 & 324.712240 & 57.475642 & -13.1 & ---  & ---    & MMT & M2.6 & 0.00 & 3.57  & 4.2 $\pm$ 2.7 & P & PN & Y/P & - & P & 4 & H$_\alpha$ \\ 
2179215832763067648 & 323.715379 & 57.861243 & -126.7 & ---  & ---    & MMT & M2.6 & 0.17 & 0.96  & 1.5 $\pm$ 0.1 & P & Y & -/Y & - & Y & 6 & H$_\alpha$+V \\ 
2178417278089773952 & 324.145473 & 57.436727 & -24.9 & ---  & ---    & MMT & M2.8 & 0.00 & 1.08  & 1.7 $\pm$ 0.2 & P & P & -/ND & - & P & 7/12 & PNL(12) \\
2178433938251317632 & 323.880654 & 57.524380 & -84.0 & ---  & ---    & MMT & M3.2 & 0.00 & 0.96  & 2.2 $\pm$ 0.6 & P & Y & -/Y & - & Y & 6/7 & H$_\alpha$ \\ 
2178398070997784960 & 324.749987 & 57.514765 & -60.6 & ---  & ---    & MMT & M3.2 & 0.97 & 0.96  & 1.2 $\pm$ 1.1 & P & Y & Y/P & - & Y & 4 & H$_\alpha$+V \\ 
2179224727641060352 & 324.033004 & 58.069681 & -58.8 & ---  & ---    & MMT & M3.4 & 0.00 & 1.11  & 1.6 $\pm$ 0.3 & P & Y & -/Y & - & Y & 6 & H$_\alpha$ \\ 
2179216824887982592 & 324.011633 & 57.920189 & -83.7 & ---  & ---    & MMT & M3.6 & 0.30 & 2.00  & 1.1 $\pm$ 0.3 & P & Y & Y/Y & - & Y & 6 & H$_\alpha$+V \\ 
2178496919651892224 & 325.187327 & 57.803815 & -148.5 & ---  & ---    & MMT & M3.6 & 0.00 & 1.08  & 1.9 $\pm$ 1.4 & P & Y & -/Y & - & Y & 6 & H$_\alpha$ \\ 
2178418167132388736 & 324.070784 & 57.444386 & -25.1 & ---  & ---    & MMT & M3.8 & 0.00 & 1.05  & 1.4 $\pm$ 0.2 & P & Y & Y/? & - & Y & 5/10/12 & H$_\alpha$ \\ 
2178351307389803008 & 324.459489 & 57.152589 & -16.2 & ---  & ---    & MMT & M3.8 & 0.00 & 1.11  & 5.8 $\pm$ 4.3 & P & PN & Y/ND & - & P & 12 & H$_\alpha$ \\
2179212328069456640 & 323.313996 & 57.995097 & -13.5 & ---  & ---    & MMT & M3.8 & 0.00 & 0.98  & 2.1 $\pm$ 0.4 & P & P & -/- & - & P & 7/10 & --- \\
2178369857340398208 & 324.172312 & 57.367809 & -31.8 & ---  & ---    & MMT & M3.8 & 0.00 & 1.06  & 2.5 $\pm$ 0.5 & P & P & -/ND & - & P & 12 & PL \\
2178417106293215104 & 324.092363 & 57.391083 & -22.0 & ---  & ---    & MMT & M4.0 & 0.00 & ---   & 4.7 $\pm$ 3.1 & P & PN & -/ND & - & P & 7/10/12 & --- \\ 
2178434767196685440 & 323.941123 & 57.611124 & -236.9 & ---  & ---    & MMT & M4.0 & 0.53 & 2.12  & 1.0 $\pm$ 0.9 & P & Y & Y/Y & - & Y & 6 & H$_\alpha$+V \\ 
2179224212235705344 & 324.197649 & 58.078482 & -88.7 & ---  & ---    & MMT & M4.2 & 0.13 & 1.63  & 3.4 $\pm$ 1.6 & P & PN & Y/Y & - & P & 6 & H$_\alpha$+V \\ 
2178480122032875904 & 325.169676 & 57.537420 & -74.5 & ---  & ---    & MMT & M4.8 & 0.00 & 1.03  & 2.4 $\pm$ 0.6 & P & Y & -/Y & - & Y & 6 & H$_\alpha$ \\ 
2178441428675371136 & 324.292226 & 57.524045 & -96.1 & ---  & ---    & MMT & M5.2 & 0.13 & 0.91  & 2.4 $\pm$ 1.3 & P & P & -/Y & - & P & 5/10 & H$_\alpha$+V \\
2178395425294441344 & 324.407814 & 57.479645 & -2.0 & 0.3: & ---    & GTC & K7.2 & 0.00 & 1.02  & 1.9 $\pm$ 0.1 & P & P & ND/ND & Y & P & 7 & A \\ 
2178421366898744320 & 324.053528 & 57.498982 & -2.1 & 0.4: & ---    & GTC & K8.8 & 0.00 & 3.15  & 2.3 $\pm$ 1.3 & P & P & ND/ND & - & P & 7 & --- \\ 
2178441089388913920 & 324.322583 & 57.490908 & -21.3 & ---  & ---    & GTC & K9.6 & 0.00 & 1.03  & 1.9 $\pm$ 0.3 & P & Y & Y/Y & Y & Y & 6/5/10 & H$_\alpha$/A \\
2178393398069567488 & 324.239518 & 57.389290 & -4.4 & 0.3: & ---    & GTC & M0.0 & 0.00 & 1.12  & 1.1 $\pm$ 0.2 & P & P & -/ND & - & P & 7 & --- \\ 
2178442974872913152 & 324.349750 & 57.548867 & -4.7 & 0.2: & ---    & GTC & M2.0 & 0.00 & 0.95  & 1.7 $\pm$ 0.4 & P & P & -/ND & - & P & 7 & --- \\ 
2178393088819611392 & 324.410292 & 57.435372 & -11.5 & ---  & ---    & GTC & M2.8 & 0.00 & 1.03  & 1.7 $\pm$ 0.8 & P & P & -/ND & - & P & 7 & --- \\ 
2178441669197172608 & 324.220037 & 57.495467 & -75.8 & ---  & ---    & GTC & M3.2 & 0.00 & 1.03  & 3.1 $\pm$ 3.9 & P & U & -/Y & - & P & 7 & H$_\alpha$ \\ 
2178393569868318208 & 324.287895 & 57.430134 & -129.9 & ---  & ---    & GTC & M3.4 & 0.00 & 1.08  & 1.3 $\pm$ 0.3 & P & P & -/ND & - & P & 6/7 & --- \\ 
2178395592793858944 & 324.386841 & 57.474184 & -5.8 & 0.4: & ---    & GTC & M3.6 & --- & 1.05  & --- & P & U & -/Y & - & P & 7 & NoV \\ 
2179213083983775744 & 323.371746 & 58.047405 & -27.0 & ---  & ---    & GTC & M4.0 & 0.00 & 1.72  & 2.8 $\pm$ 0.8 & P & Y & Y/- & - & Y & 10 & H$_\alpha$ \\
2178394188343600896 & 324.353173 & 57.485799 & -23.4 & ---  & ---    & GTC & M4.4 & 0.00 & 1.19  & 2.5 $\pm$ 0.7 & P & Y & Y/Y & - & Y & 6 & H$_\alpha$ \\ 
2178394080959535488 & 324.356203 & 57.462558 & -11.4 & ---  & ---    & GTC & M4.6 & 0.00 & 1.94  & 3.1 $\pm$ 1.3 & P & P & -/Y & - & P & 6 & H$_\alpha$ \\ 
2178385460970709888 & 324.993708 & 57.431154 & ---  & 0.2: & ---    & MMT & G5.2 & 0.00 & 1.40  & 1.3 $\pm$ 0.1 & PN & P & -/N & - & PN & 12 & UML \\
2178443528930624512 & 324.428842 & 57.608732 & -2.8 & ---  & ---    & MMT & K2.3 & 0.00 & 0.90  & 0.4 $\pm$ 0.0 & PN & PN & Y/Y & Y & P & 7 & A \\ 
2178397177641430784 & 324.822232 & 57.502357 & ---  & ---  & ---    & MMT & K7.2 & 0.00 & 0.96  & 1.9 $\pm$ 0.2 & PN & P & -/P & Y & P & 4 & A \\ 
2178397864836164096 & 324.636885 & 57.488417 & -5.1 & ---  & ---    & MMT & M2.2 & 0.00 & 1.02  & 1.2 $\pm$ 0.1 & PN & P & ND/ND & - & PN & 4 & --- \\ 
2178385460970715904 & 324.986803 & 57.415679 & ---  & 0.3: & ---    & MMT & M0.8 & 0.00 & 1.39  & 1.3 $\pm$ 0.4 & PN & P & ND/- & - & PN & EDR3 & --- \\ 
2178398070994610432 & 324.738872 & 57.507029 & -4.6 & ---  & ---    & MMT & M1.8 & 0.00 & 1.00  & 1.6 $\pm$ 0.3 & PN & P & -/P & - & P & 4 & --- \\ 
2178397796116441728 & 324.631094 & 57.483656 & -6.0 & ---  & ---    & MMT & M2.0 & 0.00 & 1.05  & 1.3 $\pm$ 0.1 & PN & P & ND/P & - & P & 4 & --- \\ 
2178403190595837440 & 324.829289 & 57.635192 & -4.9 & ---  & ---    & MMT & M3.4 & 0.00 & 1.11  & 1.0 $\pm$ 0.2 & PN & P & ND/ND & - & PN & 12 & NML \\
2179214076109611520 & 323.397300 & 58.113108 & ---  & ---  & ---    & GTC & A7.3 & 0.00 & 1.09  & 8.3 $\pm$ 1.9 & PN & PN & ND/- & - & PN & 9 & --- \\ 
------------------- & ---------- & --------- & -4.7 & ---  & ---    & GTC & G8.7 & --- & ---  & --- & PN & U & -/ND & - & PN & 7 & --- \\ 
2178440573992846080 & 324.268223 & 57.440456 & ---  & ---  & ---    & GTC & G0.3 & 0.00 & 1.04  & 1.8 $\pm$ 0.1 & PN & P & -/- & - & PN & 7 & UML \\ 
2178442669938739712 & 324.341791 & 57.512066 & -1.5 & ---  & ---    & GTC & M0.2 & 0.00 & 0.95  & 1.7 $\pm$ 0.2 & PN & P & ND/ND & Y & P & 7 & A \\ 
2178441226827857920 & 324.256374 & 57.479491 & -3.6 & ---  & ---    & GTC & M1.0 & 0.00 & 1.03  & 1.3 $\pm$ 0.1 & PN & P & ND/ND & Y & P & 7 & A \\ 
2178394050904650752 & 324.314410 & 57.454728 & -8.2 & ---  & ---    & GTC & M2.0 & 0.00 & 0.99  & 2.3 $\pm$ 0.2 & PN & P & Y/Y & - & P & 7 & --- \\ 
2178416934494524416 & 324.185214 & 57.398771 & -4.8 & ---  & ---    & GTC & M2.2 & 0.00 & ---   & 4.2 $\pm$ 2.8 & PN & PN & ND/ND & - & PN & 7 & --- \\ 
2179212465508486528 & 323.357388 & 58.027358 & -5.9 & ---  & ---    & GTC & M2.8 & 0.00 & 1.08  & 1.5 $\pm$ 0.3 & PN & P & -/- & - & PN & 8 & --- \\ 
2178441226827857792 & 324.262816 & 57.483628 & -5.7 & ---  & ---    & GTC & M2.8 & 0.00 & 1.02  & 1.5 $\pm$ 0.2 & PN & P & -/ND & - & PN & 7 & --- \\ 
2178393982185185920 & 324.342839 & 57.445544 & -4.6 & ---  & ---    & GTC & M2.8 & 0.00 & 0.97  & 1.4 $\pm$ 0.1 & PN & P & ND/ND & - & PN & 7 & --- \\ 
2179209093946918016 & 323.422074 & 57.954868 & -4.0 & ---  & ---    & GTC & M2.8 & 0.00 & 1.01  & 0.6 $\pm$ 0.0 & PN & PN & -/- & - & PN & 8 & --- \\ 
2178444903321664384 & 324.106343 & 57.526956 & -5.3 & ---  & ---    & GTC & M3.2 & 0.00 & ---   & 4.9 $\pm$ 4.4 & PN & PN & -/- & - & PN & 7/10 & UML \\
\end{longtable}
\tablefoot{Columns 4, 5, and 6 show the equivalent width (EW). Uncertain values are marked by (:). Columns 12,13,14 and 15 show the membership criteria. Column 12: Identification of youth spectral lines (Li I and CaII), confirmed member=Y, probable member=P, probable non-member=PN. Column 13: Membership according to the extinction value A$_0$. Some sources have uncertain values (labeled "U"), and it is corrected by the mean extinction value (1.6 mag). Column 14: Membership according to the presence of disk. We give two results, the first one is derived in this work (Y=object with disk, ND=No disk, "-"=Incomplete or uncertain data, no information on disks.), considering the typical error at 3$\sigma$, which are dominated by calibration. The second one is given by the literature. Column 15: Membership according to the Mahalanobis method. Column 16: The final membership combining all criteria.
Column 17: References (4) \citet{Mercer2009A}; (5) \citet{Morales2009}; (6) \citet{Barentsen2011}; (7) \citet{Getman2012};(8) Getman (private communication); (10) \citet{Nakano2012}; (12) \citet{Sicilia2013}; (14) \citet{Meng2019}. Column 18: Comments, including more information about the source. If the source has a disk (column 14), we specify if it has accretion lines (AL), Veiling (V), No Veiling (NoV), or only strong H$\alpha$ emission, CaII, or Li I lines. If the source has been identified by the Mahalanobis method (column 15), we specify the subcluster to which it belongs: A=Subcluster A, B=Subcluster B, and F=Subcluster F. If the source has been labeled as probable member, probable non-member, uncertain member or non-member-by the literature: PL=probable member, probable non-member=PNL, UML=uncertain member, NML=non member.}
}
\end{landscape}
}

\section{Maximum likelihood function} \label{appe:MLF}

The presence of subclusters in the region (Section \ref{subsec:likelihood}) was studied in a way similar to
\citet{Roccatagliata2018,Roccatagliata2020}, using the complete astrometric data for each object, including their uncertainties and correlated errors, and estimating which subcluster distribution and properties maximize the likelihood function. In this appendix, we include the details of the terms involved in the calculation of the individual likelihood function, L$_{i,j}$.

Each i-th star is defined by its astrometric data, which includes the trigonometric parallax ($\varpi_i$) and the proper motion in right ascension and declination ($\mu_{\alpha,i}$, $\mu_{\delta,i}$). Each quantity has its associated uncertainty, but because the uncertainties are not independent, we also have the cross-correlated uncertainties, $\rho_i$. In the multiparameter space, the position of star $i$ will be represented by the vector a$_i$=[$\varpi_i$, $\mu_{\alpha,i}$, $\mu_{\delta,i}$] and each j-th subcluster by the vector a$_{j}$=[$\varpi_{j}$, $\mu_{\alpha,j}$, $\mu_{\delta,j}$], where $\varpi_{j}$, $\mu_{\alpha,{j}}$ and $\mu_{\delta,{j}}$ are the astrometric positions for subcluster $j$.

In Eq. (\ref{eq_likelihood}), $a_{ij}$=($a_i$-$a_{j}$) is the vector that represents the difference between the astrometric parameter of the star vs each subcluster, and $(a_i$-$a_{j})^T$ is the transposed vector. Therefore, 
\begin{equation}
a_i-a_{j}=
    \begin{bmatrix}
    \varpi_i - \varpi_{j}\\
    \mu_{\alpha,i} - \mu_{\alpha,j}\\
    \mu_{\delta,i} - \mu_{\delta,j}\\
    \end{bmatrix}
\end{equation}

Following the formulation from \citet{Lindegren2000}, the analog of the Gaussian standard deviation in the multivariate Gaussian space is given by the covariance matrix C$_{i,j}$ (see equation (\ref{eq_likelihood}). The covariance matrix contains the standard uncertainties derived from the observational data, which will be $\sigma_{\varpi,i}$, $\sigma_{\mu_\alpha,i}$ and $\sigma_{\mu_\delta,i}$ for the errors associated to the parallax and proper motion measurements for the i-th star. For each j-th subcluster, $\sigma_{\varpi,j}$,  $\sigma_{\mu_\alpha,j}$ and $\sigma_{\mu_\delta,j}$ are the intrinsic dispersion of the parallax and proper motions. These, together with the correlated uncertainties between each pair of quantities, $\rho_i(\varpi,\mu_\alpha)$, $\rho_i(\varpi,\mu_\delta)$ and $\rho_i(\mu_\alpha,\mu_\delta)$, define
the covariance matrix C$_{i,j}$ given by
\begin{equation}
C_{i,j}=
    \begin{bmatrix}
    C_{ij,11} & C_{ij,12}&C_{ij,13}\\
    C_{ij,21} & C_{ij,22}&C_{ij,23}\\
    C_{ij,31} & C_{ij,32}&C_{ij,33}\\
    \end{bmatrix}
\end{equation}
where the various terms correspond to
\begin{multline}
[C_{ij,11}] = \sigma^2_{\varpi,i} + \sigma^2_{\varpi,j}\\
[C_{ij,22}] = \sigma^2_{\mu_\alpha,i} + \sigma^2_{\mu_\alpha,j}\\
[C_{ij,33}] = \sigma^2_{\mu_\delta,i} + \sigma^2_{\mu_\delta,j}\\
[C_{ij,12}] = [C_{ij,21}]= \sigma_{\varpi,i}\times \sigma_{\mu_\alpha,i}\times \rho_i(\varpi,\mu_\alpha)\\
[C_{ij,13}] = [C_{ij,31}]= \sigma_{\varpi,i}\times \sigma_{\mu_\delta,i}\times \rho_i(\varpi,\mu_\delta)\\
[C_{ij,23}] = [C_{ij,32}]= \sigma_{\mu_\alpha,i}\times \sigma_{\mu_\delta,i}\times \rho_i(\mu_\alpha,\mu_\delta).\\
\end{multline} 
The astrometric data, uncertainties, and correlated uncertainties for each individual object are provided by Gaia Archive, while the values for the subclusters are derived from the standard deviation of subcluster members during the iterations.

Starting with the first-guess subcluster parameters, we derive the likelihood using as initial positions those initial ones 
shifted along a grid to cover a large parameter space. The grid points are defined to sample the parameter space within 2$\sigma$ of the initial values. The subcluster values are refined on each loop by re-deriving it again using the data from the stars that are found to belong to a cluster with a probability $>$80\% to calculate a final likelihood for each refined grid point. The final subcluster positions will be those that achieve the maximum likelihood.


\section{Table of maximum likelihood results}\label{appe:TabML_EDR3}

Table \ref{tab:tableML_EDR3} shows the probabilities of the known members to belong to each subcluster, resulting from the maximum likelihood analysis from the Gaia EDR3 data. 
Objects with probabilities $>$80\% are considered members of one of the subclusters. Sources with a lower probability are probable members of one of the subclusters. Sources with probability $>$40\% for two subclusters are considered probable members of both subclusters. Objects with very small probabilities may still be members of the region but may belong to the disperse population, and they are labeled as members that do not belong to any subcluster.

\longtab[1]{
\begin{landscape} 
{\tiny
\begin{longtable}{ccccccccccccccccc}
\caption{Subcluster probabilities for known members} \label{tab:tableML_EDR3}\\
\hline\hline
ID Gaia EDR3 &  RA & DEC & $\varpi$ $\pm$ $\varpi_e$ & $\mu_\alpha$ $\pm$ $\mu_{\alpha_e}$ & $\mu_\delta$ $\pm$ $\mu_{\delta_e}$ & RUWE &  P$_A$  & P$_B$  & P$_C$ & P$_D$ & P$_E$ & P$_F$ & P$_G$ &Subcluster & Mem &   Refs.\\                
&  (deg)     & (deg)  & (mas) & (mas/yr)  & (mas/yr) &        &        &        &        &        &        &        &     &     &   \\
\hline
\endfirsthead
\caption{continued.}\\
\hline\hline
ID Gaia EDR3&  RA & DEC & $\varpi$ $\pm$ $\varpi_e$ & $\mu_\alpha$ $\pm$ $\mu_{\alpha_e}$ & $\mu_\delta$ $\pm$ $\mu_{\delta_e}$ & RUWE &  P$_A$  & P$_B$  & P$_C$ & P$_D$  & P$_E$ & P$_F$ & P$_G$ &Subcluster & Mem &   Refs.\\                
&  (deg) & (deg)  & (mas) & (mas/yr) & (mas/yr) &        &        &        &        &        &        &        &     &     &   \\
\hline
\endhead
\hline
\endfoot
\hline
\endlastfoot
2178420537960651264 & 323.817830 & 57.472808 & 1.065$\pm$0.052 & -3.117$\pm$0.067 & -5.102$\pm$0.064 & 0.973 & 0.99 & 0.00 & 0.00 & 0.00 & 0.00 & 0.00 & 0.01 & A & Y & 2 \\ 
2178434938995215232 & 323.827559 & 57.569180 & 1.101$\pm$0.032 & -2.990$\pm$0.039 & -4.903$\pm$0.036 & 1.060 & 1.00 & 0.00 & 0.00 & 0.00 & 0.00 & 0.00 & 0.00 & A & Y & 2 \\ 
2178434217440720512 & 323.852133 & 57.550292 & 1.180$\pm$0.058 & -3.065$\pm$0.070 & -4.882$\pm$0.061 & 0.968 & 1.00 & 0.00 & 0.00 & 0.00 & 0.00 & 0.00 & 0.00 & A & Y & 2 \\ 
2178417656046880384 & 323.957301 & 57.401123 & 1.142$\pm$0.051 & -4.139$\pm$0.066 & -5.210$\pm$0.060 & 1.020 & 0.01 & 0.00 & 0.00 & 0.00 & 0.00 & 0.00 & 0.99 & G & Y & 2 \\ 
2178446174630143488 & 324.031103 & 57.574794 & 0.650$\pm$0.057 & -2.818$\pm$0.071 & -2.139$\pm$0.062 & 0.952 & 0.00 & 0.02 & 0.00 & 0.00 & 0.00 & 0.00 & 0.98 & G & Y & 11 \\ 
2178391886229562624 & 324.614155 & 57.455895 & 1.414$\pm$0.301 &  0.208$\pm$0.738 & -2.928$\pm$0.494 & 1.206 & 0.10 & 0.06 & 0.00 & 0.00 & 0.41 & 0.00 & 0.43 & E/G & P2 & 2 \\
2179201053771428480 & 323.106599 & 57.827847 & 1.153$\pm$0.086 & -3.981$\pm$0.099 & -4.161$\pm$0.092 & 1.033 & 0.00 & 0.00 & 0.00 & 0.00 & 0.00 & 0.01 & 0.01 & - & NS & 8 \\ 
2178465759681247104 & 323.981837 & 57.844343 & 0.722$\pm$0.043 & -3.409$\pm$0.050 & -3.357$\pm$0.046 & 0.944 & 0.00 & 0.01 & 0.00 & 0.42 & 0.00 & 0.00 & 0.57 & D/G & P2 & 8 \\ 
2178483072676684928 & 325.340837 & 57.681176 & 1.297$\pm$0.147 & -2.101$\pm$0.170 & -2.097$\pm$0.167 & 1.390 & 0.00 & 0.79 & 0.00 & 0.00 & 0.00 & 0.00 & 0.21 & B & P & 8 \\ 
2178487681188556928 & 325.649571 & 57.747945 & 1.000$\pm$0.058 & -1.462$\pm$0.074 & -3.911$\pm$0.071 & 1.000 & 0.14 & 0.06 & 0.00 & 0.00 & 0.61 & 0.00 & 0.19 & E & P & 8 \\ 
\end{longtable}
\tablefoot{Column 1: ID Gaia EDR3. Columns 2, 3, 4, 5, and 6 are the astrometric parameters of each source, position, parallax, and proper motion, along with errors, respectively. Column 7: RUWE value. Columns 8 to 14 show the probability for each source belonging to one of the subclusters, A, B, C, D, E, F, and the extended population G. Column 15: Subcluster to which the source belongs. Column 16: Membership labeled as "Y" (member, with probability $>$80\%), "P" (probable member to belong to one of the subclusters, with probability $<$80\%), "P2" (probable member of belonging to two subclusters with probabilities $>$40\%), "NS" (no subcluster, low probability of not belonging to any cluster). Column 19: References of previous studies,
(1) \citet{Sicilia2004}; (2) \citet{Sicilia2005}; (3) \citet{Sicilia2006}; (4) \citet{Mercer2009A}; (5) \citet{Morales2009}; (6)\citet{Barentsen2011}; (7) \citet{Getman2012};(7up) Getman (unpublished); (8) \citet{Nakano2012}; (9) \citet{Rebull2013}; (10) \citet{Sicilia2013}; (11) \citet{Sicilia2015}; (12) \citet{Meng2019}. The complete version of 578 known members is available in electronic format via CDS.} 
}
\end{landscape}
}


\section{New members table} \label{appe:TableMahalanobis}

Table \ref{tab:tableMD_EDR3} shows the new members found using the Mahalanobis distance method. These new members have a probability of 95\% or higher to belong to one of the four reliable subclusters, A, B, E, and F, and members with lower probabilities of belonging to two subclusters. 

\longtab[1]{
\begin{landscape} 
{\small
\begin{longtable}{cccccccccccc}
\caption{New members found from the Mahalanobis distance analysis, using the Gaia EDR3} \label{tab:tableMD_EDR3}\\
\hline\hline
ID Gaia EDR3 & RA & DEC & $\varpi$ $\pm$ $\varpi_e$ & $\mu_\alpha$ $\pm$ $\mu_{\alpha_e}$ & $\mu_\delta$ $\pm$ $\mu_{\delta_e}$ & RUWE & P$_A$ & P$_B$  & P$_E$ & P$_F$  & Subcluster \\
& (deg) & (deg)  & (mas)  & (mas/yr) & (mas/yr) &  & & & & &   \\
\hline
\endfirsthead
\caption{continued.}\\
\hline\hline
ID Gaia EDR3 &  RA & DEC & $\varpi$ $\pm$ $\varpi_e$ & $\mu_\alpha$ $\pm$ $\mu_{\alpha_e}$ & $\mu_\delta$ $\pm$ $\mu_{\delta_e}$ & RUWE & P$_A$ & P$_B$  & P$_E$ & P$_F$  & Subcluster \\
& (deg) & (deg)  & (mas)  & (mas/yr) & (mas/yr) &  & & & & &   \\
\hline
\endhead
\hline
\endfoot
\hline
\endlastfoot
2203402614670559232 & 326.063135 & 59.360206 & 1.112 $\pm$ 0.016 & -2.230 $\pm$ 0.019 & -4.181 $\pm$ 0.016 & 0.975 & 0.95 & ---- & ---- & ---- & A \\ 
2203398697660400768 & 326.168794 & 59.323927 & 1.069 $\pm$ 0.031 & -2.507 $\pm$ 0.038 & -4.368 $\pm$ 0.030 & 1.027 & 0.95 & ---- & ---- & ---- & A \\ 
2203412888232208640 & 325.827953 & 59.366154 & 1.123 $\pm$ 0.013 & -2.880 $\pm$ 0.015 & -2.966 $\pm$ 0.014 & 1.094 & ---- & 0.95 & ---- & ---- & B \\ 
2203406188083035776 & 325.664257 & 59.244763 & 1.046 $\pm$ 0.015 & -2.560 $\pm$ 0.017 & -2.837 $\pm$ 0.015 & 1.169 & ---- & 0.95 & ---- & ---- & B \\
2202645188605787648 & 326.367554 & 59.141696 & 1.084 $\pm$ 0.055 & -1.320 $\pm$ 0.069 & -3.031 $\pm$ 0.060 & 1.232 & ---- & ---- & 0.95 & ---- & E \\ 
2203405565302940416 & 325.580452 & 59.249619 & 1.112 $\pm$ 0.011 & -1.309 $\pm$ 0.013 & -3.185 $\pm$ 0.011 & 0.906 & ---- & ---- & 0.95 & ---- & E \\
2179429344178082048 & 324.761669 & 59.134190 & 1.042 $\pm$ 0.039 & -3.645 $\pm$ 0.047 & -3.731 $\pm$ 0.052 & 0.988 & ---- & ---- & ---- & 0.95 & F \\ 
2178408683842112896 & 323.753814 & 57.204424 & 1.160 $\pm$ 0.017 & -3.513 $\pm$ 0.019 & -3.944 $\pm$ 0.018 & 1.114 & ---- & ---- & ---- & 0.95 & F \\
2178328355081854976 & 326.554329 & 57.902130 & 1.104 $\pm$ 0.080 & -1.558 $\pm$ 0.087 & -3.230 $\pm$ 0.090 & 0.998 & ---- & 0.34 & 0.61 & ---- & B/E \\ 
2179397355261072768 & 324.037597 & 58.792397 & 1.095 $\pm$ 0.030 & -1.543 $\pm$ 0.034 & -3.209 $\pm$ 0.034 & 1.055 & ---- & 0.38 & 0.57 & ---- & B/E \\ 
\end{longtable}
\tablefoot{Column 1: ID Gaia EDR3. Columns 2, 3, 4, 5, and 6 are the astrometric parameters of each source, position, parallax, and proper motion, along with their errors. Column 7: RUWE value. Columns 8 to 11 show the probability for each source belonging to one of the subclusters, or in some cases, to two subclusters. Column 12: The subcluster to which the source belongs. The complete version of 334 new members is available in electronic format via CDS.}
}
\end{landscape}
}
\section{Three color images of the BRCs} \label{appe:3colorimages}

In this appendix, we include some detailed images of the star-cloud interactions discussed in Section \ref{subsec:feedbackBRCs}. Figures \ref{fig:F1}, \ref{fig:F2}, and \ref{fig:F3} show evidence of feedback in the BRCs IC1396A, IC1396N, and IC1396B, respectively.

\begin{figure*}
    \centering
    \begin{tabular}{c}
    \includegraphics[height=0.58\textwidth]{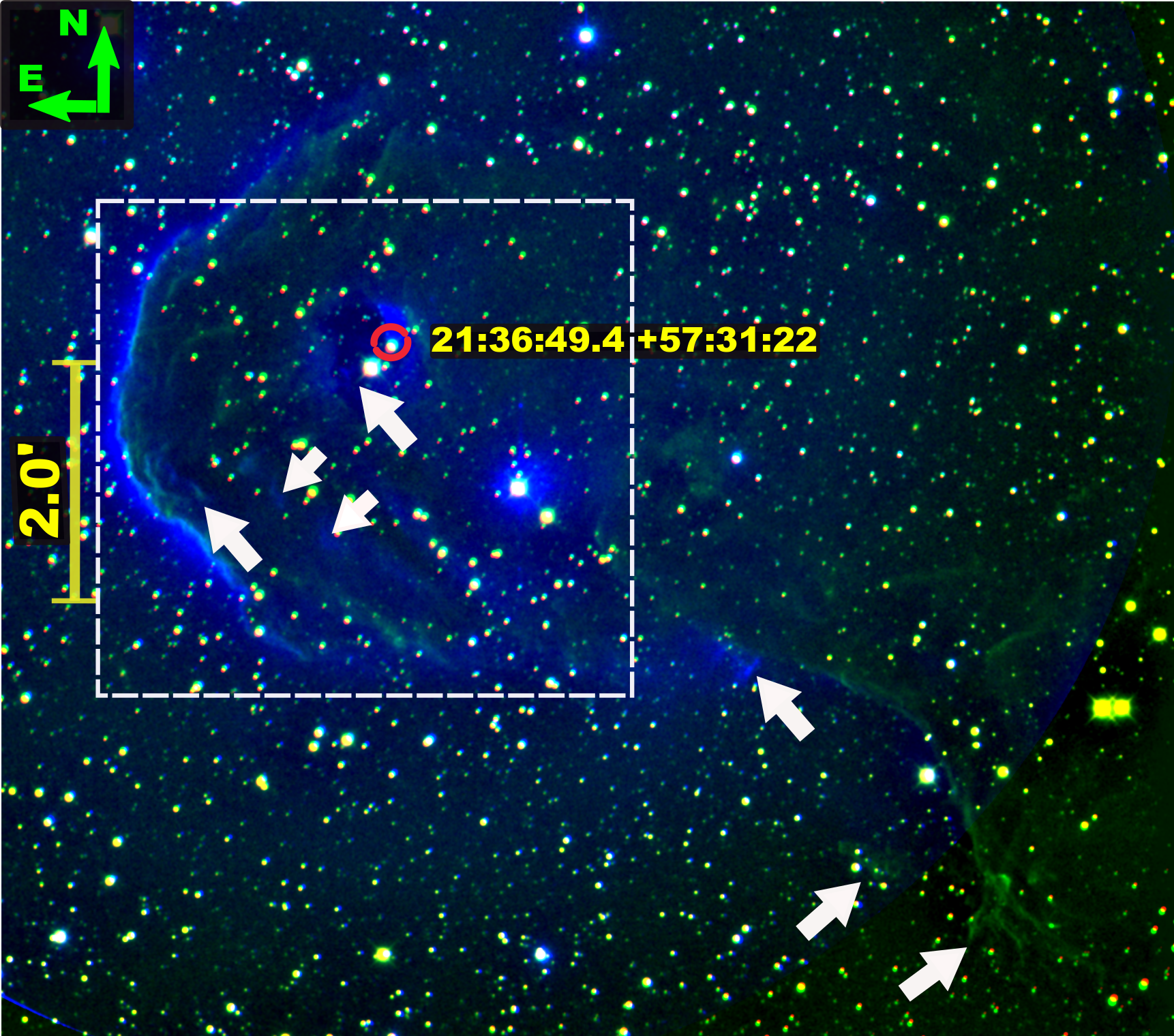}\\
    \includegraphics[width=0.66\textwidth]{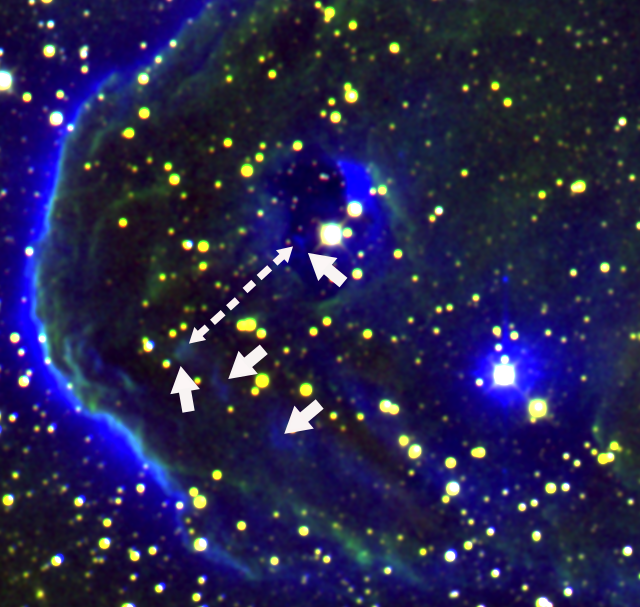}\\
    \end{tabular}
    \caption{Three-color images (Ks, H$_2$, and S[II] as red, green, and blue, respectively) of the BRCs. This figure shows IC1396A (top, zoomed on the bottom). The white dotted-line box is zoomed at the bottom. The shocks and knots are visible in H$_2$ and S[II] as green and blue structures and marked with white arrows. Dotted white arrows show the trajectory of jets following the shocks or knots (see text section \ref{subsec:feedbackBRCs}). The remaining globules are shown in Figures \ref{fig:F2} and \ref{fig:F3}}
    \label{fig:F1}
\end{figure*}

\begin{figure*}
    \centering
    \begin{tabular}{c}
    \includegraphics[width=0.76\textwidth]{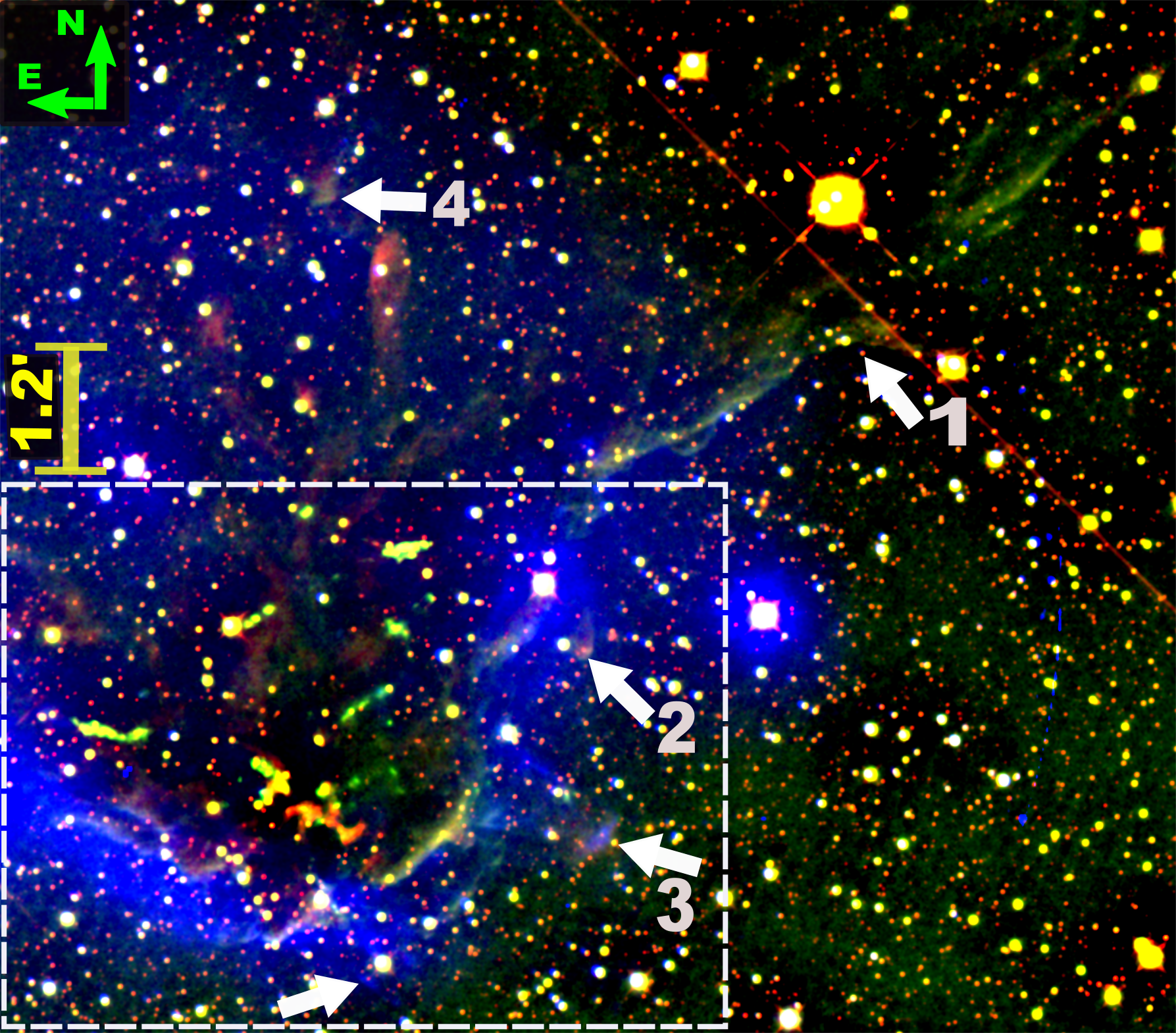}\\
    \includegraphics[height=0.6\textwidth]{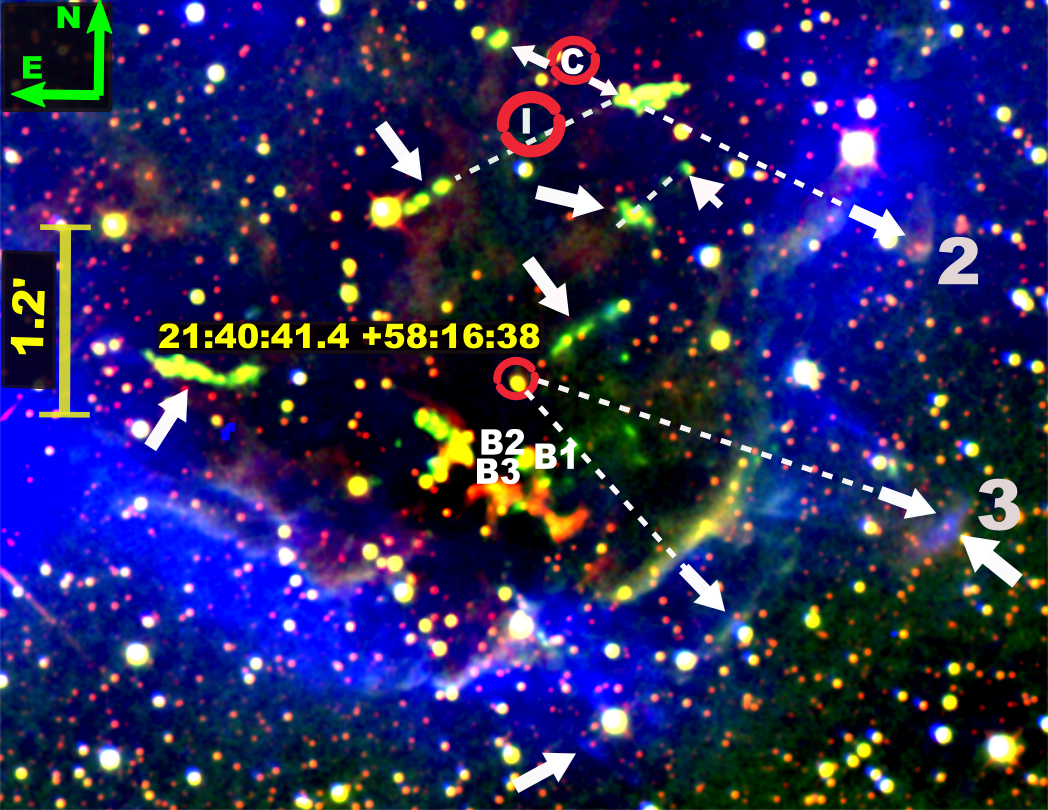}\\
    \end{tabular}
    \caption{continued. IC1396N (top, zoomed on the bottom). The strong shocks are numbered from 1-4. Sources B1, B2, and B3 from \citet{Beltran2009} are located in the middle of three chains of knots. Red circles enclose sources C, I, and 331A (21:40:41.4 +58:16:38).}
    \label{fig:F2}
\end{figure*}


\begin{figure*}
    \centering
    \begin{tabular}{c}
    \includegraphics[width=0.9\textwidth]{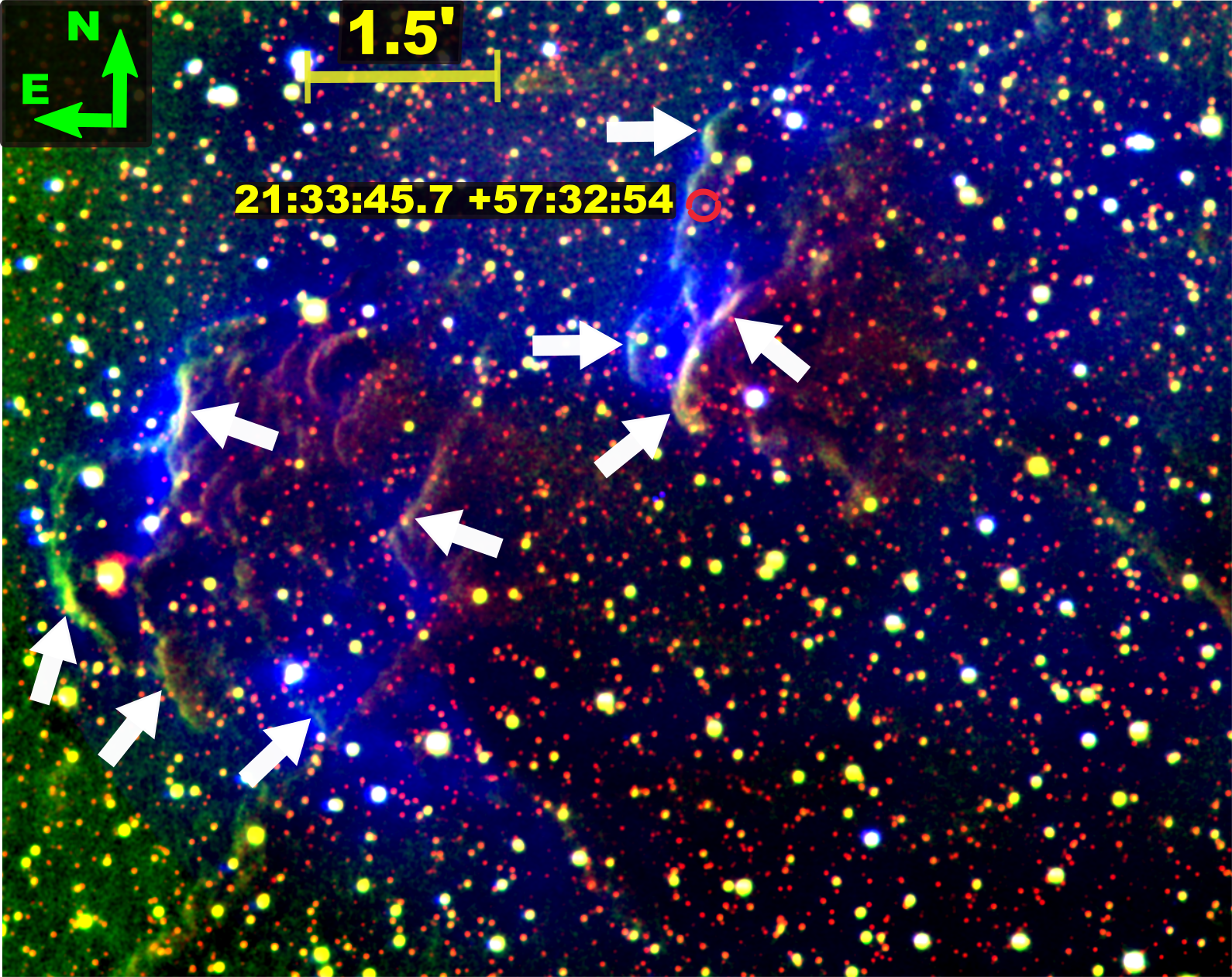}\\
    \end{tabular}
    \caption{continued. Shocked material on the small irregular edges of IC1396B. A source is labeled as a coordinate reference.}
    \label{fig:F3}
\end{figure*}


\end{appendix}

\end{document}